\documentclass[12pt]{article}%
\usepackage{float}
\usepackage{epsfig, graphicx}
\usepackage{amsmath}
\usepackage{graphicx}
\usepackage{amsfonts}
\usepackage{amssymb}%
\newtheorem{Defi}{Definition}
\newtheorem{theorem}{Theorem}
\newtheorem{lem}{Lemma}

\newtheorem{prop}{Proposition}
\newtheorem{coro}{Corollary}
\font\bb=msbm10 at 12pt

\def\QED{\quad\hbox{\hskip 4pt\vrule width 5pt height 6pt depth 1.5pt}}

\font\bb=msbm10 at 12pt

\def\QED{\quad\hbox{\hskip 4pt\vrule width 5pt height 6pt depth 1.5pt}}

\font\bb=msbm10 at 12pt

\begin{document}

\title{Singular perturbation for the first eigenfunction and blow up analysis}
\author{David Holcman \thanks{Permanent address: Weizmann Institute of
Science, department of Mathematic, Rehovot 76100, Israel. D.H is incumbent of the Madeleine Hass Russel Chair.}  \and Ivan Kupka\thanks{ Universit\'{e} Paris VI, department of Mathematic, 175 rue du Chevaleret 75013 Paris, France.}}
\date{}
\maketitle

\begin{abstract}
On a compact Riemannian manifold $(V_{m},g)$, we consider the second order
positive operator $L_{\epsilon} = \epsilon\Delta_{g} +(b,\nabla) +c$, where
$-\Delta_{g}$ is the Laplace-Beltrami operator and $b$ is a Morse-Smale (MS)
field, $\epsilon$ a small parameter. We study the measures which are the
limits of the normalized first eigenfunctions of $L_{\epsilon}$ as $\epsilon$
goes to the zero.

In the case of a general MS field $b$, such a limit measures is the sum of a
linear combination of Dirac measures located at the singular point of $b$ and
a linear combination of measures supported by the limit cycles of $b$.

When $b$ is a MS-gradient vector field, we use a Blow-up analysis to determine
how the sequence concentrates on the critical point set. We prove that the set
of critical points that a critical point belongs to the support of a limit
measure only if the Topological Pressure defined by a variational problem (see
\cite{Kifer90}) is achieved there. Also if a sequence converges to a measure
in such a way that every critical points is a limit point of global maxima of
the eigenfunction, then we can compute the weight of a limit measure.This
result provides a link between the limits of the first eigenvalues and the
associated eigenfunctions . We give an interpretation of this result in term
of the movement of a Brownian particle driven by a field and subjected to a
potential well, in the small noise limit.

\end{abstract}
\tableofcontents

\section{Introduction}

Let $(V_{m},g)$ denote a compact Riemannian manifold of dimension $m\geq2$,
with no boundary. $-\Delta_{g}$ denotes the associated Laplace-Beltrami
operator . In this paper we consider a second order elliptic operator
depending on the parameter $\varepsilon>0$,%
\begin{equation}
L_{\epsilon}=\epsilon\Delta_{g}+\theta(b)+c \label{najaf},%
\end{equation}
 where $b$ denotes a $C^{\infty}$vector field on $V,c$ a strictly positive
C$^{\infty}$ function on $V_{m}$ and $\theta(b)$ the Lie derivative operator
associated to $b$: $\theta(b)u=$ $du(b)$ for any function $u$ on $V$.

$L_{\epsilon}$ is a positive operator to which the Krein-Rutman \cite{pinsky}
theorem can be applied. Hence the smallest eigenvalue $\lambda_{\epsilon}$ of
$L_{\epsilon}$ is simple and strictly positive. The associated eigenspace is
generated by a strictly positive function $u_{\epsilon}$, normalized in
$L_{2}(V_{m})$. The behavior of $\lambda_{\epsilon}$ as $\epsilon$ goes to
zero has been extensively studied \cite{Kifer80,Kifer88,Kifer90,DF,DF1}.

In the case when the $\omega$-set is a disjoint union of compact invariant
hyperbolic set, under a mild additional assumption Y. Kifer has proved that
the limit of $\lambda_{\epsilon}$ as $\epsilon$ goes to zero is the
topological pressure (TP), associated to the flow of $b$ and the potential $c$
(for definitions and details see \cite{Kifer90}).

Unfortunately, much less is known about the weak limits of the eigenfunctions
$u_{\epsilon}$, as $\epsilon$ goes to zero. By weak limits, we mean the limits
of the measures $u_{\epsilon}^{2}dvol_{g}$ in the weak topology of measures.
Past results include the case when $V_{m}$ is a bounded domain of
$\mathbb{R}^{n}$, the limit sets of the vector field $b$ is reduced to a
single point, $\lambda_{\epsilon}$ and $u_{\epsilon}$ are related to the zero
Dirichlet boundary condition.

When this singular point is attractive, $u_{\epsilon}$ as $\epsilon$ goes to
zero, converges to a Dirac measure supported by the attractor. When it is
repulsive $u_{\epsilon}$ converges to a constant on every compact sub-domain
of the domain of definition (see \cite{DEF,DF}). When the attractive set
consists of a single limit cycle, the first eigenvalue converges to the
average of the potential $c$ of the PDE, along the cycle (see \cite{H} ),
$\lim_{\epsilon\rightarrow0}\lambda_{\epsilon}=\frac{\int_{0}^{T}%
c(x_{0}(t))dt}{T}$, where $x_{0}$ parametrizes the limit cycle of period T,
the proof uses a stochastic approach. It is interesting to note that this
approach and the deterministic methods give complementary results.

When $b=0$,the equation (\ref{najaf}) reduces to:%
\[
\epsilon\Delta_{g}u_{\varepsilon}+cu_{\varepsilon}=\lambda_{\varepsilon
}u_{\varepsilon}%
\]
The limits of the $u_{\epsilon}^{2}dvol_{g}$ (normalized by:$\int
_{V}u_{\varepsilon}^{2}dvol_{g}=1$) has been studied for example in
(\cite{Simon1,Si,Helffer}) and is known as the semi-classical limit. In
particular when the potential $c$ has a double well in $P_{1}$ and $P_{2}$ (
that is an absolute non-degenerate minimum at each points), it is well known
that the limits of the measures $u_{\epsilon}^{2}dvol_{g}$ as $\varepsilon$
goes to $0$  can concentrate on those points only and in the distribution
sense,
as $\epsilon$ goes to $0$
\[
u_{\epsilon}^{2}dvol_{g}\rightarrow c_{1}\delta_{P_{1}}+c_{2}\delta_{P_{2}},
\]
where $c_{1}+c_{2}=1$. As far as we know, nobody has addressed the question of
computing explicitly the coefficients $c_{1}$ and $c_{2}$ in general. Of
course, under additional assumptions, for example, when the $u_{\epsilon}$ are
invariant by a group of isometries, $c_{1}=c_{2}=\frac{1}{2}$. It is not clear
at present time whether or not those coefficients are unique. We will see here
that those coefficients depend only on the Hessian of the potential at the
points $P_{1}$, $P_{2}$ and on the limit of the ratio (called the modulating
ratio) of the local maximum to the global maximum of the $u_{\epsilon}$, but
they do not depend on anything else.

Sometimes, as explained in \cite{Simon1}, one of the coefficient is zero and
the limits of the measures $u_{\epsilon}^{2}dvol_{g}$ concentrate on the
remaining point. In that case, we use the terminology of \cite{Simon1} and say
that the degeneracy is removed. The degeneracy can be removed by looking at
the expansion of $\lambda_{\epsilon}$ in terms of $\epsilon$ in the following
way. $\lambda_{\epsilon}$ depends on the values of the potential and of the
derivatives of the metric tensor at the minimum points. The Taylor expansions
of $\lambda_{\epsilon}$ have to be the same to each order at $P_{1}$ and
$P_{2}$.  If this is not the case then only one point will be charged. In this
paper we compute the necessary conditions up to order 4, in order that the
degeneracy can be removed in terms of the geometry.  These conditions narrow
down the set of points which can be charged (\ i.e.the set of points where the
coefficient $c_{k}$ is strictly positive).

We conjecture that the coefficients $c_{k}$ are unique and when the degeneracy
cannot be removed they are charged by a global maximum sequence (the modulating 
ratio is equal to 1).

When $b$ is a gradient $b= \nabla \phi$, the limits of the measures $u_{\epsilon}^{2}dvol_{g}$
as $\varepsilon$ goes to $0$ can concentrate on the critical points of $b$
only. The concentration results are proved for the normalized measures
\begin{equation} \label{newmeasure}
\frac{e^{-\frac{\phi}{\epsilon}}u_{\epsilon}^{2}dvol_{g}}{\int_{V_m} e ^{-\frac{\phi}{\epsilon}}u_{\epsilon}^{2}dvol_{g}   }
\end{equation}
and we shall see how the potential $c$ interacts with the field $b$ to select the points that will be charged. This set is the set of points where the Topological Pressure attains its minimum (see \cite{Kifer90}). 

When $b$ is a not necessarily the gradient of function, very few results are known about the behavior of the measure $u_{\epsilon}^{2}dvol_{g}$ as $\epsilon$ goes to zero, because it not clear by which function if any, the function $\phi$ should be replaced in expression \ref{newmeasure}. 
Recall that the operator $L_{\epsilon}$ with general drifts have not been considered before, especially in the context of Quantum Mechanics, because it has an interpretation only in the context of diffusion. Actually, on the striking results in this paper is that $\phi$ can be replaced by any global Lyapunov function $L$ associated to the field $b$. Using a weight which is a Gaussian in the Lyapunov function, is a crucial input into the problem because it enables us to "filter" the limits in order to get results about the concentration.  Using a Lyapunov type function, as  presented in section \ref{general}, is by far more general than using a solution of the Hamilton-Jacobi equation, 
\begin{equation}
|\nabla L|^2+(b,\nabla L) =0,
\end{equation}
as it is currently used in the formal expansion of the WKB theory. Another striking result is that the possible limit measures are supported by specific limit cycles of the field, which are not necessarily the attractors. On a compact Riemannian manifold, the existence of a solution to the Hamilton-Jacobi equation is not always guarrante to be smooth, while in appendix 2, we prove the existence of a smooth Lyapunov function. In this context, we prove that the measure 
\begin{equation} \label{newmeasure1}
\frac{e^{-\frac{L}{\epsilon}}u_{\epsilon}^{2}dvol_{g}}{\int_{V_m}  e^{-\frac{L}{\epsilon}}u_{\epsilon}^{2}dvol_{g}   }
\end{equation}
can concentrate on the limit cycle of a Morse-Smale dynamical system, without any additional assumptions. Finally, we give explicit results about the decay of the eigenfunction sequence near the concentration set and analyze the influence of the Riemannian geometry.\par
\bigskip

{\bf Our main results in this paper are:}

\begin{itemize}
\item In theorem \ref{thfdtpr} of section \ref{general}, we prove 
that the limits of the normalized eigenfunctions $u_{\epsilon}$ as $\epsilon$ 
tends to zero, are measures concentrated on the limit sets of a Morse-Smale field $b$.
The possible limit measures are supported by specific limit cycles of the field, 
which are not necessarily the attractors. 

\item In theorem \ref{thnf} we prove that the blown up function is the
standard solution of the Harmonic Oscillator. This lead us to theorem
\ref{th4} to give a precise expression of the coefficient of the limit
measures and to characterize the set of minimum of $c$ which are in the
support of the limit measures. As a byproduct of the analysis, we obtain some 
estimates of the velocity at which the sequence of local maximum $P_{\epsilon}$ 
converges to a critical point of the potential. 

\item In theorem\ref{Th-hk copy(1)}, we give a geometric interpretation of the
coefficients of the expansion of $\lambda_{\epsilon}$ in terms of the power of
$\sqrt{\epsilon}$. These coefficients are invariant of the couple (Riemannian
metric, potential).
The value of the second term of the expansion of $\lambda_{\varepsilon}$ in
powers of $\sqrt{\varepsilon}$ is computed, using minimax procedures, but this
value has to be compared to the one obtained in \cite{Helffer2}, based on the
WKB formula. The result obtained here are based on the variational approach and seems 
to lead to results that have to be compare with the results obtained by the formal WKB expansion of the first eigenfunction.

\item In theorem \ref{th-final}, we study the case where the field b is a
gradient of a Morse-Smale function. We prove that the concentration of the
first eigenfunction occurs at the critical points of $b$ where the Topological
Pressure is attained. In addition, the weights of the limit measures is given
under specific assumptions.
\end{itemize}

{\bf Remarks.}

Some of the results presented here complete and extend also some previous work in analysis \cite{K,K1,DEF,DF,DF1}. In addition, at the time the results of this paper were announced \cite{HK1}, we were not aware of any reports about the fine selection by the limit of the eigenfunction sequence of a subset of critical points of the potential. The detail of this selection process is presented in the subsection \ref{expanse} and \ref{dege} and should clarify how the metric and the potential are involved. This results extend in particular the work of \cite{Si,Si2,Simon1}.


\subsection{Notations}

 ($x^{1}$,...,$x^{m}$):$U\longrightarrow\mathbb{R}$, is a coordinate patch
on $V$ and $f$ a function on $V$:%

\begin{align*}
d_{g}  &  :V\times V\text{---%
$>$%
}\mathbb{R}_{+}:=\text{ distance associated to }g\\
(,)_{g}  &  :=\text{ scalar product associated to g}\\
\exp_{x}  &  :T_{x}V\longmapsto V\text{:=exponential map of \ }g\text{ with
pole }x\\
vol_{g}  &  :=\text{volume measure associated to }g\\
L^{2}(E)  &  :=L^{2}\text{ space associated to vol}_{g}\text{ on the subset
}E\text{ of }V\\
L^{2}  &  \text{:=}L^{2}(V)\\
\Delta_{g}  &  :=\text{ negative Laplacian associated to the metric }\\
b  &  :=\text{ vector field on }V\\
\theta(b)  &  :=\text{ Lie derivation operator associated to }b\\
\nabla &  :=\text{ gradient associated to }g\\
M(V)  &  :=\text{ space of all probability measures on }V\\
C^{\infty}\text{topology}  &  \text{:=uniform convergence of all the
derivatives on compact sets} \\
g  &  :=\sum_{ij=1}^{m}g_{ij}dx_{i}dx_{j}\\
\Delta_{g}  &  :=-\frac{1}{\sqrt{\det(g)}}\sum_{ij=1}^{m}\frac{\partial
}{\partial x_{i}}\sqrt{\det(g)}g^{ij}\frac{\partial}{\partial x_{j}}\\
g^{ij}  &  :=\text{ inverse matrix of }g_{ij}\\
\det(g)  &  :=\text{det (}g_{ij})\\
\Gamma_{ij}^{k}  &  :=\text{Christoffel symbols of }g_{ij}\\
\Gamma_{ij}^{k}  &  =\frac{1}{2}g^{kl}\left(  \frac{\partial g_{il}}{\partial
x_{j}}+\frac{\partial g_{jl}}{\partial x_{i}}-\frac{\partial g_{ij}}{\partial
x_{l}}\right) \\
R_{ijk\cdot}^{\cdot\cdot\cdot l}  &  :=\frac{\partial\Gamma_{jk}^{l}}{\partial
x_{i}}-\frac{\partial\Gamma_{ik}^{l}}{\partial x_{j}}+\sum_{n=1}^{m}\left[
\Gamma_{in}^{l}\Gamma_{jk}^{n}-\Gamma_{jn}^{l}\Gamma_{ik}^{n}\right] \\
R_{ijkl}  &  :=\sum_{n=1}^{m}g_{\ln}R_{ijk\cdot}^{\cdot\cdot\cdot n}\\
Ric_{kl}  &  :=\sum_{j=1}^{m}R_{jklj}\\
R  &  :=\sum_{j=1}^{m}Ric_{jj}=\sum_{i,j=1}^{m}R_{ijji}\\
\theta(b)  &  :=\sum_{i=1}^{m}b^{i}\frac{\partial}{\partial x^{i}}\\
dx^{i}(\nabla f)  &  :=\sum_{i=1}^{m}g^{ij}\frac{\partial f}{\partial x^{j}%
}\text{,}1\leq i\leq m
\end{align*}

\subparagraph{\bigskip}
\subsection{Normal coordinates}
For each P$\in$C$_{\min}$,we choose a normal coordinate system ($x_{1}%
$,...$x_{m}$):$U_{P}\longrightarrow\mathbb{R}$, centered at P, defined on a
domain U$_{P}$ such that:
\begin{enumerate}
\item x$_{1}\times$...$\times$x$_{m}(U_{P})$ contains the closed ball
B$_{P}(\delta)$ centered at P and having radius $\delta>0.$
\item for all i,j, 1$\leq$ i,j$\leq$ m, $\frac{\partial^{2}c}{\partial
x_{i}\partial x_{j}}(P)=\lambda_{i}(P)\delta_{ij}$.
\item U$_{P}\cap$U$_{Q}=\varnothing$ for all P,Q$\in C_{\min}$,P$\neq$Q.
\end{enumerate}
In the following we will identifie U$_{P}$ with the open neighborhood of O in
$\mathbb{R}^{m}$,$x_{1}\times$...$\times x_{m}(U_{P})$. For r such that
B$_{P}%
$(r)$\subset$U$_{P}$,B$_{P}$(r) will denote both the geodesic ball centered at
P and of radius r in V or its image by the mapping $x_{1}\times$...$\times
x_{m}.$ For the sake of streamlining the notations, we shall commit the abuse,
when working with the coordinate system ($x_{1}$,...$x_{m}$):$U_{P}%
\longrightarrow\mathbb{R},$ of denoting by $B_{P}$(r) the ball of center O and
radius r whatever the value of r($\geq0$ of course).

\bigskip Now a few words about the blow-up procedure. On a manifold $V$ let
$P$
be a point and a chart ($U$,$x_{1}$,...$x_{m})$ of V centered at P: $x_{i}%
$(P)=0, $1\leq$i$\leq m$. The blow-up of power t
$>$%
0\ associated to $P$ and the chart ($U$,$x_{1}$,...$x_{m})$ is the
diffeomorphism $Bl_{t}:U\rightarrow\mathbb{R}^{m}$, $Bl_{t}$($Q$)=($\frac
{1}{t}$x$_{1}(Q),...,\frac{1}{t}x_{m}(Q)$). All functions, tensors,
differential operators can be transported to the open subset $Bl_{t}$(U).
Suitably normalized by a power of t, they will have limits when t goes to 0
which will be defined on $\mathbb{R}^{m}$. These limits contain a trove of
information about the behaviour of the original objects in the neighborhood
(more precisely in the infinitesimal neighborhood) of $P$. To simplify the
notations we shall write:$\frac{1}{t}Q$ instead of $Bl_{t}$($Q$), $\frac{1}%
{t}A$ instead of $Bl_{t}$($A$) if \ $A$ is a subset of $U$ and so on.

\bigskip

Recall that a field $b$ is Morse-Smale MS if : (i)the recurrent set of $b$
consists of a finite number of hyperbolic points and periodic orbits (ii) each
pair of stable or unstable manifolds of these points or orbits intersect
transversally. If moreover $b$ is a gradient of a function with respect to the
metric $g$, $b$ will be called a MS gradient field.\par


\subsection{The self-adjoint case}

In the self adjoint case the vector field $b$ is zero. This assumption
simplifies the problem because it can be handled by variational methods.
Theorem 1 below seems well-known but we could not find a proof for Riemannian
manifolds, using deterministic techniques in the literature. Hence as a
starting point, we provide a simple one here.
\bigskip

\subsubsection*{Notations.}

\noindent For each P$\in$C$_{\min}$,we choose a normal coordinate system
($x_{1}%
$,...$x_{m}$):$U_{P}\longrightarrow\mathbb{R}$ , centered at P, defined on a
domain $U_{P}$ such that:
\begin{enumerate}
\item $x_{1}\times$...$\times x_{m}(U_{P})$ contains the closed ball B$_{P}%
(\delta)$ centered at P and having radius $\delta>0.$
\item for all i,j, 1$\leq$ i,j$\leq$ m, $\frac{\partial^{2}c}{\partial
x_{i}\partial x_{j}}(P)=\lambda_{i}(P)\delta_{ij}$.
\item $U_{P}\cap U_{Q}=\varnothing$ for all $P,Q\in C_{\min}$,$P\neq Q$.
\end{enumerate}
In the following we will identifie $U_{P}$ with the open neighborhood of O in
$\mathbb{R}^{m}$, $x_{1}\times$...$\times x_{m}(U_{P})$. For r such that
$B_{P}%
$(r)$\subset U_{P}$,  $B_{P}$(r) will denote both the geodesic ball centered
at
P and of radius r in V or its image by the mapping $x_{1}\times$...$\times
x_{m}.$ For the sake of streamlining the notations,  we shall commit the abuse
while working with the coordinate system (x$_{1}$, ...x$_{m}$):$U_{P}%
\longrightarrow\mathbb{R}$, of denoting by $B_{P}$(r) the ball of center O and
radius r whatever the value of r($\geq0$ of course).
\begin{theorem}
\label{Potential} Consider the first eigenvalue problem for the operator
$\Delta_{g}+c$ where $c$ is a function with a finite set of minimum
points, C$_{\min}$,  which are not degenerate (in the sense of Morse)$.$
Assume
that the first eigenvalue of the operator $\lambda_{\varepsilon}$ is
positive. $\lambda_{\varepsilon}$ has the following variational expression:%
\[
\lambda_{\epsilon}=\inf_{u\in H^{1}(V)-\{0\}}\frac{\int_{V}\left[
\epsilon{||\nabla u||}_{g}^{2}+cu^{2}\right]  dvol_{g}{\ }}{\int_{V}%
u^{2}dvol_{g}}%
\]
Then,  when $\epsilon$ converges to zero,  $\lambda_{\epsilon}$ converges to
the
minimum of the function $c$ and the set of \ weak limits,  when $\epsilon$
goes
to zero,  of the family of measures $\frac{u_{\epsilon}^{2}dvol_{g}}{\int
_{V}u_{\epsilon}^{2}dvol_{g}}$ defined by the positive solutions $u_{\epsilon
}$ of the PDE,
\begin{equation}
\epsilon\Delta_{g}u_{\epsilon}+cu_{\epsilon}=\lambda_{\epsilon}u_{\epsilon
}\text{ on }V \label{edpfdt}%
\end{equation}
is contained in the simplex
\begin{equation}
M=\{\nu=\sum\left(  \gamma_{P}\delta_{P}|\text{
P}\in C_{\min}\right)  ||\sum\gamma_{P}=1, \gamma_{P}\geq0\}
\end{equation}
 of all probability measures with support in the finite set C$_{\min}$ where
$\delta_{P}$ denotes the Dirac measure at the point $P$.
\end{theorem}
\bigskip

\noindent \textbf{ Remarks. } $u_{\epsilon}$ is uniquely defined up to a
multiplicative
constant by the Krein-Rutman theorem (see \cite{Kra}).

In the following proofs,  over and over,  we will chose appropriate
sub-sequences of \{$u_{\epsilon}$,  $\epsilon>0$\}, \{$\lambda_{\varepsilon
}|\varepsilon>0\}$ and so on,  without saying so explicitly: In order to keep
the notations simple we will write $u_{\epsilon}$,  $\lambda_{\varepsilon},
...$
instead of a sequence $(u_{\epsilon_{k}}|k=1..), (\lambda_{\epsilon_{k}%
}|k=1..)$.. \medskip\noindent

\textbf{Proof:} without restricting the generality\textbf{\ }we can assume
that
$c\geq0.$ \noindent To start note that $\lambda_{\epsilon}\geq\underset
{V}{\min}c\geq0.$ Using a constant as test function in the functional that
define the first eigenvalue,  we get:%
\[
\lambda_{\epsilon}\leq\frac{\int_{V}c\text{ }dvol_{g}}{vol_{g}(V)}\leq\sup
_{V}c.
\]
$\lambda_{\epsilon}$ is a decreasing function of $\epsilon$ because the
functional decreases with $\epsilon$. $u_{\epsilon}$ being bounded,  there
exists a sub-sequence which converges weakly and
\[
\forall\phi\in H^{1}(V)\, , \int_{V}[\epsilon(\Delta\phi)u_{\epsilon}+c\phi
u_{\epsilon}]=\lambda_{\epsilon}\int_{V}u_{\epsilon}\phi
\]
To obtain an upper estimate for $\lambda_{\epsilon}$,  consider the following
radial function defined on $V_{m}$ for $m>2$:
\begin{align}
\phi_{\mu}(r)=\frac{\mu^{(m-2)/2}}{(r^{2}+\mu^{2})^{(m-2)/2}}-\frac
{\mu^{(m-2)/2}}{(\delta^{2}+\mu^{2})^{(m-2)/2}}  &  \text{,  on \ }B_{P}%
(\delta)\nonumber\\
=0\,   &  , \text{on }V-B_{P}(\delta)
\end{align}
Then $\phi$ belongs to $H^{1}(V)$. \noindent We use this function in the
energy function $I(u)=\frac{\epsilon\int_{V}({\ \mid|\nabla u\mid|}_{g}%
^{2}+cu^{2})dvol_{g}}{\int_{V}u^{2}dvol_{g}}$. The following standard
computation gives an estimate of $I(\phi_{\mu})$:
\[
\epsilon\int_{V}{||\nabla\phi_{\mu}||}_{g}^{2}dvol_{g}=\epsilon(\omega
_{m-1}\int_{B_{\delta/\mu}(P)}r^{m-1}dr\frac{1}{(1+r^{2})^{m}}+o(\mu
))=\epsilon c(m)+o(\mu)\epsilon
\]
where the constant $c(m)$ depends only on the dimension n and $\omega_{m-1}$
is the volume of the unit (n-1)-sphere of $\mathbb{R}^{m}$,  $r=d(P, Q)$ is
the
geodesic distance and $\delta$ is less than the injectivity radius. We have to
evaluate the other quantities in the functional. Using the change of variable
$x=y\mu$,  we get:
\[
\int_{V}c\phi_{\mu}^{2}dvol_{g}=\int_{B_{\delta/\mu}(P)}a\phi_{\mu}%
^{2}dvol_{g}=\mu^{2}V(P)\omega_{m-1}J(\delta/\mu)+o(\mu),
\]
where
\[
J(\delta/\mu)=\int_{B_{\delta/\mu}(P)}(\frac{1}{(r^{2}+1)^{(m-2)/2}}-\frac
{1}{(\delta^{2}+1)^{(m-2)/2}})^{2}dvol_{g}.
\]
We have $J(\delta/\mu)=O(1)$ for $n\geq5$ and for n=4,  $J(\delta/\mu
)=O(\ln\delta/\mu)$
\[
\int_{B_{\delta/\mu}(P)}c\phi_{\mu}^{2}dvol_{g}=\int_{B_{\delta/\mu}(P)}%
c\mu^{2}(\frac{1}{(r^{2}+1)^{(m-2)/2}}-\frac{1}{(\delta^{2}+1)^{(m-2)/2}}%
)^{2}dvol_{g}%
\]
\noindent then after some computations:
\[
\int_{V}\phi_{\mu}^{2}dvol_{g}=\mu^{2}\omega_{m-1}J(\delta/\mu)+o(\mu).
\]
Taking $\mu=\epsilon^{1/3}$,  we get the expansion
\[
I(\phi_{\epsilon})=c(P)+o(1),
\]
where P is a minimal point of $c$.

In the 2-dimensional case,  we can consider the test function
\begin{align*}
\phi_{\mu}(r)=\frac{\mu^{p}}{(r^{2}+\mu^{2})^{p}}-\frac{\mu^{p}}{(\delta
^{2}+\mu^{2})^{p}}  &  \text{,  on }B_{P}(\delta)\\
&  =0\, \text{,  on }V-B_{P}(\delta)
\end{align*}
where $p$ satisfies $0<p\leq1/2$. The same computations as before,  taking
$\mu^{4}=\epsilon$,  gives the result. Finally
\begin{align*}
\underset{V}{\min}c  &  \leq\lambda_{\epsilon}\leq c(P)+o(1)\\
\lim_{\epsilon\rightarrow+\infty}\lambda_{\epsilon}  &  =\underset{V}{\min}c
\end{align*}
\bigskip

\noindent Using the energy equation,  we have $\lambda_{\epsilon}\geq
\epsilon\int_{V}{|[|\nabla u_{\epsilon}||}_{g}^{2}+\underset{V}{\min}c]$,
which forces $\epsilon\int_{V}{||\nabla u_{\epsilon}||}_{g}^{2}$ $dvol_{g}$ to
tend to zero as $\epsilon$ goes to zero:%
\begin{equation}
\underset{\varepsilon->0}{\lim}\epsilon\int_{V}{||\nabla u_{\epsilon}||}%
_{g}^{2}dvol_{g}=0 \label{inegra1}%
\end{equation}
Also for any $\phi\in$C$^{2}$(V) multiplying equation (\ref{edpfdt}) by the
function $\phi u_{\epsilon}$ and integrating by part gives :%
\begin{equation}
\int_{V}\left[  \epsilon{||\nabla u_{\epsilon}||}_{g}^{2}\phi+cu_{\epsilon
}^{2}\phi+\epsilon(\Delta\phi)\frac{u_{\epsilon}^{2}}{2}\right]
dvol_{g}=\lambda_{\epsilon}\int_{V}\phi u_{\epsilon}^{2}dvol_{g}
\label{intbypar}%
\end{equation}
where $\int_{V}u_{\epsilon}^{2}=1$. Because the first term and the last term
on the left hand-side are converging to zero (see \ref{inegra1}),  we obtain
that
\[
\lim_{\epsilon\rightarrow0}\int_{V}\phi u_{\epsilon}^{2}(c-\lambda_{\epsilon
})dvol_{g}=0.
\]
But:%
\[
0\leq\int_{V}(c-\underset{V}{\min}c)\phi u_{\varepsilon}^{2}dvol_{g}\leq
\int_{V}(c-\lambda_{\epsilon})\phi u_{\epsilon}^{2}dvol_{g}=0
\]
\begin{equation}
\underset{\varepsilon->0}{\lim}\int_{V}(c-\underset{V}{\min}c)\phi
u_{\varepsilon}^{2}=0 \label{ltwo1}%
\end{equation}
Moreover if $u$ denotes a weak limit of $u_{\epsilon}$ as $\epsilon$ goes to
zero, relation (\ref{ltwo1}) implies that $\forall\phi\in C^{1}(V)\, $:%
\begin{equation}
\int_{V}(c-\underset{V}{\min}c)\phi u^{2}dvol_{g}=0 \label{hun}%
\end{equation}
Hence if the set $\{P\in V|$ $c(P)=\underset{V}{\min}c\}$ is of measure 0,  we
conclude that $u=0$. Hence all weak limits of $u_{\epsilon}$ are zero and the
sequence $u_{\epsilon}$ concentrates,  as we shall see shortly.

Note that relation (\ref{ltwo1}) implies that for any $\psi\in C(V)$,  which
is
zero in a neighborhood of C$_{\min}$:%
\begin{equation}
\underset{\varepsilon->0}{\lim}\int_{V}\psi u_{\varepsilon}^{2}dvol_{g}=0
\label{null}%
\end{equation}
If A is a measurable subset of V such that $\overline{A}\cap C_{\min
}=\varnothing, $ $\overline{A}$ denoting the closure of A,  applying
(\ref{null}) to a positive continuous function $\psi$ with support disjoint
from C$_{\min}$ and $\psi\geq1$ on A we get:%
\[
\underset{\varepsilon->0}{\lim}\int_{A}u_{\varepsilon}^{2}dvol_{g}=0
\]
In fact,  we can prove that any sequence \{$\varepsilon_{n}|n\in\hbox{\bb
N}$\}
converging to 0,  contains a subsequence \{$\varepsilon_{n_{k}}|k\in\hbox{\bb
N}$\} such that the corresponding $u_{\epsilon_{n_{k}}}$ converges to a convex
sum of Dirac distributions located at the minimum points of $c$.We assume the
u$_{\varepsilon_{n}}$ normalized so that $\int_{V}u_{\epsilon_{n}}^{2}=1.$
Consider the following decomposition
\begin{equation}
\int_{V}\phi u_{\epsilon}^{2}dvol_{g}=\sum_{P\in C_{\min}}\int_{B_{P}(\delta
)}\left(  (\phi-\phi(P))+\phi(P)\right)  u_{\epsilon}^{2}dvol_{g}+\int
_{V-\cup_{P\in C_{\min}}B_{P}(\delta)}\phi u_{\epsilon}^{2}dvol_{g}
\label{cretin}%
\end{equation}
Relation (\ref{cretin})implies that:
\begin{equation}
\int_{V}\phi u_{\epsilon}^{2}dvol_{g}-\phi(P)\sum_{P\in C_{\min}}\int
_{B_{P}(\delta)}u_{\epsilon}^{2}dvol_{g}=\sum_{P\in C_{\min}}\int
_{B_{P}(\delta)}(\phi-\phi(P))u_{\epsilon}^{2}dvol_{g}+\int_{V-\cup_{P\in
C_{\min}}B_{P}(\delta)}\phi u_{\epsilon}^{2}dvol_{g} \label{idiot}%
\end{equation}
By the continuity of $\phi$, given an $\eta>0, $ one can find a
$\delta(\eta)>0$
such that
$\vert$%
$\phi(x)-\phi(P)|\leq\eta$ \ if x$\in B_{P}(\delta(\eta))$,  for all i. Hence
with N=card C$_{\min}$:%
\[
\left\vert \sum_{P\in C_{\min}}\int_{B_{P}(\delta(\eta))}(\phi-\phi
(P))u_{\epsilon}^{2}dvol_{g}\right\vert \leq N\eta
\]
Relation (\ref{null}) implies that:%
\[
\underset{\varepsilon->0}{\lim}\int_{V-\cup_{P\in C_{\min}}B_{P}(\delta
(\eta))}\phi u_{\epsilon}^{2}dvol_{g}=0
\]
After choosing a subsequence of \{$\varepsilon_{n}|n\in\hbox{\bb N}$\},  still
called \{$\varepsilon_{n}|n\in\hbox{\bb N}$\}, if necessary,  we can assume
that
all the limits $\underset{n->\infty}{\lim}$ $\int_{V}\phi u_{\epsilon_{n}}%
^{2}dvol_{g}$ ,  $\underset{n->\infty}{\lim}\int_{B_{P}(\delta(\eta
))}u_{\epsilon_{n}}^{2}dvol_{g}$, $P\in C_{\min}, $ exist.
Relation (\ref{idiot}) implies that after choosing a subsequence of:%
\begin{equation}
\underset{n->\infty}{\lim}\left\vert \int_{V}\phi u_{\epsilon_{n}}^{2}%
dvol_{g}-\phi(P)\sum_{i}\underset{n->\infty}{\lim}\int_{B_{P}(\delta(\eta
))}u_{\epsilon_{n}}^{2}dvol_{g}\right\vert \leq N\eta\label{escr}.%
\end{equation}
Now note that $\underset{n->\infty}{\lim}\int_{B_{P}(\delta)}u_{\epsilon_{n}%
}^{2}$ does not depend on $\delta.$Let $\delta_{1}, \delta_{2}, \delta_{1}%
\leq\delta_{2}$. Then
\[
\int_{B_{P}(\delta_{2})}u_{\epsilon_{n}}^{2}dvol_{g}=\int_{B_{P}(\delta_{1}%
)}u_{\epsilon_{n}}^{2}dvol_{g}+\int_{B_{P}(\delta_{2})-B_{P}(\delta_{1}%
)}u_{\epsilon_{n}}^{2}dvol_{g}%
\]
Relation (\ref{null}) implies that:%
\[
\underset{n->\infty}{\lim}\int_{B_{P}(\delta_{2})-B_{P}(\delta_{1}%
)}u_{\epsilon_{n}}^{2}dvol_{g}=0
\]
Hence if one of the limits $\underset{n->\infty}{\lim}\int_{B_{P_{i}}%
(\delta_{1})}u_{\epsilon_{n}}^{2}dvol_{g}$,  $\underset{n->\infty}{\lim}$
$\int_{B_{P_{i}}(\delta_{2})}u_{\epsilon_{n}}^{2}dvol_{g}$ exists,  so does
the
other and is equal to it. Set:%
\[
\underset{n->\infty}{\lim}\int_{B_{P}(\delta(\eta))}u_{\epsilon_{n}}%
^{2}dvol_{g}=\gamma_{P_{i}}%
\]
Relation (\ref{escr}) implies that for any $\eta>0$:%
\[
\left\vert \underset{n->\infty}{\lim}\int_{V}\phi u_{\epsilon_{n}}^{2}%
dvol_{g}-\sum_{i}\phi(P)\gamma_{P}\right\vert \leq N\eta
\]
\begin{subequations}
\begin{equation}
\underset{n->\infty}{\lim}\int_{V}\phi u_{\epsilon_{n}}^{2}dvol_{g}=\sum
_{i}\phi(P)\gamma_{P} \label{bar}%
\end{equation}
$\phi\in C(V)$ being arbitrary, (\ref{bar}) shows that the sequence of
measures
u$_{\varepsilon_{n}}$vol$_{g}$ converges weakly (in the measure sense) to the
measure $\sum_{P\in C_{\min}}\gamma_{P}\delta_{P}.$ Finally note that if we
apply (\ref{bar}) to the constant function 1, we get:
\end{subequations}
\[
\sum_{P\in C_{\min}}\gamma_{P}=1
\]
\medskip Now we prove that $\underset{V}{\sup}u_{\epsilon}$ diverges to
infinity. Because $\int_{V}u_{\epsilon}$ tends to zero,  if $\ $ $\underset
{V}{\sup}u_{\epsilon}$ were bounded,  then using the following inequality,  we
get a contradiction with%
\[
1=\int_{V}u_{\epsilon}^{2}\leq\ \underset{V}{\sup}u_{\epsilon}\int
_{V}u_{\epsilon},
\]
as the right-hand side would converge to zero. Let $P_{\epsilon}$ be a maximum
point of $u_{\epsilon}$. Because the manifold is compact,  it is possible to
find a subsequence of $P_{\epsilon}$ which converges to a point $P$.
centered at the point $P$.
$v_{\epsilon}(x)=\frac{u_{\epsilon}(\sqrt{\epsilon}x+P_{\epsilon})}{\underset{
V}{\sup }u_{\epsilon}}$ and $g_{\epsilon}$ denotes the rescaled \ metric,
then the
At a maximum point $P_{\epsilon}$,  using the maximum principle,
$c(P_{\epsilon
}) \leq\lambda_{\epsilon}$. Since $\lambda_{\epsilon}$ converges to the
minimum of $v$,  at the limit,  $c(P) \leq\min c$. This proves that P is a
minimum point. Using the fact that $P$ is a nondegenerate minimum point,
$d(P_{\epsilon}, P) \leq C \epsilon^{1/2}$. \quad\hbox{\hskip4pt\vrule width
5pt height 6pt depth 1.5pt}
\bigskip

\noindent We will use the following definition.

\begin{Defi}
The coefficient $\gamma_{P}$ is given by,  for all $\delta$ small enough,
\[
\gamma_{P}={\lim_{n\rightarrow\infty}}\int_{B_{P}(\delta)}u_{\epsilon_{n}}%
^{2}dvol_{g}%
\]
is called the concentration coefficient or the weight of the limit measure at
point $P\in C_{\min}$.
\end{Defi}
The coefficient  depends on the subsequence,  but not on $\delta$.
This coefficient characterizes the concentration measure at point $P$.
\subsection{The first eigenvalue problem for the gradient case}
We consider the limits of the first eigenfunctions as $\epsilon$ goes to zero
when the vector field $b$ is the gradient of a Morse function. We establish a
result similar to the one obtained in the last paragraph : when $\epsilon$
converges to zero the limits of first eigenfunctions in the weak topology of
measures concentrate at the critical points of the field $b$.

Consider a Morse function $\phi$ and the vector field,  $b=\nabla\phi$ and a
function $c$ chosen such that the eigenvalue $\lambda_{\epsilon}$ of the
operator $\epsilon\Delta_{g}+<b, \nabla.>+c$ is positive on the manifold. To
study the solutions of the PDE
\begin{equation}
\epsilon\Delta_{g}u_{\epsilon}+<b, \nabla u_{\epsilon}>+cu_{\epsilon}%
=\lambda_{\epsilon}u_{\epsilon}, \text{on }V \label{edpfdtpp}%
\end{equation}
we use the transformation $b=\nabla\phi=-2\epsilon\nabla\ln\psi_{\epsilon}$
(it is defined up to a constant) and consider the new variable $v_{\epsilon
}=u_{\epsilon}\psi_{\epsilon}$. Equation (\ref{edpfdtpp}) is transformed into
the following PDE where the first order term disappeared.
\[
\epsilon^{2}\Delta_{g}v_{\epsilon}+c_{\epsilon}v_{\epsilon}=\epsilon
\lambda_{\epsilon}v_{\epsilon}, \text{on }V
\]
where $c_{\epsilon}=c\epsilon+\frac{\epsilon\Delta\phi}{2}+\frac{(\nabla
\phi)^{2}}{4}$.

Using the theorem of the preliminary section,  we obtain the following results
:
\begin{prop}
Suppose that the following condition is satisfied: at the critical points $P$
of the function $\phi$,  $c(P)+\Delta\phi(P)/2\geq0$. Let $v_{\epsilon}$ be a
minimizer of the following variational problem
\[
\epsilon\lambda_{\epsilon}=\inf_{v\in H^{1}(V_{n})-\{0\}}\frac{\epsilon
^{2}\int_{V}[{\ ||\nabla v||}_{g}^{2}+c_{\epsilon}v^{2}]dvol_{g}}{\int
_{V}v^{2}dvol_{g}}%
\]
then $\lim_{\epsilon\rightarrow0}\epsilon\lambda_{\epsilon}=\underset{V}{\min
}||\nabla\phi||_{g}^{2}=0$. The weak limits of the normalized measures
$\frac{e^{-\phi/\epsilon}v_{\epsilon}^{2}dvol_{g}}{\int_{V}e^{-\phi/\epsilon
}v_{\epsilon}^{2}dvol_{g}}$ have their support in the set of critical points
of $\phi$. $\underset{V}{\sup}v_{\epsilon}$ tends to $+\infty$ as $\epsilon$
goes to zero.
\end{prop}

\noindent \textsc{Proof.} The proof is very similar to the proof of Theorem
\ref{Potential}. Considering the one parameter family of eigenfunctions,  we
obtain that:
\[
\lim_{\epsilon\rightarrow0}\frac{\int_{V}e^{-\phi/\epsilon}u_{\epsilon}%
^{2}c_{\epsilon}dvol_{g}}{\int_{V}e^{-\phi/\epsilon}u_{\epsilon}^{2}dvol_{g}%
}=\underset{V}{\min}||\nabla\phi||_{g}^{2}=0
\]
and for all function $\psi\in C(V)$ we have
\[
\lim_{\epsilon\rightarrow0}\frac{\int_{V}e^{-\phi/\epsilon}u_{\epsilon}%
^{2}\psi dvol_{g}}{\int_{V}e^{-\phi/\epsilon}u_{\epsilon}^{2}dvol_{g}}%
=\sum_{i=1}\left\{  \gamma_{P_{i}}\psi(P)|\text{ P}\in sing(b)\right\}
\]
where the concentration coefficient $\gamma_{P_{i}}$ is now defined by
\[
\gamma_{P_{i}}=\lim_{\epsilon\rightarrow0}\frac{\int_{B_{P_{i}}(\delta
)}e^{-\frac{\phi}{\varepsilon}}u_{\epsilon}^{2}dvol_{g}}{\int_{V}%
e^{-\frac{\phi}{\varepsilon}}u_{\epsilon}^{2}dvol_{g}},
\]
(the limit is independent of $\delta$). The measure $\frac{e^{-\phi/\epsilon
}u_{\epsilon}^{2}dvol_{g}}{\int_{V}e^{-\phi/\epsilon}u_{\epsilon}^{2}dvol_{g}%
}$ converges weakly to $\sum_{i=1}^{m}c_{i}^{2}\delta_{P_{i}}$ where $P_{i}$
are the critical points of the function $\phi$ or the zeros of the vector
field $b=\nabla\phi$. The proof follows exactly the same steps of the previous
theorem. \QED
\bigskip

\noindent {\bf Remark.}

\noindent The gradient case teaches two things: one is that the concentration
occurs on some specific sets, related to the vector field and not to $c$ and
second that the role of $c$ is to select the subset of concentration.

\subsection{The radial case : an example}

In this paragraph we give an example where the recurrent sets of the vector
field consists of a limit cycle and the sequence of eigenfunction concentrates
along this limit cycle. In fact,  we obtain in presence of radial symmetry a
uniform distribution for the limit.

Consider an annulus $A$ of $\mathbb{R}^{n}$ $(A=\{x\in\mathbb{R}%
^{n}|\, \, 1/2<||x||_{\mathbb{R}^{n}}<3/2\})$,  and the radial function
$u_{\epsilon}$,  solution of the partial differential equation
\begin{equation}
\epsilon\Delta_{g}u_{\epsilon}+<b, \nabla u_{\epsilon}>+cu_{\epsilon}%
=\lambda_{\epsilon}u_{\epsilon}, \text{ on }A
\end{equation}
\[
u_{\epsilon}=0, \text{ on the boundary }\partial A,
\]
where the field $b$ is given by :
\begin{align}
b_{r}  &  =(1-r)\nonumber\\
b_{\theta}  &  =1
\end{align}
and the function $a$ is radial and positive. The field $b$ has an attractive
limit cycle at $r=1$. The problem reduces to:
\begin{align}
\epsilon(-\partial_{rr}u_{\epsilon}-\frac{\partial_{r}u_{\epsilon}}{r})+  &
b_{r}.\partial_{r}u_{\epsilon}+au_{\epsilon}=\lambda_{\epsilon}u_{\epsilon
}, \text{on }A\label{edpr}\\
u_{\epsilon}  &  =0, \text{ on }\partial A\nonumber
\end{align}
$b_{r}$ is the gradient of a function of $r$. Hence the results of the
previous paragraph can be applied here. The presence of a boundary does not
invalidate these results because the limit cycle is an attractor. As
$\epsilon$ tends to zero,  $u_{\epsilon}$ tends to a limit entirely supported
by the limit cycle (see also Friedman \cite{DEF}).


\section{Blow up analysis with no vector fields\label{blow}}

In the next sections,  the limit measures are analyzed using a blow-up
procedure. We shall prove that as $\varepsilon$ goes to 0,  the eigenfunctions
blow up in the neighborhood of some points that are determined by the
potential $c$ and the vector field b. The speed with which these
eigenfunctions blow up can also be determined when b=0 or when b is a gradient
field,  using the Lyapunov functions associated to the field.

The results differ substantially in the two cases. It appears that the correct
scaling is not the same in the case when there is only a potential $c$ and the
case where there are a potential $c$ and a vector field $b$.

The general case,  where the field can have recurrent sets of integer
dimension
$n\geq1$ will be considered elsewhere. The concentration phenomenon is much
more complicated,  depending on the set and on the chose of Lyapunov function.
More important,  it cannot be studied by variational techniques,  see
\cite{HK3}.

In this section we determine exactly all the possible limits of the
eigenfunctions as $\varepsilon$ tends to 0,  when there is no field. The main
result says that the limit measure is concentrated on a subset of the minimum
point of the potential $c$. This limit set is useful in the study in the small
noise limit,  the movement of a random particle moving on a Riemannian
manifold
in the presence of a killing potential c \cite{HMS}.

Moreover,  we can explain the assumption 4,  p.93 made by B. Simon (\cite{Si})
to study the double-well potential problem when $\epsilon$ is small. The
blow-up method provides a method for the explicit computation of the
concentration near a bottom well. This generalizes also the results obtained
in part 9 of \cite{Dobro} about the concentration of the eigenfunctions in the
case of $\mathbb{R}^{m}$.

Let us recall the eigenfunction problem,
\begin{align}
\epsilon\Delta_{g}u_{\epsilon}+cu_{\epsilon}  &  =\lambda_{\epsilon
}u_{\epsilon}\label{edpV}\\
\int_{V_{n}}u_{\epsilon}^{2}dvol_{g}  &  =1, \nonumber
\end{align}
$c$ is a Morse function and $C_{\min}$,  denotes the subset of minimal points.
Recall the quotient
\[
Q_{\epsilon}(v)=\frac{\int_{V}\left[  \epsilon||\nabla u||_{g}^{2}%
+cu^{2}\right]  dvol_{g}}{\int_{V}u^{2}dvol_{g}}.
\]
We shall now state and prove the main theorems of section \ref{blow}. We
introduce some concepts which will be used in the proofs of these theorems.

Now a few words about the blow-up procedure. For each $P$ $\in C_{\min}$, we
choose a normal coordinate system ($x_{1}$, ...$x_{m}$):$U_{P}\longrightarrow
\mathbb{R}$ ,  centered at $P$,  defined on a domain $U_{P}$ such that:
\begin{enumerate}
\item $x_{1}\times$...$\times x_{m}(U_{P})$ contains the closed ball $B_{P}%
(\delta)$ centered at $P$ and having radius $\delta>0.$
\item for all i, j,  1$\leq$ i, j$\leq$ m,  $\frac{\partial^{2}c}{\partial
x_{i}\partial x_{j}}(P)=\lambda_{i}(P)\delta_{ij}$.
\item $U_{P}\cap U_{Q}=\varnothing$ for all $P, Q\in C_{\min}$, $P\neq Q$.
\end{enumerate}
In the following we will identifie $U_{P}$ with the open neighborhood of 0 in
$\mathbb{R}^{m}$, $x_{1}\times$...$\times x_{m}(U_{P})$. For r such that
$B_{P}%
$(r)$\subset U_{P}$, $B_{P}$(r) will denote both the geodesic ball centered at
$P$ and of radius r in $V$ or its image by the mapping $x_{1}\times$...$\times
x_{m}.$ For the sake of streamlining the notations,  we shall commit the abuse
while working with the coordinate system ($x_{1}$, ...$x_{m}$):$U_{P}%
\longrightarrow\mathbb{R}$, of denoting by $B_{P}$(r) the ball of center O and
radius r whatever the value of r($\geq0$ of course).On a manifold V let P be a
point and a chart ($U$, $x_{1}$, ...$x_{m})$ of V centered at $P$: $x_{i}$%
($P$)=0, $1\leq i\leq m$. The blow up of power t
$>$%
0\ associated to P and the chart ($U$, $x_{1}$, ...$x_{m})$ is the
diffeomorphism $Bl_{t}:U\rightarrow\mathbb{R}^{m}$, $Bl_{t}$(Q)=($\frac{1}%
{t}x_{1}(Q), ..., \frac{1}{t}x_{m}(Q)$). All functions,  tensors,
differential
operators can the be transported to the open subset $Bl_{t}$($U$). Suitably
normalized by a power of t,  they will have limits when t goes to 0 which will
be defined on $\mathbb{R}^{m}$. These limits contain a trove of information
about the behaviour of the original objects in the neighborhood (more
precisely in the infinitesimal neighborhood) of P. To simplify the notations
we shall write:$\frac{1}{t}Q$ instead of $Bl_{t}$($Q$),  $\frac{1}{t}A$
instead
of $Bl_{t}$($A$) if $A$ is a subset of U and so on.

\bigskip In the following all the blow-ups will be associated to geodesic
charts ($U$, $x_{1}$, ...$x_{m})$ with pole at $P$. On the magnified set
$\frac{1}{\sqrt[4]{\varepsilon}}x_{1}\times$...$\times x_{m}(U_{P})$,  we can
define the function w$_{P, \varepsilon}$ blow up of the function$\frac
{v_{\varepsilon}}{\overline{v_{\varepsilon}}}$:%

\[
w_{P, \varepsilon}(y)=\frac{u_{\varepsilon}(y\sqrt[4]{\varepsilon})}%
{\overline{u}_{\varepsilon}}%
\]
where $\overline{u}_{\varepsilon}$=$\underset{V}{\max}$ u$_{\varepsilon}.$

\subsection{Main theorem}

\subparagraph{}
\begin{Defi}
We define $\Lambda$ as
\[
\Lambda=\inf\left[  \text{ }\sum_{n=1}^{m}\sqrt{\lambda_{n}(R)}|R\in C_{\min
}\right]
\]

\end{Defi}
\begin{theorem}{\bf Selection-Concentration.}
\label{thnf}

\begin{itemize}
\item (i)For any P$\in C_{\min, }$any sequence of v$_{\varepsilon}$'s,  with
$\varepsilon$ tending to 0,  contains a subsequence \{u$_{\varepsilon_{n}}$\},
such that the sequence of blown-up functions $w_{P, \varepsilon_{n}}$ at P
converges to a function w$_{P}$:$\mathbb{R}^{m}$--%
$>$%
$\mathbb{R}_{+}$,  both in the L$^{2}$norm and the C$^{\infty}$ topology.

\item (ii) w satisfies the equation and inequality:%
\[
\Delta_{E}w+\sum_{i=1}^{m}\lambda_{i}(P)x_{i}^{2}w=\lambda w
\]

\[
0<w\leq\max_{\mathbb{R}^{m}}w\leq1
\]
where $\Delta_{E}$ is the negative standard Euclidean Laplacian on
$\mathbb{R}^{m}$.

\item (iii)If $\sum_{n=1}^{m}\sqrt{\lambda_{n}(P)}>\Lambda:$ then
\[
w_{P}=0
\]

\item (iv) If $\sum_{n=1}^{m}\sqrt{\lambda_{n}(P)}=\Lambda$ and there exists a
sequence S $\subset\mathbb{N}, $ such that each u$_{\varepsilon_{n}}, n\in$S,
has a maximum point Q$_{n}$ with the property that the sequence \{Q$_{n}$%
$\vert$%
n$\in S$\} converges to P,  then this function w$_{P}$ is:%
\[
w_{P}(x)=\prod_{n=1}^{m}\exp\left(  -\frac{x_{n}^{2}\sqrt{\lambda_{n}(P)}}%
{2}\right)
\]
and%
\[
\lambda=\Lambda
\]
\item (v) If $\sum_{n=1}^{m}\sqrt{\lambda_{n}(P)}=\Lambda$ and no such
subsequence S exists,  then%
\begin{align*}
\lambda &  =\Lambda\\
w_{P}  &  =f_{P}\prod_{n=1}^{m}\exp\left(  -\frac{x_{n}^{2}\sqrt{\lambda
_{n}(P)}}{2}\right)
\end{align*}
where $f_{P}$ is a factor $\geq0, $which depends on the sequence
\{u$_{\varepsilon_{n}}$\}.
\item (vi) For any sequence $\epsilon^{\prime}$s,  there exists at least one
$P\in C_{min}$ and at least one subsequence $S$ for which the case (iv)
occurs.
\end{itemize}
\end{theorem}

\noindent \textbf{Remark.} Note that in case (iv) the limit w is independent
of the
sequence \{u$_{\varepsilon_{n}}$\}.

\bigskip In order to prove the main theorem,  we need two propositions. The
first gives estimates of the first eigenvalue and the second,  estimates of
the
decay of the eigenfunctions.

\subsection{Auxiliary propositions}


\begin{prop}
\label{lm1} The first eigenvalue $\lambda_{\epsilon}$ satisfies the following
inequality
\[
\underset{V}{\min}c\leq\lambda_{\epsilon}\leq\underset{V}{\min}c+\Lambda
\epsilon^{1/2}%
\]
where $\Lambda=\inf\left[  \sum_{n=1}^{m}\sqrt{\lambda_{n}(P)}|P\in C_{\min
}\right]  .$
\end{prop}

\bigskip

\textbf{\noindent Proof.} \noindent For all u$\in$H$^{1}$(V),  $Q_{\epsilon
}(u)\underset{V}{\geq\min}c\int_{V}u^{2}$. Hence$\ \underset{V}{\min}%
c\leq\lambda_{\epsilon}$. To prove the right hand-side inequality,  we will
use
a test function in the variational quotient $Q$. In the neighborhood of a
point $P\in C_{\min}$,  consider the function:
\begin{align*}
\phi_{\epsilon}  &  =e^{-\sum\mu_{i}x_{i}^{2}/{2}}-e^{-\frac{\rho}%
{2\sqrt{\varepsilon}}}, \text{ on }\mathcal{N}_{P}(\rho)\\
&  =0, \text{ on }V\text{--}\mathcal{N}_{P}(\rho)
\end{align*}
\bigskip$\mathcal{N}_{P}(\rho)$ is the connected component of the set \{x%
$\vert$%
$\sum_{i=1}^{m}\sqrt{\lambda_{i}}x_{i}^{2}\leq\rho\}$ containing 0, where
$\rho$ is taken so small that $\mathcal{N}_{P}(\rho)$ is contained in B$_{P}%
$($\delta$). We take as coefficients $\mu_{i}=\sqrt{\frac{\lambda_{i}%
}{\epsilon}}$ where $\lambda_{i}=\lambda_{i}(P)$ for simplicity. In the
coordinate system at $P$,
\begin{align*}
g_{ij}(P)  &  =\delta_{ij}+O(||x||_{\mathbb{R}^{m}}^{2}),\\
\sqrt{detg}  &  =1-\sum_{i, j=1}^{m}\frac{Ric_{ij}}{6}x^{i}x^{j}%
+O(||x||_{\mathbb{R}^{m}}^{3}),
\end{align*}
$Ric$ denotes the Ricci tensor. We denote $\prod_{1}^{m}\mu_{i}$ by $\mu$ and
the quadratic form $\sum\mu_{i}(x^{i})^{2}$ by $q$. We recall that:
\begin{align*}
\int_{\mathbb{R}}e^{-\mu_{i}x^{2}}dx  &  =\sqrt{\frac{\pi}{\mu_{i}}},\\
\int_{\mathbb{R}}x^{2}e^{-\mu_{i}x^{2}}dx  &  =\frac{\sqrt{\pi}}{2\mu
_{i}^{3/2}}.%
\end{align*}
To evaluate the quotient $Q_{\varepsilon}(u_{\varepsilon})$,  we compute the
leading terms in $\epsilon$ of the integrals $\mu$%
\begin{align*}
&  \int_{\mathcal{N}_{P}(\rho)}||\nabla\phi_{\epsilon}||_{g}^{2}\, dvol_{g}\,
\\
&  \int_{\mathcal{N}_{P(\rho)}}c\phi_{\epsilon}^{2}dvol_{g},%
\end{align*}%
\[
\int_{\mathcal{N}_{P}(\rho)}(||\nabla\phi_{\epsilon}||_{g}^{2})dvol_{g}%
=\int_{\mathcal{N}_{P}(\rho)}\sum_{i, j=1}^{m}g^{ij}\frac{\partial
\phi_{\varepsilon}}{\partial x_{i}}\frac{\partial\phi_{\varepsilon}}{\partial
x_{j}}\sqrt{\det g}dx,
\]
where $g^{ij}$ is the matrix inverse of g$_{ij}$ and dx is the Lebesgue
volume.%
\[
\int_{\mathcal{N}_{P}(\rho)}(||\nabla\phi_{\epsilon}||_{g}^{2})dvol_{g}%
=\int_{\mathcal{N}_{P}(\rho)}\left[  \sum_{i=1}^{m}\mu_{i}^{2}x_{i}^{2}%
+\sum_{i, j, k=1}^{m}a_{ijk}x_{i}x_{j}x_{k}\right]  e^{-q}dx,
\]
where the functions $a_{ijk}$ are defined C$^{\infty}$ and bounded on
$B_{P}(\delta).$ Performing the blow-up at $P$ i.e. the change of variable
$z_{i}=x_{i}\sqrt{\mu_{i}}, $
\[
\int_{\mathcal{N}_{P}(\rho)}(||\nabla\phi_{\epsilon}||_{g}^{2})dvol_{g}%
=\int_{B(0, \frac{\rho}{\sqrt[4]{\varepsilon}})}\left[  \sum_{i, j=1}^{m}\mu
_{i}z_{i}^{2}+\sum_{i, j, k=1}^{m}a_{ijk}(x)\frac{z_{i}z_{j}z_{k}}{\sqrt{\mu
_{i}\mu_{j}\mu_{k}}}\right]  \exp\left(  -||z||_{\mathbb{R}^{m}}^{2}\right)
\frac{dz}{\sqrt{\mu}}%
\]
where $\mu=\prod_{i=1}^{m}\mu_{i}.$%
\[
\varepsilon\int_{\mathcal{N}_{P}(\rho)}(||\nabla\phi_{\epsilon}||_{g}%
^{2})dvol_{g}=\frac{\pi^{m/2}}{\sqrt{\mu}}\left[  \sum_{i=1}^{n}\frac
{\sqrt{\epsilon}\lambda_{i}}{2}+O(\varepsilon^{\frac{7}{4}}).\right]
\]
We will now evaluate the potential term:
\begin{align*}
\int_{\mathcal{N}_{P}(\rho)}c\phi_{\epsilon}^{2}dvol_{g}  &  =\int
_{\mathcal{N}_{P}(\rho)}\phi_{\epsilon}^{2}\left[  \left\{  c(P)+\sum
_{k=1}^{m}\lambda_{i}x_{i}^{2}+O(||x||_{\mathbb{R}^{m}}^{3})\right\}  \left(
1-\frac{Ric_{ij}(P)}{6}x^{i}x^{j}+O(||x||_{\mathbb{R}^{m}}^{3}\right)
\right]  dx\\
&  =c(P)\int_{\mathcal{N}_{P}(\rho)}\phi_{\epsilon}^{2}dx+\int_{\mathcal{N}%
_{P}(\rho)}\phi_{\epsilon}^{2}\left[  \sum_{k=1}^{m}\left(  \lambda_{i}%
-\frac{c(P)Ric_{ii}(P)}{6}\right)  x_{i}^{2}+\sum_{i, j, k=1}^{m}b_{ijk}%
x_{i}x_{j}x_{k}\right]  dx
\end{align*}
where the functions a$_{ijk}$ are defined C$^{\infty}$ and bounded on
B$(\delta)$. With the same change of variables, after expanding the square
$u_{\epsilon}^{2}(x)=e^{-q}-2e^{-q/2-\rho/2\sqrt{\varepsilon}}+e^{-\rho
/\sqrt{\varepsilon}}$ we get:
\[
\int_{\mathcal{N}_{P}(\rho)}{x}_{i}^{2}\phi_{\epsilon}^{2}dx=\frac{\pi^{m/2}%
}{2\sqrt{\mu}}\sqrt{\frac{\varepsilon}{\lambda_{i}(P)}}+O_{w}(e^{-\rho
/2\sqrt{\epsilon}}).
\]

O$_{w}(\exp-\frac{\rho}{\sqrt{\varepsilon}})$ means that for any $\eta
\in\lbrack0, $1[,  there exists a constant K($\eta$) independant of
$\varepsilon$ sucht that the error is at most equal in absolute value to
K($\eta$)$\exp\left[  -\frac{\eta\rho}{\sqrt{\varepsilon}}\right].$ An easy
symmetry argument shows that: \bigskip%
\[
\int_{\mathcal{N}_{P}(\rho)}{x}_{i}x_{j}\phi_{\epsilon}^{2}dx=0\text{ if
i}\neq\text{j}%
\]
The potential term in the integral becomes:
\[
\int_{\mathcal{N}_{P}(\rho)}c\phi_{\epsilon}^{2}=\frac{\pi^{m/2}}{\sqrt{\mu}%
}\left(  c(P)+\frac{1}{2}\sum_{i=1}^{m}(\lambda_{i}-c(P)\frac{Ric_{ii}(P)}%
{6})\sqrt{\frac{\epsilon}{\lambda_{i}}}+O(\varepsilon^{\frac{3}{2}})\right)
\]

Also:%
\[
\int_{V}\phi_{\varepsilon}^{2}dvol_{g}=1-\sum_{i=1}^{m}\frac{Ric_{ii}(P)}%
{12}\sqrt{\frac{\epsilon}{\lambda_{i}}}+O_{w}(e^{-\rho/2\epsilon^{1/2}})
\]

\bigskip

The quotient can now be evaluated:
\[
Q_{\epsilon}(u)=\frac{\epsilon\int_{V}\left[  ||\nabla u||_{g}^{2}%
+cu^{2}\right]  dvol_{g}}{\int_{V}u^{2}dvol_{g}}%
\]%
\[
Q_{\varepsilon}(\phi_{\varepsilon})=\frac{\frac{\pi^{m/2}}{\sqrt{\mu}}\left(
\sum_{i=1}^{m}\frac{\sqrt{\varepsilon\lambda_{i}}}{2}+c(P)+\sum_{i=1}%
^{m}(\lambda_{i}-c(P)\frac{Ric_{ii}(P)}{6})\sqrt{\frac{\epsilon}{\lambda_{i}}%
}+O(\varepsilon^{\frac{3}{2}})\right)  }{\frac{\pi^{m/2}}{\sqrt{\mu}}%
+\sum_{i=1}^{m}\frac{Ric_{ii}(P)}{6}\sqrt{\frac{\epsilon}{\lambda_{i}}%
}+o(e^{-\rho/4\epsilon^{1/2}})}%
\]%
\[
Q_{\varepsilon}(\phi_{\varepsilon})=\frac{c(P)\left[  1-\sum_{i=1}^{m}%
\frac{Ric_{ii}(P)}{12}\sqrt{\frac{\epsilon}{\lambda_{i}}}\right]  +\sum
_{i=1}^{m}\sqrt{\varepsilon\lambda_{i}}+O(\varepsilon^{\frac{3}{2}}))}%
{1-\sum_{i=1}^{m}\frac{Ric_{ii}(P)}{12}\sqrt{\frac{\epsilon}{\lambda_{i}}%
}+O_{w}(e^{-\rho/2\epsilon^{1/2}})}%
\]
\bigskip If we use the fact that $c(P)=\underset{V}{\min}c$,  the quotient can
be simplified as follows
\[
Q_{\epsilon}(\phi_{\varepsilon})=\underset{V}{\min}c+\frac{\sum_{i=1}^{m}%
\sqrt{\lambda_{i}\epsilon}+O(\varepsilon^{\frac{3}{2}})}{1+\sum_{i=1}^{m}%
\frac{Ric_{ii}(P)}{12}\sqrt{\frac{\epsilon}{\lambda_{i}}}+O_{w}(e^{-\rho
/2\epsilon^{1/2}})}%
\]
If we take the minimum over all test functions centered at any critical points
of the set of minimal points $C_{m}$,  we obtain the estimate:%
\[
\underset{V}{\min}c\leq\lambda_{\epsilon}\leq Q_{\epsilon}(\phi_{\varepsilon
})\leq\underset{V}{\min}c+\inf\left\{  \sum_{i=1}^{m}\sqrt{\lambda
_{i}(P)\epsilon}|P\in C_{\min}\right\}  +O(\varepsilon)\leq\underset{V}{\min
}c+\Lambda\epsilon^{1/2}+O(\varepsilon).
\]
\QED

\bigskip{\noindent\textbf{Remark:}} We expect in general for smooth potential
that there exists an asymptotic expansion:%
\[
\lambda_{\epsilon}=\sum_{k=0}^{n}c_{k}\epsilon^{k/2}+o(\epsilon^{n/2}).
\]
If it does exist can one find a systematic procedure to compute the
coefficients c$_{k}?$ From the previous result,  we have that
\[
c_{0}=\min c,
\]
which is the Topological Pressure. We will see in the following results that
\[
c_{1}=\inf\left\{  \sum_{i=1}^{m}\sqrt{\lambda_{i}(P)}|P\in C_{\min}\right\}
.
\]
Proposition \ref{lm1} provides an estimate of the velocity of convergence of
the sequence of maximum points $Q_{\epsilon}$ of u$_{\varepsilon}$ to an
element of $C_{\min}.$

\begin{lem}
\label{creetin} (i)If for a sequence $\epsilon_{n}$ tending to zero,
$\lim_{n\rightarrow\infty}\int_{V}u_{\epsilon_{n}}^{2}dvol_{g}>0$ , then

\[
\lim_{n\rightarrow\infty}\sup_{V}u_{\epsilon_{n}}=+\infty.
\]
\end{lem}

{\noindent\textbf{Proof.}} (i)Suppose that for a subsequence $\varepsilon_{n}%
$,  still denoted by $\varepsilon_{n}$,  $\lim_{n\rightarrow\infty}\sup
_{V}u_{\epsilon_{n}}<+\infty$. Then for all n,  $\sup_{V}u_{\epsilon_{n}}\leq
N$,  a constant. For any $\eta>0$,  chose an open \ neighborhood $K$ of
$C_{min}$, such that $vol_{g}(K)$ is smaller than $\frac{\eta}{2N}$. Now
\[
\int_{V}u_{\epsilon_{n}}^{2}dvol_{g}=\int_{K}u_{\epsilon_{n}}^{2}dvol_{g}%
+\int_{V-K}u_{\epsilon_{n}}^{2}dvol_{g}%
\]

\[
\int_{V}u_{\epsilon_{n}}^{2}dvol_{g}=\int_{K}u_{\epsilon_{n}}^{2}dvol_{g}%
+\sup_{V}u_{\epsilon_{n}}\int_{V-K}\frac{u_{\epsilon_{n}}^{2}}{\underset
{V}{\sup}u_{\epsilon_{n}}}dvol_{g}%
\]

By Appendix \textrm{II,} $\frac{u_{\epsilon_{n}}^{2}}{\sup_{V}u_{\epsilon_{n}%
}}$--%
$>$%
0 uniformly on $V-K$. Hence $\lim_{n\rightarrow\infty}\int_{V}u_{\epsilon_{n}%
}^{2}dvol_{g}\leq\eta.$ Because $\eta$ is arbitrary,
$\lim_{n\rightarrow\infty
}\int_{V}u_{\epsilon_{n}}^{2}dvol_{g}=0.$ A contradiction.

Recall that $d_{g}$ denotes the Riemmanian distance associated to the metric
$g$,
\begin{lem} \label{max}
For each $\varepsilon$,  let us denote by $\mathcal{M}%
_{\varepsilon}$ the set of all $\max$imum points of v$_{\varepsilon}$. There
exists a constant $A$ depending only on $c$,  such that:
\[
\underset{q\in\mathcal{M}_{\varepsilon}}{\sup}d_{g}^{2}(q, C_{\min})\leq
A\epsilon^{1/2}%
\]
It follows from this that the set of limit points of the set of maximum points
of v$_{\varepsilon}$ is contained in C$_{\min}.$
\end{lem}

\bigskip

{\noindent\textbf{Proof.}} \noindent Let q$\in\mathcal{M}_{\varepsilon}$.
Recall $\epsilon\Delta_{g}u_{\epsilon}(q)+c(q)u_{\epsilon}(q)=\lambda
_{\epsilon}u_{\epsilon}(q).$ Because the solution u$_{\varepsilon}$ is
positive and $\Delta_{g}u_{\epsilon}(q)\geq$ 0,  at the maximum point q of
$u_{\varepsilon}, $ $c(q)\leq\lambda_{\varepsilon}$. By Proposition \ref{lm1},
$\ 0\ \leq\ c(q)-\underset{V}{\min}c\leq\lambda_{\epsilon}-\underset{V}{\min
}c\leq\Lambda\epsilon^{1/2}.$ Because the critical points of c are
non-degenerate,  it is easy to see that there exists a constant $\Gamma
$\ depending only on $c$ such that for $P\in$ $V$,  d$_{g}^{2}$($P$,
$C_{\min}%
$)$\leq$ $\Gamma$($c$($P$)--$\underset{V}{\min}$ $c$). Take $A$=$\Gamma
\Lambda.$
\QED
\bigskip

\noindent \textbf{Remark.} This result proves that any sequence of
$\varepsilon^{\prime
}$s converging to 0,  contains a subsequence \{$\varepsilon_{k}|k\in
\mathbb{N\}}$ such that there exists a $P\in C_{\min}$ and a vector $P$*$\in
T_{P}V$ with the property:%
\[
P_{\epsilon}=\text{exp}_{P}\text{(}\epsilon^{1/4}P^{\ast}+o(\epsilon^{1/4})).
\]
The length $||P^{\ast}||$ of the vector $P^{\ast}$ is the distance between the
peak of concentration and the set $C_{\min}$ in the blow up space. It can be
considered as a measure of the convergence velocity.

\bigskip

We have so far computed an estimate of the rescaled eigenvalue $\frac
{\lambda_{\epsilon}-\min_{V_{m}}c}{\epsilon^{1/2}}$. Now we will provide an
estimate of the eigenfunction in the neighborhood of the points in $C_{\min}.$
Let $P$ be a point in $C_{\min}$. Recall that w$_{\varepsilon
}=\frac{u_{\varepsilon}}{\overline{u_{\varepsilon}}}$ and $\overline
{u_{\varepsilon}}=\underset{V}{\max}$ u$_{\varepsilon}.$

\begin{prop}
\label{estim} For all $\varepsilon_{0}\in]0, 1[:$

\begin{itemize}
\item (i) $\underset{]0, \varepsilon_{0}]}{\sup}$ $\int_{B_{P}(\delta
/\sqrt[4]{\varepsilon})}w_{\varepsilon}(y)^{2}dy<+\infty.$ More generally,
for
any continuous function $f:[0, 1]\times\mathbb{R}^{m}$---%
$>$%
$\mathbb{R}, (\epsilon, y)$---%
$>$%
$f(\epsilon, y), $\ having at most polynomial growth at infinity,
\[
\underset{]0, \varepsilon_{0}]}{\sup}\int_{B_{P}(\delta/\sqrt[4]{\varepsilon}%
)}f(\epsilon, y)w_{\varepsilon}(y)^{2}dy<+\infty
\]

\item (ii) the set of restrictions $w_{\varepsilon}^{2}$%
$\vert$%
$B_{P}(\delta/\sqrt[4]{\varepsilon})$, $\varepsilon$ $\in]0, 1]$,  of the
$w_{\varepsilon}^{2}$ to the balls $B_{P}(\delta/\sqrt[4]{\varepsilon})$
satisfies the following condition: for any $\eta>0, $ there exists a compact
$K\subset\mathbb{R}^{m}$ and a $\varepsilon(\eta)>0$ such that
\[
\int_{B_{P}(\delta/\sqrt[4]{\varepsilon})-K}f(\epsilon, y)w_{\varepsilon
}(y)^{2}dy\leq\eta,
\]
for all $\varepsilon$ $\in]0, \varepsilon(\eta)].$
\end{itemize}
\end{prop}

\bigskip

\subsection{Proof of the Proposition \ref{estim}}

\noindent To start with,  rewrite equation (\ref{edpV}) as follows
\[
\sqrt{\epsilon}\Delta_{g}u_{\epsilon}+\frac{c-\min_{V_{m}}c}{\epsilon^{1/2}%
}u_{\epsilon}=\frac{\lambda_{\epsilon}-\min_{V_{m}}c}{\epsilon^{1/2}%
}u_{\epsilon}%
\]
Introducing the notations $c_{\epsilon}=\frac{c-\min_{V_{m}}c}{\epsilon^{1/2}%
}$ and $\mu_{\epsilon}=\frac{\lambda_{\epsilon}-\min_{V_{m}}c}{\epsilon^{1/2}%
}$ for simplicity,  we have:%
\[
\sqrt{\epsilon}\Delta_{g}u_{\epsilon}+c_{\varepsilon}u_{\epsilon}%
=\mu_{\varepsilon}u_{\epsilon}%
\]
The function e$^{-t\mu_{\varepsilon}\text{ }}\upsilon_{\varepsilon}(x)$ is the
solution of the parabolic Cauchy problem:%
\begin{align*}
\frac{\partial p}{\partial t}  &  =-\sqrt{\epsilon}\Delta_{g}p-c_{\epsilon}p\\
p(0, x)  &  =u_{\epsilon}(x).
\end{align*}
Note that the sequence $\mu_{\epsilon}$ is bounded. We estimate
$w_{\varepsilon}$ in the ball $B_{P}(\delta), $ using the fact that the
restriction of the function $\ e^{-\mu_{\epsilon}t}v_{\epsilon}(x)$ to
B$_{P}(\delta)$ is the solution of the parabolic initial- boundary value
problem:%
\begin{align*}
\frac{\partial p}{\partial t}  &  =-\sqrt{\epsilon}\Delta_{g}p-c_{\epsilon}p\\
p(0, x)  &  =u_{\epsilon}(x)\\
p(t, x)_{|\partial B_{P}(\delta)}  &  =u_{\epsilon}(x)_{|\partial B_{P}%
(\delta)}e^{-\mu_{\epsilon}t}.
\end{align*}
\bigskip In the coordinate system at $P$, %
\[
\frac{\partial p}{\partial t}=\sqrt{\epsilon}\sum_{i, j=1}^{m}g^{ij}%
\frac{\partial^{2}p}{\partial x_{i}\partial x_{j}}-\sqrt{\epsilon}\sum
_{k=1}^{m}B^{k}\frac{\partial p}{\partial x_{k}}-c_{\epsilon}p,
\]
where
\[
B_{k}=-\sum_{i, j=1}^{m}g^{ij}\Gamma_{ij}^{k}.
\]
For x$\in B_{P}(\delta)$, the solution is given by the Feynman-Kac formula
with
boundary term:
\begin{equation}
e^{-\mu_{\epsilon}t}u_{\epsilon}(x)=E_{x}\left(  e^{-\mu_{\epsilon}%
t}u_{\epsilon}(X_{\epsilon}(t))\chi_{(t<\tau_{\epsilon}^{x})}e^{-\int_{0}%
^{t}c_{\epsilon}(X_{\epsilon}(s)ds)}\right)  +E_{x}\left(  e^{-\mu_{\epsilon
}\tau_{\varepsilon}^{x}}u_{\epsilon}(X_{\epsilon}(\tau_{\varepsilon}^{x}%
))\chi_{(t>\tau_{\epsilon}^{x})}e^{-\int_{0}^{\tau_{\epsilon}^{x}}c_{\epsilon
}(X_{\epsilon}(t))ds}\right)  \label{fk}%
\end{equation}
where $X_{\varepsilon}(t)$ is the process starting at x at time 0 and
satisfying the It\^{o} equation:%
\begin{equation}
dX_{\varepsilon}(t)=\sqrt{\varepsilon}B(X_{\varepsilon}(t))+\sqrt[4]%
{\varepsilon}\sigma(X_{\varepsilon}(t))dW(t)\text{ for t}\leq\text{ }%
\tau_{\varepsilon}^{U} \label{sde}%
\end{equation}
where $W$(t) is a standard m dimensional Brownian motion and $\sigma
$:$U\longrightarrow$End($\mathbb{R}^{m}$) is the positive definite square root
of the matrix function ($2g^{ij}$). $\tau_{\varepsilon}^{x}$ is the first exit
time from the ball $B_{P}(\delta)$,  of the process $X_{\varepsilon}, $
starting
at x. Let $\overline{u}_{\varepsilon}=\underset{V}{\max}$ $u_{\varepsilon}%
$. Then equation (\ref{fk}) implies the inequality:
\begin{equation}
\frac{u_{\epsilon}(x)}{\overline{u_{\varepsilon}}}\leq\left[  E_{x}\left(
\chi_{(t<\tau_{\epsilon}^{x})}e^{-\int_{0}^{t}c_{\epsilon}(X_{\epsilon}%
(s))ds}\right)  +E_{x}\left(  e^{\mu_{\varepsilon}(t-\tau_{\varepsilon}^{x}%
)}\frac{u_{\epsilon}(X(\tau_{\varepsilon}))}{\overline{u_{\varepsilon}}}%
\chi_{(t>\tau_{\epsilon}^{x})}e^{-\int_{0}^{\tau_{\epsilon}^{x}}c_{\epsilon
}(X_{\epsilon}(s))ds}\right)  \right]  \label{estimp}%
\end{equation}

\bigskip for $x\in B_{P}(\delta).$ Let us define:%
\begin{align*}
\text{\textrm{I}}  &  =e^{\mu_{\epsilon}t}E_{x}\left(  \frac{u_{\epsilon
}(X(\tau_{\varepsilon}))}{\overline{u_{\varepsilon}}}\chi_{(t>\tau_{\epsilon
}^{x})}e^{-\int_{0}^{\tau_{\epsilon}^{x}}c_{\epsilon}(X_{\epsilon}%
(s))ds}\right) \\
\text{\textrm{II}}  &  =E_{x}\left(  \chi_{(t<\tau_{\epsilon}^{x})}%
e^{-\int_{0}^{t}c_{\epsilon}(X_{\epsilon}(s))ds}\right)
\end{align*}
We estimate the terms $\mathrm{I, II}$ independently.

\subsubsection*{Estimate of $I$}

\bigskip

\noindent To estimate the boundary term $\mathrm{I}$ in inequality
(\ref{estimp}),  we apply the results proved in Appendix 2 to the
equation(\ref{edpV}) taking $\psi$=$c$--$\underset{V}{\min}c, $ $b$%
=0, $c_{\varepsilon}=\underset{V}{\min}$ $c$--$\lambda_{\varepsilon}$ and
$\varepsilon=\varepsilon$. Then for any integer k, any compact subset
$\mathcal{C}$ disjoint from the set C$_{\min}, $ there exists a positive
constant A(k, $\mathcal{C}$) such that for $\varepsilon\in]0, 1]$:%
\[
\underset{\mathcal{C}}{\max}u_{\epsilon}\leq A(k, \mathcal{C})\varepsilon
^{k}\overline{v}_{\varepsilon}%
\]
Taking $\mathcal{C=}$V-$\cup\{B_{P}(\delta)|P\in C_{\min}\}:$
\[
\text{\textrm{I}}=e^{\mu_{\epsilon}t}E_{x}\left(  \frac{u_{\epsilon
}(X_{\varepsilon}(\tau_{\varepsilon}^{x}))}{\overline{u}_{\varepsilon}%
}e^{-\int_{0}^{\tau_{\epsilon}^{x}}c_{\epsilon}(X_{\varepsilon}(s))ds}\right)
\leq e^{\mu_{\epsilon}t}A(k, \delta)\epsilon^{k}%
\]

\subsubsection*{Estimate of $\overline{\underline{\text{II}}}$}

Let $\alpha$ be a number in $]0$, $\frac{1}{6}[$. Assume that $\varepsilon$%
$<$%
1. We split \textrm{II }as follows:%
\[
\text{\textrm{II}}=E_{x}\left(  \chi_{1}\chi_{(t<\tau_{\epsilon}^{x})}%
e^{-\int_{0}^{t}c_{\epsilon}(X_{\epsilon}(s))ds}\right)  +E_{x}\left(
(1-\chi_{1})\chi_{(t<\tau_{\epsilon}^{x})}e^{-\int_{0}^{t}c_{\epsilon
}(X_{\epsilon}(s))ds}\right)
\]

\bigskip where $\chi_{1}$ is the characteristic function of the set $\left\{
\underset{[0, t]}{\sup}||X_{\varepsilon}(s)||_{\mathbb{R}^{m}}\geq
\delta\varepsilon^{\alpha}\right\}  $.
\[
\mathrm{II=III}+\mathrm{IV}%
\]

Because c$_{\varepsilon}$ is non negative and $\alpha<\frac{1}{6}$:%
\begin{equation}
\mathrm{III}=E_{x}\left(  \chi_{1}\chi_{(t<\tau_{\epsilon}^{x})}e^{-\int
_{0}^{t}c_{\epsilon}(X_{\epsilon}(s))ds}\right)  \leq P_{x}\text{ }\left[
\underset{[0, t\wedge\tau_{\varepsilon}^{x}]}{\sup}||X_{\varepsilon
}(t)||_{\mathbb{R}^{m}}\geq\delta\varepsilon^{\alpha}\right] \nonumber
\end{equation}
We have:%
\[
X_{\varepsilon}(t\wedge\tau_{\varepsilon}^{x})=\sqrt{\varepsilon}\int
_{0}^{t\wedge\tau_{\varepsilon}^{x}}B(X_{\varepsilon}(s))+\sqrt[4]%
{\varepsilon}\int_{0}^{t\wedge\tau_{\varepsilon}^{x}}\sigma(X_{\varepsilon
}(s))dW(s)\text{ }%
\]
Now $\underset{[0, t\wedge\tau_{\varepsilon}^{x}]}{\sup}||X_{\varepsilon
}(t)||_{\mathbb{R}^{m}}\leq\delta.$ Hence
$\vert$%
$\vert$%
$\sqrt{\varepsilon}\int_{0}^{t\wedge\tau_{\varepsilon}^{x}}B(X_{\varepsilon
}(s))||_{\mathbb{R}^{m}}$ $\leq tM_{1}\sqrt{m\varepsilon\text{ }},$ where
$M_{1}=\underset{x\in B(\delta)}{\sup}||B(x)||_{\mathbb{R}^{m}}.$
Then a well known lemma (see \cite{St}) shows that if $\delta\varepsilon
^{\alpha}>\ ||x||_{\mathbb{R}^{m}}+tM_{1}\sqrt{m\varepsilon},$%
\[
P_{x}\text{ }\left[  \underset{[0,
t\wedge\tau_{\varepsilon}^{x}]}{\sup}\text{%
$\vert$%
$\vert$%
}X_{\varepsilon}(t)\text{%
$\vert$%
$\vert$%
}_{\mathbb{R}^{m}}\geq\delta\varepsilon^{\alpha}\right]  \leq2m\exp
-\frac{[\delta\varepsilon^{\alpha}-\ ||x||_{\mathbb{R}^{m}}-tM_{1}%
\sqrt{m\varepsilon}]^{2}}{2mM^{2}t\sqrt{\varepsilon}},%
\]
where $M$=$\underset{B_{P}(\delta)}{\sup}$%
$\vert$%
$\vert$%
$\sigma(x)||_{\mathbb{R}^{m}}.$ It is clear that:%
\[
E_{x}\left(  \chi_{1}\chi_{(t<\tau_{\epsilon}^{x})}e^{-\int_{0}^{t}%
c_{\epsilon}(X_{\epsilon}(s))ds}\right)  \leq E_{x}(\chi_{1}\chi
_{(t<\tau_{\epsilon}^{x})}).
\]
Hence if $\ ||x||_{\mathbb{R}^{m}}+tM_{1}\sqrt{\varepsilon}<\delta
\varepsilon^{\alpha}$:%
\[
E_{x}\left(  \chi_{1}\chi_{(t<\tau_{\epsilon}^{x})}e^{-\int_{0}^{t}%
c_{\epsilon}(X_{\epsilon}(s))ds}\right)  \leq2m\exp-\frac{[\delta
\varepsilon^{\alpha}-(||x||_{\mathbb{R}^{m}}+tM_{1}\sqrt{m\varepsilon})]^{2}%
}{2mM^{2}t\sqrt{\varepsilon}}.%
\]
\bigskip

\noindent To estimate the second integral $\mathrm{IV}=E_{x}\left(  (1-\chi
_{1})\chi_{(t<\tau_{\epsilon}^{x})}e^{-\int_{0}^{t}c_{\epsilon}(X_{\epsilon
}(s))ds}\right), $ let us\ define the process $Y_{\epsilon}(t)$ for $t$
$<\tau_{\epsilon}^{U}=$ first exit time of the process $X_{\varepsilon}$ from
$U_{P}$:%
\[
Y_{\epsilon}(t)=\frac{X_{\varepsilon}(t)}{\sqrt[4]{\varepsilon}}%
\]
Then:
\[
dY_{\varepsilon}(t)=\sqrt[4]{\varepsilon}B(X_{\varepsilon}(t))+\sigma
(X_{\varepsilon}(t))dW(t)
\]

\[
Y_{\varepsilon}(t)=Y_{\varepsilon}(0)+\sqrt[4]{\varepsilon}\int_{0}%
^{t\wedge\tau_{\epsilon}^{U}}B(X_{\varepsilon}(s))ds+\int_{0}^{t\wedge
\tau_{\epsilon}^{U}}\sigma(X_{\varepsilon}(s))dW(s)
\]
for $t$ $\leq\tau_{\varepsilon}^{U}.$But using the definition of $\sigma$ and
the system of coordinates ($x_{1}$, ...$x_{m}$):%
\[
\sigma(x)=\sqrt{2}(Id_{m}+\Phi(x))
\]

where $\Phi$:$U$---%
$>$%
$End$($\mathbb{R}^{m})$ is a $C^{\infty}$matrix function such that :%
\[
\Phi_{ij}(x)=\sum_{k, l=1}^{m}\Phi_{ijkl}(x)x_{k}x_{l}%
\]

Its value at P is:%

\[
\Phi_{ijkl}(P)=-\frac{1}{6}(R_{iklj}(P)+R_{ilkj}(P)),
\]%
\[
\sigma(X_{\varepsilon}(t))=\sqrt{2}(Id_{m}+\sqrt{\varepsilon}\widehat{\Phi
}(\varepsilon, Y_{\varepsilon}(t)))
\]
where $\widehat{\Phi}(\varepsilon, Y_{\varepsilon})$ is a matrix function such
that:%
\[
\widehat{\Phi}_{ij}(\varepsilon, Y_{\varepsilon}(t))=\sum_{k, l=1}^{m}%
\Phi_{ijkl}(\sqrt[4]{\varepsilon}Y_{\varepsilon}(t))Y_{\varepsilon
, k}(t)Y_{\varepsilon, l}(t)
\]
Also the components of the field $B$ :%
\[
B_{i}(x)=\sum_{j=1}^{m}B_{ij}(x)x_{j}%
\]
Hence:%
\[
B(X_{\varepsilon}(t))=\sqrt[4]{\varepsilon}\widehat{B}(\varepsilon
, Y_{\varepsilon}(t))
\]%
\[
\widehat{B}_{i}(\varepsilon, y)=\sum_{j=1}^{m}B_{ij}(y\sqrt[4]{\varepsilon
})y_{j}%
\]
Finally:%
\[
Y_{\varepsilon}(t)=Y_{\varepsilon}(0)+\sqrt{2}W(t)+Z_{\varepsilon}(t)
\]%
\[
Z_{\varepsilon}(t)=\sqrt{\varepsilon}\int_{0}^{t\wedge\tau_{\epsilon}^{U}%
}\widehat{B}(\varepsilon, Y_{\varepsilon}(s))ds+\sqrt{2\varepsilon}\int
_{0}^{t\wedge\tau_{\epsilon}^{U}}\widehat{\Phi}(\varepsilon, Y_{\varepsilon
}(s))dW(s)
\]
To estimate \textrm{IV},  we split it into \textrm{V} and \textrm{VI},
choosing
a $\beta\in$[0, $\alpha$[:%
\[
\mathrm{V}=E_{x}\left(  (1-\chi_{1})\chi_{(t<\tau_{\epsilon}^{x})}e^{-\int
_{0}^{t}c_{\epsilon}(X_{\epsilon}(s))ds};\underset{[0, t]}{\sup}%
||Z_{\varepsilon}(s)||_{\mathbb{R}^{m}}\geq\varepsilon^{\beta}\right)
\]

\[
\mathrm{VI}=E_{x}\left(  (1-\chi_{1})\chi_{(t<\tau_{\epsilon}^{x})}%
e^{-\int_{0}^{t}c_{\epsilon}(X_{\epsilon}(s))ds};\underset{[0, t]}{\sup
}||Z_{\varepsilon}(s)||_{\mathbb{R}^{m}}<\varepsilon^{\beta}\right)
\]

\subsubsection*{Estimate of V}
We shall use a variant of the inequality (37.9) \ stated in (\cite{RW} p.78):%
\begin{equation}
P\{\underset{[0, t]}{\sup}\text{ }||M_{s}||\geq y;[M]_{t}\leq K\}\leq
2m\exp\left(  -\frac{y^{2}}{2mK}\right),  \label{rw}%
\end{equation}
where $M$ is a continuous local martingale, 0 at $t$=0. We apply this formula
to $M_{t}=Z_{\varepsilon}(t)-\sqrt[4]{\varepsilon}\int_{0}^{t}\widehat
{B}(\varepsilon, Y_{\varepsilon}(s))ds=\sqrt{2}\sqrt[4]{\varepsilon}\int
_{0}^{t}\widehat{\Phi}(\varepsilon, Y_{\varepsilon}(s))dW(s)$ stopped at
$\tau_{\varepsilon}^{x}$, the exit time of the process $X_{\varepsilon}$ from
$B_{P}(\delta).$

Note that :%
\[
(1-\chi_{1})||\sqrt{\varepsilon}\widehat{\Phi}(\varepsilon, Y_{\varepsilon
}(s))||_{End(\mathbb{R}^{m})}\leq C_{2}\delta^{2}\varepsilon^{2\alpha}%
\]
and that:%
\[
(1-\chi_{1})||\sqrt{\varepsilon}\widehat{B}(\varepsilon, Y_{\varepsilon
}(s))||_{\mathbb{R}^{m}}\leq C_{3}\delta\varepsilon^{\alpha+\frac{1}{4}},%
\]
for some cons$\tan$ts $C_{2}$ depending only on the values of $\ \sigma$ on
$B_{P}$($\delta$),  $C_{3}$ depending only on the values of the vector field
$B$ on $B_{P}$($\delta$). Thus we have:%
\[
\{\underset{[0, t]}{\sup}\text{
$\vert$%
$\vert$%
}M_{s}||\geq y;\chi_{1}=0\}\subset\{\underset{[0, t]}{\sup}\text{
$\vert$%
$\vert$%
}M_{s}||\geq y;[M]_{t}\leq t(C_{2}\delta^{2}\varepsilon^{2\alpha})^{2}\}
\]
Recall that:
\begin{align*}
\lbrack M]_{t}  &  =2\sqrt{\varepsilon}\int_{0}^{t\wedge\tau_{x}^{\varepsilon
}}tr[\widehat{\Phi}(\varepsilon, Y_{\varepsilon}(s))\widehat{\Phi}%
(\varepsilon, Y_{\varepsilon}(s))^{\ast}]ds,\\
\lbrack M]_{t}  &  \leq2\int_{0}^{t\wedge\tau_{x}^{\varepsilon}}%
||\sqrt[4]{\varepsilon}\widehat{\Phi}(\varepsilon, Y_{\varepsilon
}(s))||_{End(\mathbb{R}^{m})}^{2}ds.
\end{align*}
Hence:%
\begin{align*}
\mathrm{V}  &  \leq P_{x}\{Z_{\varepsilon}(t)\geq\varepsilon^{\beta};\chi
_{1}=0\}\\
\mathrm{V}  &  \leq P\{\underset{[0, t]}{\sup}\text{
$\vert$%
$\vert$%
}M_{s}||\geq\varepsilon^{\beta}-tC_{3}\delta\varepsilon^{\alpha+\frac{1}{4}%
};[M]_{t}\leq t(C_{2}\delta^{2}\varepsilon^{2\alpha})^{2}\}\\
\mathrm{V}  &  \leq2m\exp-\frac{(\varepsilon^{\beta}-tC_{3}\delta
\varepsilon^{\alpha+\frac{1}{4}})^{2}}{2mtC_{2}^{2}\delta^{4}\varepsilon
^{4\alpha}}.
\end{align*}

\subsubsection*{Estimate of VI}

By Taylor formula there exist a constant $C_{1}>0$ depending only on the
function $c$ such that for x$\in U_{P}$:%
\[
|c(x)-c(P)-\sum_{n=1}^{m}\lambda_{n}x_{n}^{2}|\leq C_{1}||x||_{\mathbb{R}^{m}%
}\sum_{1}^{m}\lambda_{i}x_{i}^{2}%
\]
Then, recalling that c(P)=$\underset{V}{\min}c$ and c$_{\varepsilon}$%
=$\frac{c-\min_{V_{m}}c}{\sqrt{\epsilon}}$ for t%
$<$%
$\tau_{\varepsilon}^{U}$:%
\[
(1-\sqrt[4]{\varepsilon}C_{1}||Y_{\varepsilon}(t)||_{\mathbb{R}^{m}})\sum
_{1}^{m}\lambda_{i}Y_{\varepsilon, i}(t)^{2}\leq c_{\varepsilon}(X_{\epsilon
}(t))\leq(1+\sqrt[4]{\varepsilon}C_{1}||Y_{\varepsilon}(t)||_{\mathbb{R}^{m}%
})\sum_{1}^{m}\lambda_{i}Y_{\varepsilon, i}(t)^{2}%
\]
For the sake of simplicity let us denote the positive definite quadratic form
$\sum_{1}^{m}\lambda_{i}X_{i}^{2}$ by $q(X)$. Given any $\sigma>0, $ any
$a, b\in\mathbb{R}^{m}$ we have:%
\begin{equation}
(1-\sigma)q(a)+(1-\frac{1}{\sigma})q(b)\leq q(a+b)\leq(1+\sigma)q(a)+(1+\frac
{1}{\sigma})q(b). \label{ung0}%
\end{equation}
Now:%
\[
q(Y_{\varepsilon}(t))=q(Y_{\varepsilon}(0)+\sqrt{2}W(t)+Z_{\varepsilon}(t)).
\]
Using the inequality (\ref{ung0}) taking $a$=$Y_{\varepsilon}(0)+\sqrt
{2}W(t), $ $b$=$Z_{\varepsilon}(t)$,  $\ \sigma=1:$%
\begin{equation}
q(Y_{\varepsilon}(t))\geq\frac{1}{2}q(Y_{\varepsilon}(0)+\sqrt{2}%
W(t))-q(Z_{\varepsilon}(t)). \label{ung1}%
\end{equation}
Hence setting y=$\frac{x}{\sqrt[4]{\varepsilon}}=Y_{\varepsilon}(0), $for
y$\in
B(\delta\varepsilon^{\alpha-\frac{1}{4}})$:%
\[
\mathrm{VI}=E_{y}\left(  (1-\chi_{1})\chi_{(t<\tau_{\epsilon}^{x})}%
e^{-\int_{0}^{t}c_{\epsilon}(\sqrt[4]{\varepsilon}Y_{\epsilon}(s))ds}%
;\underset{[0, t\wedge\tau_{\varepsilon}^{x}]}{\sup}||Z_{\varepsilon
}(s)||_{\mathbb{R}^{m}}<\varepsilon^{\beta}\right)
\]%
\begin{align*}
\mathrm{VI}  &  \leq E_{y}\left(  (1-\chi_{1})\chi_{(t<\tau_{\epsilon}^{x}%
)}\exp-(1-\delta\sqrt[4]{\varepsilon}C_{1})\int_{0}^{t}q(Y_{\varepsilon
}(s))ds;\underset{[0, t\wedge\tau_{\varepsilon}^{x}]}{\sup}||Z_{\varepsilon
}(s)||_{\mathbb{R}^{m}}<\varepsilon^{\beta}\right) \\
\mathrm{VI}  &  \geq E_{y}\left(  (1-\chi_{1})\chi_{(t<\tau_{\epsilon}^{x}%
)}\exp-(1+\delta\sqrt[4]{\varepsilon}C_{1})\int_{0}^{t}q(Y_{\varepsilon
}(s))ds;\underset{[0, t\wedge\tau_{\varepsilon}^{x}]}{\sup}||Z_{\varepsilon
}(s)||_{\mathbb{R}^{m}}<\varepsilon^{\beta}\right)
\end{align*}
Using the inequality(\ref{ung1}):%
\begin{align*}
\mathrm{VI}  &  \leq E_{y}\left[  (1-\chi_{1})\chi_{(t<\tau_{\epsilon}^{x}%
)}\exp(1-\delta\sqrt[4]{\varepsilon}C_{1})\int_{0}^{t}q(Z_{\varepsilon
}(s)ds-\right. \\
&  \left.  (1-\delta\sqrt[4]{\varepsilon}C_{1})\frac{1}{2}\int_{0}%
^{t}q(y+\sqrt{2}W(t))ds;\underset{[0, t\wedge\tau_{\varepsilon}^{x}]}{\sup
}||Z_{\varepsilon}(s)||_{\mathbb{R}^{m}}<\varepsilon^{\beta}\right]
\end{align*}%
\[
\mathrm{VI}\leq C_{4}(\delta, \varepsilon, t)E_{0}\left(  \exp-(1-\delta
\sqrt[4]{\varepsilon}C_{1})\frac{1}{2}\int_{0}^{t}q(y+\sqrt{2}W(s))ds\right)
\]
where:%
\begin{align*}
C_{4}(\delta, \varepsilon, t)  &  =\exp\left(  (1-\delta\sqrt[4]{\varepsilon
}C_{1})t\sqrt{\varepsilon}\max\{q(v)|\text{ }||v||_{\mathbb{R}^{m}%
}<\varepsilon^{\beta}\}\right) \\
C_{4}(\delta, \varepsilon, t)  &  =\exp\left(  1-\delta\sqrt[4]{\varepsilon
}C_{1})\varepsilon^{\frac{1}{2}+2\beta}\max\{q(v)|\text{ }||v||_{\mathbb{R}%
^{m}}<1\}\right).
\end{align*}
Note that $C_{4}(\delta, \varepsilon, t)$---%
$>$%
1 as $\varepsilon$---%
$>$%
0 (keeping t fixed).
\noindent Let us estimate the expectation $E_{0}\left(  \exp-(1-\delta
\sqrt[4]{\varepsilon}C_{1})\frac{1}{2}\int_{0}^{t}q(y+W(t))ds\right) .$ For
simplicity set $a(\varepsilon)=(1-\delta\sqrt[4]{\varepsilon}C_{1}).$ Then:%
\[
E_{0}\left(  \exp-\frac{a(\varepsilon)}{2}\int_{0}^{t}q(y+\sqrt{2}%
W(t))ds\right)  ={ \Pi}_{i=1}^{m}E_{0}\left(  \exp-\frac{a(\varepsilon
)}{2}\int_{0}^{t}\lambda_{i}(y_{i}+\sqrt{2}W_{i}(s))^{2}ds\right)
\]
where $y$=($y_{1}$, ...$y_{m}$) and W(t)=(W$_{1}$(t), ..., W$_{m}$(t)).

To find the value of $E_{0}\left(  \exp-\frac{a(\varepsilon)}{2}\int_{0}%
^{t}\lambda_{i}(y_{i}+\sqrt{2}W_{i}(s))^{2}ds\right)  $ we use the following
lemma:

\begin{lem}
\label{noyau} Consider $w(s)$ the Brownian motion in $\mathbb{R}$,  a positive
real number $\lambda>0$ and the function $z_{\mu}^{\lambda}$:
\[
(x, t)\in\mathbb{R}\times\mathbb{R}^{+}\rightarrow E_{x}\left[  e^{-\int
_{0}^{t}\lambda(\mu w(s)+w(s)^{2})ds}\right]
\]
where $E_{x}$ is the expectation for a process starting at a point $x$. Then
$z_{\mu}^{\lambda}$ is a bounded solution of the parabolic equation,
\begin{align}
\frac{\partial z}{\partial t}  &  =\frac{1}{2}\frac{\partial^{2}z}{\partial
x^{2}}-\lambda(x^{2}+\mu x)z, \text{for }x\in\mathbb{R}, t\geq0\label{FKF}\\
z(0, x)  &  =1.\nonumber
\end{align}
For all $x\in\mathbb{R}, $t$\geq0, $the solution is given by
\[
z_{\mu}^{\lambda}(x, t)=\frac{1}{\sqrt{\cosh(t\sqrt{2\lambda})}}\exp
{(-\frac{\sqrt{\lambda}\tanh(t\sqrt{2\lambda})}{\sqrt{2}}}\left(  {x+}%
\frac{\mu}{2}\right)  ^{2}{+\frac{\lambda\mu^{2}t}{4})}%
\]
and the value of the expectation at the origin 0 is,
\[
z_{\mu}^{\lambda}(x, t)=\frac{1}{\sqrt{\cosh(t\sqrt{2\lambda})}}\exp
{(-\frac{\sqrt{\lambda}\tanh(t\sqrt{2\lambda})}{\sqrt{2}}}\left(  \frac{\mu
}{2}\right)  ^{2}{+\frac{\lambda\mu^{2}t}{4})}.%
\]

\end{lem}

\bigskip

{\bf \noindent Assuming the lemma we pursue the evaluation of}${\ }${:}%
\begin{align*}
E_{0}\left(  \exp-\frac{a(\varepsilon)}{2}\int_{0}^{t}\lambda_{i}(y_{i}%
+\sqrt{2}W_{i}(s))^{2}ds\right)  \\=e^{-\frac{a(\varepsilon)t\lambda_{i}%
y_{i}^{2}}{2}}E_{y_{i}}\left(  \exp-\frac{a(\varepsilon)}{2}\int_{0}%
^{t}\lambda_{i}[2\sqrt{2}y_{i}w_{i}(s)+\sqrt{2}W_{i}(s)^{2}]ds\right)
\end{align*}

\begin{align*}
E_{0}\left(  \exp\left[  -\frac{a(\varepsilon)}{2}\int_{0}^{t}\lambda
_{i}(y_{i}+\sqrt{2}W_{i}(s))^{2}ds\right]  \right)   &  =& \\\frac{e^{-\frac
{a(\varepsilon)t\lambda_{i}y_{i}^{2}}{2}}}{\sqrt{\cosh(t\sqrt{2a(\varepsilon
)\lambda_{i}})}}\exp{-}\left(  {\frac{\sqrt{a(\varepsilon)\lambda_{i}}%
\tanh(t\sqrt{2a(\varepsilon)a\lambda_{i}})}{2\sqrt{2}}}y_{i}^{2}\right.
&  {+}\left.  \frac{a(\varepsilon){{\lambda}}_{i}y_{i}^{2}{t}}{2}\right),
\end{align*}
and
\[
E_{0}\left(  \exp\left[  -\frac{a(\varepsilon)}{2}\int_{0}^{t}\lambda
_{i}(y_{i}+\sqrt{2}W_{i}(s))^{2}ds\right]  \right)  =\frac{\exp\left(
{-\frac{\sqrt{a(\varepsilon)\lambda_{i}}\tanh(t\sqrt{2a(\varepsilon
)\lambda_{i}})}{2\sqrt{2}}}y_{i}^{2}\right)  }{\sqrt{\cosh(t\sqrt
{2a(\varepsilon)\lambda_{i}})}}.%
\]
Finally:%
\[
\mathrm{VI}\leq\frac{C_{4}(\delta, \varepsilon, t)}{\sqcap_{i=1}^{m}\sqrt
{\cosh(t\sqrt{2a(\varepsilon)\lambda_{i}})}}\exp\left(  {-}\sum_{i=1}%
^{m}{\frac{\sqrt{a(\varepsilon)\lambda_{i}}\tanh(t\sqrt{2a(\varepsilon
)\lambda_{i}})}{2\sqrt{2}}}y_{i}^{2}\right).
\]

\subsubsection*{End of the proof of Proposition \ref{estim}}


Now we can wrap up the proof of Proposition \ref{estim}. For any $\alpha
\in]0, \frac{1}{6}[$, any $\beta\in]0, \alpha\lbrack$ , any $\varepsilon$
$\in]0, \varepsilon_{0}]$ where $\varepsilon_{0}$%
$<$%
1  depends only on $\alpha$, $\beta, \delta$ and any integer k,  we get the
estimate if x$\in B_{P}(\frac{\delta\varepsilon^{\alpha}}{2})$:%
\begin{align*}
\frac{u_{\epsilon}(x)}{\overline{u_{\varepsilon}}}e^{-\mu_{\varepsilon}t}  &
\leq A(k, \delta)\epsilon^{k}+\frac{C_{4}(\delta, \varepsilon, t)}{\sqcap
_{i=1}^{m}\sqrt{\cosh(t\sqrt{2a\lambda_{i}})}}\exp\left(  {-}\sum_{i=1}%
^{m}{\frac{\sqrt{a(\varepsilon)\lambda_{i}}\tanh(t\sqrt{2a(\varepsilon
)\lambda_{i}})}{2\sqrt{2\varepsilon}}}x_{i}^{2}\right) \\
&  +2m\left[  \exp-\frac{(\varepsilon^{\beta}-tC_{3}\delta\varepsilon^{\alpha
})^{2}}{2mtC_{2}^{2}\delta^{4}\varepsilon^{4\alpha}}+\exp-\frac{[\frac
{\delta\varepsilon^{\alpha}}{2}-M_{1}\sqrt{\varepsilon})]^{2}}{2mtM^{2}%
\sqrt{\varepsilon}}\right]
\end{align*}
Define the function $w_{\varepsilon}$:B$_{P}$($\delta)\rightarrow\hbox{\bb
R}_{+}, $%
\[
w_{\varepsilon}(y)=\frac{u_{\epsilon}(y\sqrt[4]{\varepsilon})}{\overline
{u_{\varepsilon}}}%
\]
Then for y$\in$B$_{P}(\delta/\varepsilon^{\frac{1}{4}-\alpha})$:%
\begin{align*}
w_{\varepsilon}(y)e^{-\mu_{\varepsilon}t}  &  \leq A(k, \delta)\epsilon
^{k}+\frac{C_{4}(\delta, \varepsilon, t)}{\sqcap_{i=1}^{m}\sqrt{\cosh
(t\sqrt{2a\lambda_{i}})}}\exp\left(  {-}\sum_{i=1}^{m}{\frac{\sqrt
{a(\varepsilon)\lambda_{i}}\tanh(t\sqrt{2a(\varepsilon)\lambda_{i}})}%
{2\sqrt{2}}}y_{i}^{2}\right) \\
&  +2m\left[  \exp-\frac{(\varepsilon^{\beta}-tC_{3}\delta\varepsilon^{\alpha
})^{2}}{2mtC_{2}^{2}\delta^{4}\varepsilon^{4\alpha}}+\exp-\frac{[\frac{\delta
}{2}-tM_{1}\sqrt{\varepsilon})]^{2}}{2mtM^{2}\sqrt{\varepsilon}}\right]
\end{align*}
We want to estimate $\sup\{w_{\varepsilon}(y)|$ y$\in B_{P}(\delta
/\varepsilon^{\frac{1}{4}})\}.$%

\begin{align*}
\sup\{w_{\varepsilon}(y)|y  &  \in B_{P}(\delta/\varepsilon^{\frac{1}{4}%
})\}\leq\sup\{w_{\varepsilon}(y)|\text{ y}\in B_{P}(\delta/\varepsilon
^{\frac{1}{4}})-B_{P}(\delta/2\varepsilon^{\frac{1}{4}-\alpha})\}, \\
\sup\{w_{\varepsilon}(y)|y  &  \in B_{P}(\delta/\varepsilon^{\frac{1}{4}%
})\}\leq\sup\{w_{\varepsilon}(y)|y\in B_{P}(\delta/2\varepsilon^{\frac{1}%
{4}-\alpha})\})
\end{align*}

\[
\sup\{w_{\varepsilon}(y)|\text{ y}\in B_{P}(\delta/\varepsilon^{\frac{1}{4}%
})-B_{P}(\delta/2\varepsilon^{\frac{1}{4}-\alpha})\}\leq\sup\{\frac
{u_{\varepsilon}(x)}{\overline{u}_{\varepsilon}}|\text{ x}\in B_{P}%
(\delta)-B_{P}(\delta\varepsilon^{\alpha}/2)\}
\]

Using Appendix 2 as in the evaluation of the boundary integral $\mathrm{I}$,
taking $\varepsilon=\varepsilon,$ b=0,  $\psi=c-\underset{V}{\min}c$,
c$_{\varepsilon}=\underset{V}{\min}$ c--$\lambda_{\varepsilon}$ we get for all
n $\geq$ 1
\[
\sup\{\frac{u_{\varepsilon}(x)}{\overline{u}_{\varepsilon}}|\text{ x}\in
B_{P}(\delta)-B_{P}(\delta\varepsilon^{\alpha}/2)\}\leq C(n)\frac
{\varepsilon^{\frac{n}{3}}}{\left(  \min\text{ }\psi\right)  ^{n}}%
\]
where min $\psi$ is the minimum of $\psi$ on $B_{P}(\delta)-B_{P}%
(\delta\varepsilon^{\alpha}/2)$ and C(n) is a constant depending on the data
$g$, $c$ and on n\ but not on $\varepsilon$. Now there exists a constant Q%
$>$%
0,  such that for all $x\in$V,  $c(x)\geq\underset{V}{\min}$ $c$+Qd$_{g}%
(x, C_{\min})^{2}.$ Hence min $\psi\geq\frac{Q}{4}(\delta\varepsilon^{\alpha
})^{2}.$ Since $\alpha<\frac{1}{6}, $ we see that for any k$\in\mathbb{N}, $
there exists a constant $\gamma(k)>0,$ depending on $\delta, $ such that:%
\[
\sup\{\frac{u_{\varepsilon}(x)}{\overline{u}_{\varepsilon}}|\text{ x}\in
B_{P}(\delta)-B_{P}(\delta\varepsilon^{\alpha}/2)\}\leq\gamma(k)\varepsilon
^{k}%
\]
Hence:%
\[
\sup\{w_{\varepsilon}(y)|\text{ y}\in B_{P}(\delta/\varepsilon^{\frac{1}{4}%
})-B_{P}(\delta/2\varepsilon^{\frac{1}{4}-\alpha})\}\leq\gamma(k)\varepsilon
^{k}%
\]
Finally for x$\in B_{P}(\delta)$:%
\[
e^{-\mu_{\varepsilon}t}\frac{u_{\varepsilon}(x)}{\overline{u}_{\varepsilon}%
}\leq\frac{C_{4}(\delta, \varepsilon, t)}{\prod_{i=1}^{m}\sqrt{\cosh
(t\sqrt{2a(\varepsilon)\lambda_{i}})}}\exp\left(  {-}\sum_{i=1}^{m}%
{\frac{\sqrt{a(\varepsilon)\lambda_{i}}\tanh(t\sqrt{2a(\varepsilon)\lambda
_{i}})}{2\sqrt{2\varepsilon}}}x_{i}^{2}\right)  +
\]%
\[
+(A(k, \delta)+\gamma(k))\epsilon^{k}+2m\left[  \exp-\frac{(\varepsilon^{\beta
}-tC_{3}\delta\varepsilon^{\alpha})^{2}}{2mtC_{2}^{2}\delta^{4}\varepsilon
^{4\alpha}}+\exp-\frac{[\frac{\delta\varepsilon^{\alpha}}{2}-tM_{1}%
\sqrt{\varepsilon})]^{2}}{2mtM^{2}\sqrt{\varepsilon}}\right]
\]

Equivalently,  for all y$\in B_{P}(\delta/\sqrt[4]{\varepsilon})$:%
\begin{equation}
w_{\varepsilon}(y)e^{-\mu_{\varepsilon}t}\leq\frac{C_{4}(\delta, \varepsilon
, t)}{\prod_{i=1}^{m}\sqrt{\cosh(t\sqrt{2a(\varepsilon)\lambda_{i}})}}%
\exp\left(  {-}\sum_{i=1}^{m}{\frac{\sqrt{a(\varepsilon)\lambda_{i}}%
\tanh(t\sqrt{2a(\varepsilon)\lambda_{i}})}{2\sqrt{2}}}y_{i}^{2}\right)  +
\label{ung2}%
\end{equation}%
\[
+(A(k, \delta)+\gamma(k))\epsilon^{k}+2m\left[  \exp-\frac{(\varepsilon^{\beta
}-tC_{3}\delta\varepsilon^{\alpha})^{2}}{2mtC_{2}^{2}\delta^{4}\varepsilon
^{4\alpha}}+\exp-\frac{[\frac{\delta\varepsilon^{\alpha}}{2}-tM_{1}%
\sqrt{\varepsilon})]^{2}}{2mtM^{2}\sqrt{\varepsilon}}\right]
\]
Given a function f as in the statement of Theorem \ref{thnf},  for any integer
$N$ such that $\underset{\varepsilon, y}{\sup}$ $\frac{|f(\varepsilon
, y)|}{1+||y||_{\mathbb{R}^{m}}^{N}}<+\infty$,  for any integer $k>\frac{m}%
{8}+N$ inequality(\ref{ung2})implies that
\[
\underset{]0, \varepsilon_{0}]}{\sup}\int_{B_{P}(\delta/\sqrt[4]{\varepsilon}%
)}|f(\varepsilon, y)|w(y)^{2}dy<+\infty
\]
\bigskip
This is the statement (i) of Proposition \ref{estim}.
\bigskip

\noindent  (ii)Fixing t and choosing an integer k%
$>$%
$\frac{m}{8}+N$, we have:%
\begin{align*}
&  \underset{\varepsilon\text{--%
$>$%
0}}{\lim}vol(B_{P}(\delta/\sqrt[4]{\varepsilon})\left[  1+\left(  \frac
{\delta}{\sqrt[4]{\varepsilon}}\right)  ^{N}\right]  \left[  A(k, \delta
)\epsilon^{k}+\gamma(k)\epsilon^{k}+2m\exp-\frac{(\varepsilon^{\beta}%
-tC_{3}\delta\varepsilon^{\alpha})^{2}}{2mtC_{2}^{2}\delta^{4}\varepsilon
^{4\alpha}}\right. \\
&  \left.  +2m\exp-\frac{[\frac{\delta\varepsilon^{\alpha}}{2}-tM_{1}%
\sqrt{\varepsilon})]^{2}}{2mtM^{2}\sqrt{\varepsilon}}\right]  ^{2}%
e^{2t\mu_{\varepsilon}}=0
\end{align*}

\bigskip\

Hence there exists an $\varepsilon_{1}(\eta)$ such that for all $\varepsilon
\in]0, \varepsilon_{1}(\eta)]$:
\begin{enumerate}
\item $(\varepsilon^{\beta}-tC_{3}\delta\varepsilon^{\alpha})^{2}\geq
\frac{\varepsilon^{2\beta}}{2}$,
\item $[\frac{\delta\varepsilon^{\alpha}}{2}-tM_{1}\sqrt{\varepsilon})]^{2}%
\geq\frac{\delta^{2}\varepsilon^{2\alpha}}{16}, $
\item a $(\varepsilon)=(1-\delta\sqrt[4]{\varepsilon}C_{1})\geq\frac{1}{2}, $
As a consequence there exists an $\varepsilon(\eta, t, k), 0\leq\varepsilon
(\eta, t, k)\leq\varepsilon_{1}(\eta)$, such that for $\varepsilon\in
]0, \varepsilon(\eta, t, k)]$
\item$vol(B_{P}(\delta/\sqrt[4]{\varepsilon})\left(  (A(k, \delta)+\gamma
(k))\epsilon^{k}+2m\left[  \exp-\frac{(\varepsilon^{\beta}-tC_{3}%
\delta\varepsilon^{\alpha})^{2}}{2mtC_{2}^{2}\delta^{4}\varepsilon^{4\alpha}%
}+\exp-\frac{[\frac{\delta\varepsilon^{\alpha}}{2}-tM_{1}\sqrt{\varepsilon
})]^{2}}{2mtM^{2}\sqrt{\varepsilon}}\right]  \right)  ^{2}e^{2t\mu
_{\varepsilon}}\leq\frac{\eta}{4}.$
\end{enumerate}
Then we can choose a ball B$^{m}$(P, R)
such that for all $\varepsilon\in]0, \varepsilon(\eta, t, k)]:$
\[
\int_{\mathbb{R}^{m}-B^{m}(0, R)}\left(  \frac{C_{4}(\delta, \varepsilon
, t)}{\prod_{i=1}^{m}\sqrt{\cosh(t\sqrt{2a(\varepsilon)\lambda_{i}})}}%
\exp\left(  {-}\sum_{i=1}^{m}{\frac{\sqrt{a(\varepsilon)\lambda_{i}}%
\tanh(t\sqrt{2a(\varepsilon)\lambda_{i}})}{2\sqrt{2}}}y_{i}^{2}\right)
\right)  ^{2}e^{2t\mu_{\varepsilon}}dy\leq\frac{\eta}{4}%
\]
These two inequalities prove the statement (ii) of Proposition
\ref{estim}.\QED

\subsubsection{Proof of Lemma \ref{noyau}}

\bigskip{\textbf{\ }} The fact that the function $z_{\mu}^{\lambda}$ satisfies
equation (\ref{FKF}) is a consequence of Feynman-Kac formula (\cite{Schuss}).
Making the change of coordinate $x\rightarrow x-\mu/2$,  the problem becomes:
\
\[
\frac{\partial z}{\partial t}=\frac{1}{2}\frac{\partial^{2}z}{\partial x^{2}%
}-\lambda(x^{2}-\frac{\mu^{2}}{4})z
\]

\[
z(0, x)=1
\]
To solve this equation,  by the uniqueness of solutions for the Cauchy problem
for the parabolic equation (\cite{Krylov}),  it is enough to find a solution
of
the type $e^{-\phi(t)x^{2}+\psi(t)}$. A simple computation leads to the two
coupled equations:
\begin{align*}
\dot{\phi}+2\phi^{2}  &  =\lambda\\
\dot{\psi}+\phi &  =\frac{\lambda\mu^{2}}{4}%
\end{align*}
which can be solved easily using the initial condition at time zero,
\begin{align*}
\phi(t)  &  =\sqrt{\frac{\lambda}{2}}\tanh(t\sqrt{2\lambda})\\
\psi(t)  &
=-\frac{1}{2}\log\cosh(t\sqrt{2\lambda})+\frac{\lambda\mu^{2}t}{4}%
\end{align*}
so we obtain the expression
\[
z_{\mu}^{\lambda}(x, t)=\frac{1}{\sqrt{\cosh(t\sqrt{2\lambda})}}\exp\left(
{-\frac{\sqrt{\lambda}\tanh(t\sqrt{2\lambda})}{\sqrt{2}}}\left(  {x+}\frac
{\mu}{2}\right)  {^{2}+\frac{\lambda\mu^{2}t}{4}}\right).
\] \QED

\bigskip

\subsection{Proof of the main theorem}

\label{refharmonique}
\bigskip We use the notations of the preceding sections.{\textbf{\ }} Because
$x=(x_{1}, .., x_{m})$ is a normal coordinate system centered at $P\in
C_{\min}%
$,  the eigenfunction $w_{\varepsilon}$ satisfies the following renormalized
equation,  derived from equation (\ref{edpV}), in the blown up coordinates,
$y=\sqrt[4]{\epsilon}x$:
\begin{equation}
\Delta_{g_{\epsilon}}w_{\epsilon}+\frac{c(\sqrt[4]{\varepsilon}x)-\underset
{V}{\min}c}{\sqrt[2]{\varepsilon}}w_{\epsilon}=\frac{\lambda_{\epsilon
}-\underset{V}{\min}c}{\sqrt[2]{\varepsilon}}w_{\epsilon}, \text{ in }\frac
{1}{\sqrt[4]{\varepsilon}}\Omega=\frac{1}{\sqrt[4]{\varepsilon}}x_{1}%
\times...\times x_{m}(U) \label{edpV2}%
\end{equation}
where $g_{\epsilon}$ is the metric rescaled by $\epsilon$,  converging to the
Euclidean metric uniformly on every compact set of $\mathbb{R}^{m}$ in the
C$^{\infty}$ topology:%
\[
g_{\varepsilon, ij}(y)=g_{ij}(y\sqrt[4]{\varepsilon}), \text{on }\frac
{1}{\sqrt[4]{\varepsilon}}\Omega
\]

\[
\Delta_{g_{\epsilon}}w=-\sum_{i, j=1}^{m}g^{ij}(y\sqrt[4]{\varepsilon}%
)\frac{\partial^{2}w(y)}{\partial y^{i}\partial^{j}y}+\sum_{i, j=1}^{m}%
\sqrt[4]{\varepsilon}g^{ij}(y\sqrt[4]{\varepsilon})\Gamma_{ij}^{k}%
(y\sqrt[4]{\varepsilon})(\frac{\partial w(y)}{\partial y^{k}})
\]

Remark that all coefficients of the partial differential equation
\ (\ref{edpV2}) are bounded. In particular this term is bounded with
$\epsilon$. By Lemma \ref{lm1},  the quantity $\frac{\lambda_{\epsilon}%
-\min_{V} c}{\epsilon^{1/2}}$ is bounded and nonnegative.

By the classical theory of elliptic partial differential equations,  since
$w_{\epsilon}$ is bounded by 1 for all $\varepsilon$, any sequence
\{$w_{\epsilon_{n}}\}$ with $\varepsilon_{n}$ tending to 0 contains a
subsequence,  still denoted by \{$w_{\epsilon_{n}}\}$ for simplicity, \ which
converges to a solution of the following elliptic equation \ in the
C$^{\infty}$ topology:%

\begin{equation}
\Delta_{E}w+\sum_{n=1}^{m}\lambda_{n}y_{n}^{2}w=\lambda w\text{ on }%
\mathbb{R}^{m} \label{edpfdtnf}%
\end{equation}
where:%
\[
\lambda=\underset{n\text{---%
$>$%
}\infty}{\lim\text{ \ }}\frac{\lambda_{\epsilon_{n}}-\min_{V}c}{\epsilon
_{n}^{1/2}}%
\]
and for simplicity:%
\[
\lambda_{n}=\lambda_{n}(P)
\]
For every compact K$\in\mathbb{R}^{m}:$%
\[
\int_{K}w(y)^{2}dy=\underset{n-->\infty}{\lim}\int_{K}w_{\varepsilon_{n}%
}(y)^{2}dy\leq\underset{]0, \varepsilon_{0}]}{\sup}\int_{B_{P}(\delta
/\sqrt[4]{\varepsilon})}w_{\varepsilon}(y)^{2}dy<+\infty
\]
Hence w$\in L^{2}(\mathbb{R}^{m})$ and
$\vert$%
$\vert$%
w%
$\vert$%
$\vert$%
$_{L^{2}(\mathbb{R}^{m})}\leq\underset{n-->\infty}{\lim}$
$\vert$%
$\vert$%
w$_{\varepsilon_{n}}$%
$\vert$%
$\vert$%
$_{L^{2}(\mathbb{R}^{m})}.$ In fact w=$\underset{n-->\infty}{\lim}$
w$_{\varepsilon_{n}}$ in $L^{2}(\mathbb{R}^{m}).$ To see this assume that
$\underset{n}{\inf}$ \ $||w_{\varepsilon_{n}}-w||_{L^{2}(\mathbb{R}^{m})}%
\geq\xi>0.$ By (ii) of Proposition \ref{estim} and the fact that w$\in
L^{2}(\mathbb{R}^{m}), $ we can find a compact K and an integer $N_{1}$ such
that $\int_{\mathbb{R}^{m}-K}w_{\varepsilon_{n}}(y)^{2}dy\leq\left(  \frac
{\xi}{4}\right)  ^{2}$ if \ \ $n$ $\geq N_{1}, $ and $\int_{\mathbb{R}^{m}%
-K}w(y)^{2}dy\leq\left(  \frac{\xi}{4}\right)  ^{2}.$

Now there exists an integer $N_{2}$ such that $\int_{K}\left(  w_{\varepsilon
_{n}}(y)-w(y)\right)  ^{2}dy\leq\left(  \frac{\xi}{4}\right)  ^{2}$ if n$\geq
N_{2}.$ Then for n$\geq\max(N_{1}, N_{2}), $\ $||w_{\varepsilon_{n}}%
-w||_{L^{2}(\mathbb{R}^{m})}^{2}\leq3\left(  \frac{\xi}{4}\right)  ^{2}.$This
contradicts the assumption $\underset{n}{\inf}$ \
$\vert$%
$\vert$%
$w_{\varepsilon_{n}}-w$%
$\vert$%
$\vert$%
$_{L^{2}(\mathbb{R}^{m})}\geq\xi>0.$

Since $w$ is not zero,  equation (\ref{edpfdtnf}) is an eigenvalue problem. To
determine the solutions to this problem,  consider the unbounded operator
$L$:$D\longrightarrow L^{2}(\mathbb{R}^{m}), $
\begin{equation}
\label{OpeL}Lw=\Delta_{E}w+\sum_{i=1}^{m}\lambda_{i}y_{i}^{2}w
\end{equation}
and
\[
D= \left\{  u\in L^{2}(\mathbb{R}^{m})|u\in H^{2}(\mathbb{R}^{m}),  \,
[\sum_{n=1}^{m}\lambda_{n}y_{n}^{2}]u\in L^{2}(\mathbb{R}^{m}) \right\}  ,
\]
$L$ is a self adjoint operator,  the spectrum of which we want to compute.

Let us introduce the one dimensional unbounded operators $L_{n}$:$D_{1}$---%
$>$%
$L^{2}(\mathbb{R})$ where $D_{1}$=\{$u\in L^{2}(\mathbb{R})|u\in
H^{2}(\mathbb{R}), ux^{2}\in L^{2}(\mathbb{R})\}$. $L_{n}u=-\frac{d^{2}%
u}{dy^{2}}+\lambda_{n}y^{2}u$,  $L_{n}$ is a self adjoint operator,  the
Hermite
operator. It is well known that $L^{2}(\mathbb{R}^{m})$ can be identified to
the m-fold projective tensor product $L^{2}(\mathbb{R})\widehat{\otimes}%
L^{2}(\mathbb{R})\widehat{\otimes}...,
\widehat{\otimes}L^{2}(\mathbb{R}).$Then
L is the self adjoint extension of the operator $\sum_{n=1}^{m}L_{n}%
:D_{1}\otimes...\otimes D_{1}$---%
$>$%
$L^{2}(\mathbb{R}^{m}).$%

\[
\left(  \sum_{n=1}^{m}L_{n}\right)  (u_{1}(y_{1})\otimes...\otimes u_{m}%
(y_{m}))=\sum_{n=1}^{m}u_{1}(y_{1})\otimes.L_{n}(u_{n})(y_{n}).\otimes
u_{m}(y_{m})
\]
>From this it follows that the spectrum $\sigma(L)$ of L is given by the
formula:
\[
\sigma(L)=\overline{\sigma(L_{1})+...+\sigma(L_{m})}%
\]
where $\overline{\sigma(L_{1})+...+\sigma(L_{m})}$ is the closure of the
set\{$\mu_{1}+...+\mu_{m}|$ $\mu_{k}\in\sigma(L_{k}), 1\leq k\leq m\}.$

In the present case:%
\[
\sigma(L_{n})=\{\text{(}2k+1)\sqrt{\lambda_{n}}|\text{ k}\in\hbox{\bb Z}^{+}%
\}
\]
The eigenfunction (up to multiplication by a scalar) corresponding to the
eigenvalue ($2k+1)\sqrt{\lambda_{n}}$ is the function :%
\[
e^{\frac{-\sqrt{\lambda_{n}}y^{2}}{2}}H_{k}(y\sqrt[4]{\lambda_{n}})
\]
where H$_{n}$ is the k$^{th}$ Hermite polynomial.%
\[
H_{k}(x)=e^{x^{2}}\frac{d^{k}}{dx^{k}}e^{-x^{2}}%
\]

Hence the lowest eigenvalue of the operator $L$ is $\sum_{n=1}^{m}%
\sqrt{\lambda_{n}}.$\ An associated eigenfunction of $L$\ is: $\prod_{n=1}%
^{m}$ $\exp\left(  -\frac{\sqrt{\lambda_{n}}y_{n}^{2}}{2}\right)  .$ Up to
multiplication by a scalar this is the only eigenfunction of $L$ associated to
$\sum_{n=1}^{m}\sqrt{\lambda_{n}}$ and it is strictly of one sign. This
follows from Friedrichs' theorem and the fact that L is a positive self
adjoint operator (see \cite{RS} vol.4 p.207,  Thm. XIII.48). By Proposition
\ref{lm1} :%

\[
\lambda\leq\Lambda
\]
Hence we get:%
\[
\lambda=\Lambda
\]
and:
\[
w(x)=\prod_{n=1}^{m}\exp\left(  -\frac{\sqrt{\lambda_{n}}x_{n}^{2}}{2}\right)
=\exp-\sum_{n=1}^{m}\frac{\sqrt{\lambda_{n}}x_{n}^{2}}{2}%
\]

\begin{Defi}
C$_{\min\min}$ is the subset $\{C$%
$\vert$%
$P\in C_{\min}, $ $\sum_{n=1}^{m}\sqrt{\lambda_{n}(P)}=\Lambda\}$.
\end{Defi}

\quad

The following theorem  amplifies theorem \ref{Potential} and sums up all our
results in the self adjoint case.
\begin{theorem} {\bf Distribution of limit measures.} \label{th4}

(i)Using the notations of Lemma \ref{max},  $\underset
{q\in\mathcal{M}_{\varepsilon}}{\sup}d_{g}^{2}(q, C_{\min\min})\leq
A\epsilon^{1/2}$

(ii)Let $S$ be the weak limit set of the measures $u_{\epsilon}^{2}%
dvol_{g}/\int_{V}$u$_{\varepsilon}^{2}dvol_{g}$ as $\varepsilon$ goes to zero.
Then:%
\[
S=\{\mu\in M(V)|\mu=\sum\left\{  \gamma_{P}\delta_{P}|\text{ P}\in C_{\min
\min}\right\}  ,
\]

(iii)For any sequence \{u$_{\varepsilon_{n}}$\} such that the measures
u$_{\varepsilon_{n}}^{2}vol_{g}/\int_{V}$u$_{\varepsilon_{n}}^{2}dvol_{g}$
converge weakly to $\mu, $ the concentration coefficient $\gamma_{P}$,  due to
the normalization condition in the $L^{2}$ space,  is given
\[
\gamma_{P}=(2\pi)^{\frac{m}{2}}\frac{f_{P}^{2}}{K}\prod_{k=1}^{m}\lambda
_{k}^{-1/4}(P)
\]
where
\[
f_{P}=\lim_{n->\infty}\frac{\sup_{B_{P_{k}}(\delta)}u_{\varepsilon_{n}}}%
{\sup_{V_{m}}u_{\varepsilon_{n}}}%
\]
and
\[
K=(2\pi)^{\frac{m}{2}}\sum_{\text{P}\in C_{\min\min}}f_{P}^{2}\prod_{k=1}%
^{m}\lambda_{k}^{-1/4}(P)
\]

(iii)If \ P is a limit point of the sets of maximum points of the
u$_{\varepsilon_{n}}$ then f$_{P}$=1.

(iv)There is always a $\overline{P}$ such that f$_{P}$=1
\end{theorem}

{\noindent\textbf{Proof.}} We shall use the previous notations. Note that:
\[
\int_{V}{u}_{\epsilon}^{2}dvol_{g}=\sum_{P\in C_{\min}}\int_{B_{P}(\delta)}%
{u}_{\epsilon}^{2}dvol_{g}+\int_{V-C_{\min}(\delta)}{u}_{\epsilon}^{2}dvol_{g}
\]
where
\[
C_{\min}(\delta)=\cup_{P\in C_{\min}}B_{P}(\delta)
\]
>From now on we normalize the u$_{\varepsilon}$ requiring\ $\ $that $\int
_{V}{u}_{\epsilon}^{2}dvol_{g}$=1. Let \{u$_{\varepsilon_{n}}|$\} be any
sequence such that the measures u$_{\varepsilon_{n}}^{2}vol_{g}/\int
_{V\text{u}\varepsilon_{n}}^{2}dvol_{g}$ converge weakly to $\mu\in
$M(V). Equation \ref{ltwo1} \ implies that:
\[
\underset{\varepsilon->0}{\lim}\int_{V-C_{\min}(\delta)}{u}_{\epsilon_{n}}%
^{2}dvol_{g}=0,
\]
Hence:%
\begin{equation}
\underset{\varepsilon->0}{\lim}\sum_{P\in C_{\min}}\int_{B_{P}(\delta)}%
{u}_{\epsilon_{n}}^{2}dvol_{g}=1. \label{lim}%
\end{equation}
For any P$\in C_{\min}$:%
\begin{equation}
\int_{B_{P}(\delta)}{u}_{\epsilon}^{2}dvol_{g}=\overline{u}_{\epsilon}%
^{2}\epsilon_{n}^{m/4}\int_{B_{P}(\delta/\sqrt[4]{\varepsilon_{n}})}%
{w}_{P, \epsilon_{n}}^{2}dvol_{g_{\varepsilon_{n}}}, \label{prt}%
\end{equation}
where ${w}_{P, \epsilon}$ is the restriction of the function $\frac
{u_{\epsilon}(y\sqrt[4]{\varepsilon})}{\overline{u}_{\varepsilon}}$ to B$_{P}%
$($\delta$) extended by 0 outside B$_{P}$($\delta$).

Any sequence of $\epsilon$'s converging to zero contains a subsequence(
$\epsilon_{n}$) such that for any P in $C_{min}, $ $\int_{B_{P}(\delta)}%
{u}_{\epsilon_{n}}^{2}dvol_{g}$ converges as n goes to $\infty$ and
${w}_{P, \epsilon_{n}}$ converges to $w_{P}(x)=f_{P}\exp\left(  -\sum_{n=1}%
^{m}\frac{\lambda_{n}(P)x_{n}^{2}}{2}\right)  $ by Theorem \ref{thnf}. Also by
the same theorem f$_{P}$=1 for at least one P. Hence the sequence
\{$\overline{u}_{\epsilon_{n}}^{2}\epsilon_{n}^{m/4}|n\in\hbox{\bb N}\}$ will
converge to a constant K$>0$. We claim that:
\[
\underset{\varepsilon->0}{\lim}\int_{B_{P}(\delta/\sqrt[4]{_{\varepsilon}}%
)}{w}_{P, \epsilon_{n}}^{2}dvol_{g_{\varepsilon n}}=\int_{\mathbb{R}^{m}}%
w_{P}(y)^{2}dy=f_{P}^{2}\frac{(2\pi)^{\frac{m}{2}}}{\sqrt[4]{\prod_{k=1}^{m}\lambda_{k}(P)}}%
\]
To see this note that:%
\[
\int_{B_{P}(\delta/\sqrt[4]{\varepsilon})}{w}_{P, \epsilon_{n}}^{2}%
dvol_{g_{\varepsilon_{n}}}=\int_{B_{P}(\delta/\sqrt[4]{\varepsilon})}%
{w}_{P, \epsilon_{n}}^{2}(y)\sqrt{\det(g_{\varepsilon_{n}})}dy.
\]
Using Morse's lemma on $U_{P}$:%
\[
g_{ij}(x)=\delta_{ij}+\sum_{k, l=1}^{m}G_{ijkl}(x)x_{k}x_{l},%
\]%
\[
g_{\varepsilon ij}(y)=g_{ij}(y\sqrt[4]{\varepsilon})=\delta_{ij}
+\sqrt{\varepsilon}\sum_{k,
l=1}^{m}G_{ijkl}(y\sqrt[4]{\varepsilon})y_{k}y_{l}.%
\]
Hence:%
\[
\sqrt{\det(g_{\varepsilon ij})(y)}=1+\sqrt{\varepsilon}G(\varepsilon, y)
\]
where:%
\[
|G(\varepsilon, y)|\leq C(M)(1+||y||_{\mathbb{R}^{m}}^{2m})
\]
C(M) is a constant depending only on sup\{%
$\vert$%
$G_{ijkl}(x)$%
$\vert$%
$\vert$%
1$\leq i, j, k, l\leq m, $x$\in B_{p}(\delta)$\}.%
\[
\int_{B_{P}(\delta/\sqrt[4]{\varepsilon_{n}})}{w}_{P, \epsilon_{n}}^{2}%
(y)\sqrt{\det((g_{\varepsilon_{n}})_{ij})}dy=\int_{B_{P}(\delta/\sqrt[4]%
{\varepsilon_{n}})}{w}_{P, \epsilon}^{2}(y)dy+\sqrt{\varepsilon}\int
_{B_{P}(\delta/\sqrt[4]{\varepsilon_{n}})}{w}_{P, \epsilon_{n}}^{2}%
(y)G(\varepsilon_{n}, y)dy
\]
Proposition \ref{estim} (i) implies that:%
\[
\underset{n->\infty}{\lim}\sqrt{\varepsilon_{n}}\int_{B_{P}(\delta
/\sqrt[4]{\varepsilon_{n}})}{w}_{P, \epsilon_{n}}^{2}(y)G(\varepsilon
_{n}, y)dy=0
\]
and:%
\[
\underset{n->\infty}{\lim}\int_{B_{P}(\delta/\sqrt[4]{\varepsilon_{n}})}%
{w}_{P, \epsilon_{n}}^{2}(y)dy=f_{P}^{2}\int_{\mathbb{R}^{m}}\prod_{k=1}%
^{m}\exp\left(  -\frac{\sqrt{\lambda_{k}(P)}y_{k}^{2}}{2}\right)  dy
\]
Hence:%
\[
\underset{n->\infty}{\lim}\int_{B_{P}(\delta/\sqrt[4]{\varepsilon_{n}})}%
{w}_{P, \epsilon_{n}}^{2}dvol_{g_{\varepsilon}}=f_{P}^{2}\frac{(2\pi)^{\frac
{m}{2}}}{\sqrt[4]{\prod_{k=1}^{m}\lambda_{k}(P)}}%
\]%
\[
\underset{n->\infty}{\lim}\sum_{P\in C_{\min}}\int_{B_{P}(\delta)}%
{u}_{\epsilon_{n}}^{2}dvol_{g}=1
\]
We have then the following alternative for any $P\in C_{\min}$:%
\[
\underset{n->\infty}{\lim}\int_{B_{P}(\delta)}{u}_{\epsilon_{n}}^{2}%
dvol_{g}=\text{ }\left\{
\begin{array}
[c]{c}%
0\text{,  if }P\notin\text{C}_{\min\min}\\
\frac{f_{P}^{2}}{K}\frac{(2\pi)^{\frac{m}{2}}}{\prod_{n=1}^{m}\sqrt[4]%
{\lambda_{n}(P)}}, \text{if }P\in\text{C}_{\min\min}%
\end{array}
\right.
\]
Because $\underset{\varepsilon->0}{\lim}\sum_{P\in C_{\min}}\int_{B_{P}%
(\delta)}u_{\epsilon}^{2}dvol_{g}=1, $%
\[
K=\sum_{P\in\mathcal{C}_{\min\min}}\frac{f_{P}^{2}(2\pi)^{\frac{m}{2}}}%
{\prod_{k=1}^{m}\sqrt[4]{\lambda_{k}(P)}}.%
\]
\quad Finally \bigskip Lemma\ref{max} implies (iv). \QED
\bigskip

\textbf{\noindent Remark 1:} As we shall see in the next section,  the
Topological Pressure gives enough information to determine where the
concentration process occurs. This information is contained in the quantity
$c_{0}$ only. But what we actually found is that a second quantity $c_{1}$
which carries some second order information about the potential,  narrows down
the possible concentration set.
\bigskip

\textbf{\noindent Remark 2:} When c is not a Morse function,  it is an open
problem to compute all the limit and the values of the coefficients.
\bigskip

\textbf{\noindent Remark 3:} We insist that not all minimum points are
necessarily charged,  except when the equations admit some symmetries. It is
an
open problem to prove that indeed the minimum points of the potential c where
the Topological Pressure achieves its minimum are all charged.
\bigskip

\textbf{\noindent Remark 4:} The main theorem is useful to interpret the
dynamics of a particle on a compact manifold moving under the influence of
small random noise. Indeed ,  as $\epsilon$ goes to zero,  the density
probability of the particle satisfies the Fokker-Plank equation:
\[
\frac{\partial p}{\partial t}=-\epsilon\Delta p-cp.
\]
where c is a killing term \cite{HMS}. $p$ can be expanded in a series of
eigenfunctions,  $\phi_{\epsilon}^{i}$ associated to the eigenvalues
$\lambda_{i}$,
\[
p(t, x, y)=\sum_{i}e^{-\lambda_{i}t}\phi_{\epsilon}^{i}(x)\phi_{\epsilon}%
^{i}(y).
\]
The previous analysis proves that the particle will most likely be found near
some specific minimal points of $c$. In general,  it is well known by the
standard Exit problem (see \cite{Schuss}) that a particle subjected to an
attractive field can escape outside the domain of attraction,  under the
influence of small random noise.

But in the case of a compact manifold,  they are several bassins of
attraction.
The particle wanders in and out of these bassins. But,  the particle will most
likely be found in the neighborhood of a point of $C_{\min}$,  where the
topological pressure is achieved.
\begin{coro}
Let $W\subset V$ an open subset. Assume that $\partial W\cap C_{min}$ is
empty. If for a sequence $\epsilon_{n}$ tending to zero,  $\lim_{n\rightarrow
\infty}\left[  \int_{W}u_{\epsilon_{n}}^{2}dvol_{g}/\int_{V}u_{\epsilon_{n}%
}^{2}dvol_{g}\right]  >0$,  then $\lim_{n\rightarrow\infty}\underset{W}{\sup
}u_{\epsilon_{n}}=+\infty$.
\end{coro}
\bigskip\textbf{Proof.} After taking a subsequence if need be,  we can assume
that the measures $u_{\epsilon_{n}}^{2}dvol_{g}/\int_{V}u_{\epsilon_{n}}%
^{2}dvol_{g}$ converges weakly to a measure $\sum_{P\in C_{1}}a(P)\delta_{P}$,
C$_{1}$ subset of C$_{\min \min }$,  $a(P)>0$ for all P$\in C_{1}$,

\begin{eqnarray}
\sum_{P\in C_{1}}a(P)=1.\lim_{n\rightarrow\infty
}\left[  \int_{W}u_{\epsilon_{n}}^{2}dvol_{g}/\int_{V}u_{\epsilon_{n}}%
^{2}dvol_{g}\right]  =\sum_{P\in W\cap C_{1}}a(P).
\end{eqnarray}
Hence W$\cap C_{1}$ is not empty. Say Q$\in$W$\cap C_{1}$.
Using the notation of Theorem \ref{th4}
\begin{eqnarray}
f_{Q}^{2}=a(Q)\frac{\sqrt[4]{\prod_{k=1}^{m}\lambda_{k}(Q)}}{(2\pi)^{\frac
{m}{2}}}>0.
\end{eqnarray}
\begin{eqnarray}
\underset{n->\infty}{\lim}\inf\frac{\sup_{W}u_{\epsilon_{n}}}
{\sup_{V}u_{\epsilon_{n}}}\geq\underset{n->\infty}{\lim}\frac{\underset
{B_{Q}(\delta)}{\sup}u_{\epsilon_{n}}}{\sup_{V}u_{\epsilon_{n}}}=f_{Q}>0.
\end{eqnarray}
Lemma\ref{creetin} ends the proof.


\subsection{The set of limit measures}

In the previous section,  we have proved that the limit measures are
concentrated on the set $C_{\min\min}$. However we did not determine the
concentration coefficients completely. Indeed let \{ $u_{\epsilon_{n}}%
|n\in\mathbb{N\}}$ a converging sequence with $\varepsilon_{n}->0$,  If we set
$v_{\epsilon_{n}}=\frac{{u}_{\epsilon_{n}}}{\bar{u}_{\epsilon_{n}}}$,  then
for
each point $P\in C_{\min\min}$,  $\lim_{n->\infty}v_{\epsilon_{n}}%
(P)=\alpha(P)\in\lbrack0, 1]$. The concentration coefficients are given by
\[
\gamma_{P}=\frac{\alpha^2(P)\prod_{n=1}^{m}\lambda_{n}(P)^{\frac{-1}{4}}}%
{\sum_{R\in
C_{\min\min}}\alpha^2(R)\prod_{n=1}^{m}\lambda_{n}(R)^{\frac{-1}{4}%
}}.
\]
At least one of the $\alpha(P)$ is equal to one. At present time,  we do not
know if the $\alpha(P)$ are independent of the sequence \{ $u_{\epsilon_{n}%
}|n\in\mathbb{N\}}$. The modulating factors $\alpha(P)$,  defined on the set
$C_{\min\min}, $ depend only on the ratio of the relative maximum to the
global
maximum of the eigenfunction.Given a function $\beta:C_{\min\min}$---%
$>$%
[0, 1],  is it possible to find \ a converging sequence \{ $u_{\epsilon_{n}%
}|n\in\mathbb{N\}}$ such that $\alpha(P)=\beta(P)$,  for $P\in C_{\min\min}$?
Or is the function $\alpha$ unique? So far to evaluate the concentration
coefficients only the potential $c$ was needed. We shall see in the rest of
this section that actually only a subset of $C_{\min\min}$ can be charged.
The rest of this section is devoted to study the influence of the Riemannian
structure on this selection process.

We will use the following definitions:
\begin{Defi}
A point $P\in C_{\min\min}$ is called a maximally charged point if
\[
\underset{(\epsilon, Q)\rightarrow(0, P)}{\lim\sup}\frac{u_{\epsilon}(Q)}%
{\max_{V_{m}}u_{\epsilon}}=1
\]
\end{Defi}
\begin{Defi}
A point $P\in C_{\min\min}$ is called a charged point if
\begin{equation}
\underset{(\epsilon, Q)\rightarrow(0, P)}{\lim\sup}\frac{u_{\epsilon}(Q)}%
{\max_{V_{m}}u_{\epsilon}}>0
\end{equation}
\end{Defi}

\subsubsection*{Remark.}
At this stage we do not know if indeed the analysis can be pushed further to
prove that whether the limit measure is unique or not. For example can one
find a double well asymmetric potential such that sequence of the first
eigenfunctions will concentrate at only one of the two minimum points and
another sequence only at the other?
\bigskip
\subsection{Expansion of the eigenfunction}
\bigskip

Recall the blown-up function:
\[
w_{P, \varepsilon}(x)=\frac{u_{\varepsilon}(x/\sqrt[4]{\varepsilon})}%
{\overline{u_{\varepsilon}}}%
\]
\bigskip By Theorem \ref{thnf} (i),  $w_{P, \epsilon}$ converges to w$_{P}$ in
L$^{2}(\mathbb{R}^{m})$ when $\varepsilon$ goes to 0.

The divided difference
\[
w_{1, \varepsilon}=\frac{w_{P, \varepsilon}-w_{P}}{\sqrt[4]{\varepsilon}}
\]
\bigskip
as $\varepsilon$ goes to $0, $ converges to $w_{1}$ satisfying the
equation:%
\begin{eqnarray} \label{equaw1}
L_{P}w_{1}+\sum_{1\leq i\leq j\leq k\leq m}c_{ijk}y_{i}y_{j}y_{k}w_{P}=0
\end{eqnarray}
where%
\[
L_{P}=\Delta_{E}+\left[  \sum_{k=1}^{m}\sqrt{\lambda_{k}(P)}y_{k}^{2}%
-\Lambda\right]  Id
\]
and:%
\[
c-\min c=\sum_{k=1}^{m}\lambda_{k}(P)x_{k}^{2}+\sum_{1\leq i\leq j\leq k\leq
m}c_{ijk}x_{i}x_{j}x_{k}+\sum_{1\leq i\leq j\leq k\leq l\leq m}^{m}%
c_{ijkl}(x)x_{i}x_{j}x_{k}x_{l}%
\]
where the $c_{ijk}$ are constants and the $c_{ijkl}, $ $C^{\infty}$ functions
on $U_{P}$. The proof of the convergence of the sequence $w_{1, \varepsilon}$
results from various estimates that can be obtained on the sequence and its
derivatives, which can be found in \cite{Helffer2}. But a rough idea of the
convergence of the
sequence results from the WKB expansion. The estimates are based on the
exponential decay at infinity. Those estimates are not used in the remaing
parts of the article.

\subsubsection{Hermite functions}
\bigskip We need Hermite functions. For \underline{n}$\in\mathbb{Z}_{+}^{m}$
define
\[
H_{\underline{n}}(x_{1}, ..., x_{m})=%
{\displaystyle\prod\limits_{j=1}^{m}}
h_{n_{j}}(x_{j}\sqrt[4]{\lambda_{j}(P)}),
\]
where for n$\in\mathbb{Z}_{+}$%
\[
h_{n}(x)=e^{\frac{x^{2}}{2}}\frac{d^{n}}{dx^{n}}e^{-x^{2}}%
\]
Note that%
\[
L_{P}H_{\underline{n}}=2<\underline{n}, \sqrt{\lambda(P)}>H_{\underline{n}}%
\]
where $\sqrt{\lambda(P)}=(\sqrt{\lambda_1}(P),..,\sqrt{\lambda_n}(P))$.
It is well known and easy to check that%
\[
\int_{\mathbb{R}^{m}}H_{\underline{p}}H_{\underline{q}}dy=0
\]
if $\ \underline{p}, \underline{q}\in\mathbb{Z}_{+}^{m}$ and $\underline{p}%
\neq\underline{q}.$%
\[
\int_{\mathbb{R}^{m}}H_{\underline{p}}^{2}dy=\frac{\pi^{\frac{m}{2}}%
\underline{p!}2^{|\underline{p}|}}{\sqrt[4]{\lambda_1(P)..\lambda_n(P)}}%
\]
where $|\underline{p}|=\sum_{n=1}^{m}p_{n}, $ $\underline{p!}=%
{\displaystyle\prod\limits_{n=1}^{m}}
p_{n}!.$

In particular:H$_{ijk}=-8\sqrt{\lambda_{i}\lambda_{j}\lambda_{k}}y_{i}%
y_{j}y_{k}$, for $1\leq i<j<k\leq m$,
H$_{iij}=-8\lambda_{i}\sqrt{\lambda_{j}%
}y_{i}^{2}y_{j}+4\sqrt{\lambda_{i}\lambda_{j}}y_{j}$ for $1\leq i<j\leq m$ ,
H$_{iii}=12\lambda_{i}y_{i}-8\lambda_{i}^{\frac{3}{2}}y_{i}^{3}.$ To
summarize,
\begin{lem}{\bf Convergence of the renomalization sequence.}
\bigskip

(i)As $\varepsilon$ goes to 0, $w_{1, \varepsilon}$ tends to $w_{1},$ solution
of
the equation:%
\[
\Delta_{E}w_{1}(y)+\left[  \sum_{k=1}^{m}\lambda_{k}(P)y_{k}^{2}%
-\Lambda(P)\right]  w_{1}(y)+\sum_{1\leq i\leq j\leq k\leq m}c_{ijk}%
(P)y_{i}y_{j}y_{k}w_{P}(y)=0
\]
(ii)Up to a multiple of $w_{P}$%
\begin{align*}
w_{1}  &  =-\sum_{1\leq i\leq j\leq k\leq m}\frac{c_{ijk}}{16\sqrt[4]%
{\lambda_{i}(P)\lambda_{j}(P)\lambda_{k}(P)}}\frac{H_{e_{i}+e_{j}+e_{k}}%
(y)}{<e_{i}+e_{j}+e_{k}, \sqrt{\lambda(P)}>}\\
&  \sum_{1\leq i<j\leq m}\left[  \frac{c_{iij}}{8\sqrt[4]{\lambda_{i}%
^{2}(P)\lambda_{j}^{3}(P)}}H_{e_{j}}(y)+\frac{c_{ijj}}{8\sqrt[4]{\lambda
_{i}^{3}(P)\lambda_{j}^{2}(P)}}H_{e_{i}}(y)\right]
\end{align*}

\end{lem}

\subsubsection{Expansion of the eigenvalue} \label{expanse}

\bigskip

\bigskip It is well known that the eigenvalue $\lambda_{\epsilon}$ has an
asymptotic expansion of the type $ \Lambda\sqrt{\varepsilon}+\theta
\varepsilon+...$ (see \cite{Si, Simon1}). We have already seen that:%
\[
\Lambda=\left\{  \sum_{n=1}^{m}\sqrt{\lambda_{n}(P)}|P\in Crit(c)\right\}
\]
In this section we want to compute $\theta$ using minimax procedures. Recall
that:%
\[
\lambda_{\epsilon}=\inf_{u\in H^{1}(V)-\{0\}}\frac{\int_{V}\left[
\epsilon{||\nabla u||}_{g}^{2}+cu^{2}\right]  dvol_{g}{\ }}{\int_{V}%
u^{2}dvol_{g}}%
\]
Let us set:
\[
\theta_{\epsilon}=\frac{\lambda_{\epsilon}-\underset{V\ }{\min}c-\sqrt
{\epsilon}\Lambda}{\epsilon}%
\]
Then:%
\begin{equation}
\theta_{\epsilon}=\inf_{u\in H^{1}(V)-\{0\}}\frac{\int_{V}\left[  {||\nabla
u||}_{g}^{2}+\frac{c-minc-\sqrt{\epsilon}\Lambda}{\epsilon}u^{2}\right]
dvol_{g}{\ }}{\int_{V}u^{2}dvol_{g}} \label{the}%
\end{equation}
\bigskip

 We shall prove that as $\varepsilon$ goes to 0,  $\theta_{\epsilon}$
converges and compute its limit $\theta$. It will follow that the limit
measures are supported in a subset of $C_{\min\min}$. We shall estimate the
variational quotient
\[
\theta_{\epsilon}=\tilde{Q}(u_{\epsilon})=\frac{\int_{V_{m}}\left[  ||\nabla
u_{\epsilon}||_{g}^{2}+\frac{c-minc-\sqrt{\epsilon}\Lambda}{\epsilon
}u_{\epsilon}^{2}\right]  dvol_{g}}{\int_{V_{m}}u_{\epsilon}^{2}dvol_{g}}.
\]
using the previous blow up analysis. Consider a converging subsequence
\{u$_{\varepsilon_{n}}|n\in\hbox{\bb N}\}$,  u$_{\varepsilon_{n}}$ normalized
($\int_{V_{m}}u_{\epsilon_{n}}^{2}dvol_{g}=1$),  such that
\[
u_{\epsilon_{n}}^{2}dvol_{g}\rightarrow\sum_{P\in C_{\min}}\gamma_{P}%
\delta_{P},
\]
where $\gamma_{P_{1}}=1$ for at least one P$_{1}$. To simplifie the notations
we shall drop the index n for $\varepsilon_{n}$ from now on.%
\begin{align*}
\tilde{Q}(u_{\epsilon})  &  =%
{\displaystyle\sum\limits_{P\in C_{\min}}}
\int_{B_{P}(\delta)}\left[  ||\nabla u_{\epsilon}||_{g}^{2}+\frac
{c-minc-\sqrt{\epsilon}\Lambda}{\epsilon}u_{\epsilon}^{2}\right]  dvol_{g}\\
&  +\int_{V-\cup_{P\in C_{\min}}B_{P}(\delta)}\left[  ||\nabla u_{\epsilon
}||_{g}^{2}+\frac{c-minc-\sqrt{\epsilon}\Lambda}{\epsilon}u_{\epsilon}%
^{2}\right]  dvol_{g}%
\end{align*}
Let $\phi$ be a C$^{\infty}$function:V$\longrightarrow$[0, 1], 1 on
V-$\cup_{P\in C_{\min}}$B$_{P}$($\delta$) such that $support\phi\cap C_{\min}%
$=$\varnothing$. For $\varepsilon$ sufficiently small%
\[
\int_{V-\cup_{P\in C_{\min}}B_{P}(\delta)}\left[  ||\nabla u_{\epsilon}%
||_{g}^{2}+\frac{c-minc-\sqrt{\epsilon}\Lambda}{\epsilon}u_{\epsilon}%
^{2}\right]  dvol_{g}\leq\int_{V}\left[  ||\nabla u_{\epsilon}||_{g}^{2}%
+\frac{c-minc-\sqrt{\epsilon}\Lambda}{\epsilon}u_{\epsilon}^{2}\right]  \phi
dvol_{g}%
\]
Using the relation \ref{intbypar},
\[
\int_{V}\left[  ||\nabla u_{\epsilon}||_{g}^{2}+\frac{c-minc-\sqrt{\epsilon
}\Lambda}{\epsilon}u_{\epsilon}^{2}\right]  \phi dvol_{g}=\int_{V}\left[
\frac{\lambda-minc-\sqrt{\epsilon}\Lambda}{\epsilon}\phi-\Delta\phi\right]
u_{\epsilon}^{2}dvol_{g}%
\]
Appendix 2 shows that with an appropriate constant $C$%
\[
\int_{V}\left[  \frac{\lambda-minc-\sqrt{\epsilon}\Lambda}{\epsilon}%
\phi-\Delta\phi\right]  u_{\epsilon}^{2}dvol_{g}=o(1)\epsilon^{m/4}%
\bar{u_{\epsilon}}^{2}%
\]
Hence:%
\[
\tilde{Q}(u_{\epsilon})=%
{\displaystyle\sum\limits_{P\in C_{\min}}}
\int_{B_{P}(\delta)}\left[  ||\nabla u_{\epsilon}||_{g}^{2}+\frac
{c-minc-\sqrt{\epsilon}\Lambda}{\epsilon}u_{\epsilon}^{2}\right]
dvol_{g}+o(1)\epsilon^{m/4}\bar{u_{\epsilon}}^{2}%
\]

\begin{align*}
\tilde{Q}(u_{\epsilon})  &  =\epsilon^{m/4}\bar{u_{\epsilon}}^{2}%
{\displaystyle\sum\limits_{P\in C_{\min}}}
\int_{B_{P}(\delta/\sqrt[4]{\varepsilon})}\left[  \frac{1}{\sqrt{\epsilon}%
}||\nabla w_{P, \epsilon}(y)||_{g}^{2}+\frac{\left(  c(y\sqrt[4]{\varepsilon
})-minc-\sqrt{\epsilon}\Lambda\right)  w_{P, \epsilon}^{2}(y)}{\epsilon
}\right]  \sqrt{\det g_{\varepsilon}(y)}dy\\
&  +o(1)\epsilon^{m/4}\bar{u_{\epsilon}}^{2}%
\end{align*}
The normalization condition implies that
\[
1=\sum_{P\in C_{\min}}\int_{B_{P}(\delta)}u_{\epsilon}^{2}dvol_{g}%
+\int_{V-\cup_{P\in C_{\min}}B_{P}(\delta)}u_{\epsilon}^{2}dvol_{g}%
\]

\begin{equation}
\int_{V-\cup_{P\in C_{\min}}B_{P}(\delta)}u_{\epsilon}^{2}dvol_{g}%
=o(1)\epsilon^{m/4}\bar{u_{\epsilon}}^{2} \label{coco}%
\end{equation}

\begin{equation}
\int_{B_{P}(\delta)}u_{\epsilon}^{2}dvol_{g}=\epsilon^{m/4}\bar{u_{\epsilon}%
}^{2}\int_{\mathcal{B}_{P}\mathcal{(}\sqrt{\rho}/\sqrt[4]{\varepsilon
}\mathcal{)}}w_{P, \epsilon}^{2}(y)\sqrt{\det g(y\sqrt[4]{\varepsilon})}dy
\label{cucu}%
\end{equation}

\bigskip Now$\underset{\mathcal{B}_{P}\mathcal{(}\sqrt{\rho}/\sqrt[4]%
{\varepsilon}\mathcal{)}}{\sup}$%
$\vert$%
$\det g(y\sqrt[4]{\varepsilon})|<+\infty$ and on each compact subset of
$\mathbb{R}^{m}$,  $\det g(y\sqrt[4]{\varepsilon})$ converges uniformly to 1
when $\varepsilon$ goes to0. \ Hence$\int_{\mathcal{B}_{P}\mathcal{(}%
\sqrt{\rho}/\sqrt[4]{\varepsilon}\mathcal{)}}w_{P, \epsilon}^{2}(y)\sqrt{\det
g(y\sqrt[4]{\varepsilon})}dy$ tends to $\int_{\mathbb{R}^{m}}w_{P}(y)^{2}dy$
as $\varepsilon$ goes to 0. By Theorem \ref{thnf}%
\[
\int_{\mathbb{R}^{m}}w_{P}(y)^{2}dy=(2\pi)^{\frac{m}{2}}f_{P}^{2}\prod
_{k=1}^{m}\lambda_{k}^{-1/4}(P).
\]
Then relations \ref{coco}-\ref{cucu} imply:%
\begin{eqnarray}
\underset{\varepsilon->0}{\lim}\epsilon^{m/4}\bar{u_{\epsilon}}^{2}=K=1\left/
\sum_{P\in C_{\min}}(2\pi)^{\frac{m}{2}}f_{P}^{2}\prod_{k=1}^{m}\lambda
_{k}^{-1/4}(P)\right.
\end{eqnarray}
\bigskip Now we shall compute the numerator of $\tilde{Q}(u_{\epsilon})$.
\begin{eqnarray}
\underset{\varepsilon\longrightarrow0}{\lim}\tilde{Q}(u_{\epsilon})=K%
{\displaystyle\sum\limits_{P\in C_{\min}}}
\underset{\varepsilon\longrightarrow0}{\lim}\int_{B_{P}(\delta/\sqrt[4]%
{\varepsilon})}\left[  \frac{1}{\sqrt{\epsilon}}||\nabla w_{P, \epsilon
}(y)||_{g}^{2}+ \frac{\left(  c(y\sqrt[4]{\varepsilon})-minc-\sqrt{\epsilon
}\Lambda\right)  w_{P, \epsilon}^{2}(y)}{\epsilon}\right] \nonumber\\ \times
\sqrt{\det
g(y\sqrt[4]{\varepsilon})}dy .
\end{eqnarray}
To simplify the notations below set%
\[
q(y)=\sum_{k=1}^{m}\lambda_{k}(P)y_{k}^{2}-\Lambda
\]
\begin{align*}
\int_{B_{P}(\delta/\sqrt[4]{\varepsilon})}\left[  \frac{1}{\sqrt{\epsilon}%
}||\nabla w_{P, \epsilon}(y)||_{g}^{2}+\frac{\left(  c(y\sqrt[4]{\varepsilon
})-minc-\sqrt{\epsilon}\Lambda\right)  w_{P, \epsilon}^{2}(y)}{\epsilon
}\right]  \sqrt{\det g(y\sqrt[4]{\varepsilon})}dy  &  =\mathrm{I+II+III+IV}\\
&  \mathrm{+V+VI+VII}%
\end{align*}

where,  denoting by $\nabla_{E}f$ the euclidean gradient ($\frac{\partial
f}{\partial y_{1}}, ..., \frac{\partial f}{\partial y_{m}})$%
\[
\mathrm{I=}\frac{1}{\sqrt{\epsilon}}\int_{B_{P}(\delta/\sqrt[4]{\varepsilon}%
)}\left[  ||\nabla_{E}w_{P}(y)||_{\mathbb{R}^{m}}^{2}+qw_{P}(y)^{2}\right]
\sqrt{\det g(y\sqrt[4]{\varepsilon})}dy,
\]
\[
\mathrm{II=}\frac{2}{\sqrt[4]{\epsilon}}\int_{B_{P}(\delta/\sqrt[4]%
{\varepsilon})}\left[  <\nabla_{E}w_{P}, \nabla_{E}w_{1, \varepsilon}%
>+qw_{P}w_{1, \varepsilon}\right]  \sqrt{\det g(y\sqrt[4]{\varepsilon})}dy,
\]
\[
\mathrm{III=}\frac{1}{\sqrt[4]{\epsilon}}\int_{B_{P}(\delta/\sqrt[4]%
{\varepsilon})}\sum_{1\leq i\leq j\leq k\leq m}c_{ijk}y_{i}y_{j}y_{k}w_{P}%
^{2}\sqrt{\det g(y\sqrt[4]{\varepsilon})}dy,
\]
\[
\mathrm{IV=}\int_{B_{P}(\delta/\sqrt[4]{\varepsilon})}\left[  ||\nabla
_{E}w_{1, \varepsilon}(y)||_{\mathbb{R}^{m}}^{2}+qw_{1, \varepsilon}%
(y)^{2}+2\sum_{1\leq i\leq j\leq k\leq m}c_{ijk}y_{i}y_{j}y_{k}w_{P}%
w_{1, \varepsilon}\right]  \sqrt{\det g(y\sqrt[4]{\varepsilon})}dy,
\]
\[
\mathrm{V=}\sqrt[4]{\epsilon}\int_{B_{P}(\delta/\sqrt[4]{\varepsilon})}%
\sum_{1\leq i\leq j\leq k\leq m}c_{ijk}y_{i}y_{j}y_{k}w_{1, \varepsilon}%
^{2}\sqrt{\det g(y\sqrt[4]{\varepsilon})}dy,
\]
\[
\mathrm{VI=}\int_{B_{P}(\delta/\sqrt[4]{\varepsilon})}\sum_{1\leq i\leq j\leq
k\leq l\leq m}^{m}c_{ijkl}(y\sqrt[4]{\varepsilon})y_{i}y_{j}y_{k}%
y_{l}w_{P, \varepsilon}^{2}\sqrt{\det g(y\sqrt[4]{\varepsilon})}dy,
\]
\[
\mathrm{VII=}\int_{B_{P}(\delta/\sqrt[4]{\varepsilon})}\sum_{i, j, k, l=1}%
^{m}g^{ijkl}(y\sqrt[4]{\varepsilon})y_{k}y_{l}\frac{\partial w_{P, \varepsilon
}}{\partial y_{i}}\frac{\partial w_{P, \varepsilon}}{\partial y_{j}}\sqrt{\det
g(y\sqrt[4]{\varepsilon})}dy,
\]
where $g^{ij}(x)=\delta^{ij}+\sum_{k, l=1}^{m}g^{ijkl}(x)x_{k}x_{l}.$In
particular $g^{ijkl}(0)=\frac{1}{6}\left(  R_{iklj}(P)+R_{jkli}(P)\right).$
Note that%
\[
\sqrt{\det g_{\varepsilon}(y)}=1-\frac{\sqrt{\varepsilon}}{6}R_{ij}%
(y\sqrt[4]{\varepsilon})y_{i}y_{j}%
\]
where the $R_{ij}$ are $C^{\infty}$ functions on $U_{P}$ and $R_{ij}%
(0)=Ricc_{ij}(P)$. It is easy to see that%
\[
\underset{\varepsilon\longrightarrow0}{\lim}\mathrm{VII=}\int_{\mathbb{R}^{m}%
}\sum_{i, j, k, l=1}^{m}\frac{1}{6}\left(  R_{iklj}(P)+R_{jkli}(P)\right)
y_{k}y_{l}\frac{\partial w_{P}}{\partial y_{i}}\frac{\partial w_{P}}{\partial
y_{j}}dy,
\]
\[
\underset{\varepsilon\longrightarrow0}{\lim}\mathrm{VI=}\int_{\mathbb{R}^{m}%
}\sum_{1\leq i\leq j\leq k\leq l\leq m}^{m}c_{ijkl}(0)y_{i}y_{j}y_{k}%
y_{l}w_{P}^{2}dy,
\]
\[
\underset{\varepsilon\longrightarrow0}{\lim}\mathrm{V=0}%
\]
\[
\underset{\varepsilon\longrightarrow0}{\lim}\mathrm{IV=}\int_{\mathbb{R}^{m}%
}\left[  ||\nabla_{E}w_{1}(y)||_{\mathbb{R}^{m}}^{2}+qw_{1}(y)^{2}%
+2\sum_{1\leq i\leq j\leq k\leq m}c_{ijk}y_{i}y_{j}y_{k}w_{P}w_{1}\right]  dy
\]
Using equation \ref{equaw1} satisfied by $w_{1}$,
\[
\underset{\varepsilon\longrightarrow0}{\lim}\mathrm{IV=}\int_{\mathbb{R}^{m}%
}\sum_{1\leq i\leq j\leq k\leq m}c_{ijk}y_{i}y_{j}y_{k}w_{P}w_{1}dy
\]
\[
\mathrm{III=}\frac{1}{\sqrt[4]{\epsilon}}\int_{B_{P}(\delta/\sqrt[4]%
{\varepsilon})}\sum_{1\leq i\leq j\leq k\leq m}c_{ijk}y_{i}y_{j}y_{k}w_{P}%
^{2}dy-\frac{\sqrt[4]{\epsilon}}{6}\int_{B_{P}(\delta/\sqrt[4]{\varepsilon}%
)}\sum_{\substack{1\leq i\leq j\leq k\leq m\\1\leq l, n\leq m}}c_{ijk}R_{\ln
}(y\sqrt[4]{\varepsilon})y_{i}y_{j}y_{k}y_{l}y_{n}w_{P}^{2}dy.
\]
For reason of symmetry the first integral is 0 and the second tends to 0 as
$\varepsilon\longrightarrow0$.%
\[
\underset{\varepsilon\longrightarrow0}{\lim}\mathrm{III=0}%
\]
\[
\mathrm{I=}\frac{1}{\sqrt{\epsilon}}\int_{B_{P}(\delta/\sqrt[4]{\varepsilon}%
)}\left[  ||\nabla_{E}w_{P}(y)||_{\mathbb{R}^{m}}^{2}+qw_{P}(y)^{2}\right]
\left(  1-\frac{\sqrt{\varepsilon}}{6}\sum_{ij=1}^{m}R_{ij}(y\sqrt[4]%
{\varepsilon})y_{i}y_{j}\right)  dy
\]
but
\[
\frac{1}{\sqrt{\epsilon}}\int_{B_{P}(\delta/\sqrt[4]{\varepsilon})}\left[
||\nabla_{E}w_{P}(y)||_{\mathbb{R}^{m}}^{2}+qw_{P}(y)^{2}\right]  dy=\frac
{1}{\sqrt{\epsilon}}\int_{S_{P}(\delta/\sqrt[4]{\varepsilon})}w_{P}<\nabla
_{E}w_{P}, \overrightarrow{n}>dA
\]
converges to zero,  due the exponential decay. Here $S_{P}(\delta
/\sqrt[4]{\varepsilon})$ is the sphere boundary of $B_{P}(\delta
/\sqrt[4]{\varepsilon})$, $\overrightarrow{n}$ is the exterior euclidean unit
normal to $S_{P}(\delta/\sqrt[4]{\varepsilon})$ and $dA$ the area measure. It
is then clear that%
\[
\underset{\varepsilon\longrightarrow0}{\lim}\mathrm{I=-}\int_{\mathbb{R}^{m}%
}\left[  ||\nabla_{E}w_{P}(y)||_{\mathbb{R}^{m}}^{2}+qw_{P}(y)^{2}\right]
\frac{1}{6}\sum_{ij=1}^{m}Ric_{ij}(P)y_{i}y_{j}dy,
\]
\[
\mathrm{II=II}_{1}+\mathrm{II}_{2},%
\]
\begin{align*}
\mathrm{II}_{1}  &  =\frac{2}{\sqrt[4]{\epsilon}}\int_{B_{P}(\delta
/\sqrt[4]{\varepsilon})}\left[  <\nabla_{E}w_{P}, \nabla_{E}w_{1, \varepsilon
}>+qw_{P}w_{1, \varepsilon}\right]  dy,\\
\mathrm{II}_{1}  &  =\frac{2}{\sqrt[4]{\epsilon}}\int_{S_{P}(\delta
/\sqrt[4]{\varepsilon})}w_{1, \varepsilon}<\nabla_{E}w_{P}, \overrightarrow
{n}>dA,\\
\underset{\varepsilon\longrightarrow0}{\lim}\mathrm{II}_{1}  &  =0,
\end{align*}
where an intergration by part leads to a boundary term,  which converges to
zero.
\begin{align*}
\mathrm{II}_{2}  &  =-\frac{\sqrt[4]{\varepsilon}}{6}\int_{B_{P}%
(\delta/\sqrt[4]{\varepsilon})}\left[  <\nabla_{E}w_{P}, \nabla_{E}%
w_{1, \varepsilon}>+qw_{P}w_{1, \varepsilon}\right]  \sum_{ij=1}^{m}%
R_{ij}(y\sqrt[4]{\varepsilon})y_{i}y_{j}dy\\
\underset{\varepsilon\longrightarrow0}{\lim}\mathrm{II}_{2}  &  =0.
\end{align*}
Finally $\underset{\varepsilon\longrightarrow0}{\lim}\int_{B_{P}%
(\delta/\sqrt[4]{\varepsilon})}\left[  \frac{1}{\sqrt{\epsilon}}||\nabla
w_{P, \epsilon}(y)||_{g}^{2}+\frac{\left(  c(y\sqrt[4]{\varepsilon}%
)-minc-\sqrt{\epsilon}\Lambda\right)  w_{P, \epsilon}^{2}(y)}{\epsilon}\right]
\sqrt{\det g(y\sqrt[4]{\varepsilon})}dy$ is equal to $\underset{\varepsilon
\longrightarrow0}{\lim}\mathrm{I+}$ \ $\underset{\varepsilon\longrightarrow
0}{\lim}\mathrm{IV+}$ \ $\underset{\varepsilon\longrightarrow0}{\lim
}\mathrm{V+}$ \ $\underset{\varepsilon\longrightarrow0}{\lim}\mathrm{VI.}%
$ Lengthy but straightforward computations show that with $\sqrt[4]{\lambda
(P)}=%
{\displaystyle\prod\limits_{n=1}^{m}}
\sqrt[4]{\lambda_{n}(P)}$:%
\[
\underset{\varepsilon\longrightarrow0}{\lim}\mathrm{I+}\underset
{\varepsilon\longrightarrow0}{\lim}\mathrm{V=-}f_{P}^{2}\frac{\pi^{\frac{m}%
{2}}}{\sqrt[4]{\lambda(P)}}\left[  \frac{R(P)}{4}+\frac{1}{12}\sum_{1\leq
i, j\leq m}R_{ijij}(P)\sqrt{\frac{\lambda_{i}(P)}{\lambda_{j}(P)}}\right]
\]
\[
\ \underset{\varepsilon\longrightarrow0}{\lim}\mathrm{VI=}\frac{1}{4}f_{P}%
^{2}\frac{\pi^{\frac{m}{2}}}{\sqrt[4]{\lambda(P)}}\sum_{1\leq i\leq j\leq
m}\frac{c_{iijj}(P)}{\sqrt{{\lambda_{i}(P)}{\lambda_{j}(P)}}}+\frac{1}{2}%
\sum_{i=1}^{m}\frac{c_{iiii}(P)}{\lambda_{i}(P)}.
\]
The computation of \ \ $\ \underset{\varepsilon\longrightarrow0}{\lim
}\mathrm{IV}$\ is more involved, where%
\[
L_{P}=\Delta_{E}+\sum_{n=1}^{m}x_{n}^{2}\lambda_{n}(P)-\Lambda(P).
\]
Then%
\begin{align*}
\sum_{1\leq i\leq j\leq k\leq m}c_{ijk}y_{i}y_{j}y_{k}w_{P}  &  =f_{P}%
\sum_{1\leq i\leq j\leq k\leq m}\frac{c_{ijk}}{8\sqrt[4]{\lambda_{i}%
(P)\lambda_{j}(P)\lambda_{k}(P)}}H_{e_{i}+e_{j}+e_{k}}(y)+\\
&  f_{P}\sum_{1\leq i<j\leq m}\left[  \frac{c_{iij}}{4\sqrt[4]{\lambda_{i}%
^{2}(P)\lambda_{j}(P)}}H_{e_{j}}(y)+\frac{c_{ijj}}{4\sqrt[4]{\lambda
_{i}(P)\lambda_{j}^{2}(P)}}H_{e_{i}}(y)\right].
\end{align*}
After lengthy computations we get%
\[
\int_{\mathbb{R}^{m}}\sum_{1\leq i\leq j\leq k\leq m}c_{ijk}y_{i}y_{j}%
y_{k}w_{P}w_{1}dy=-\frac{\pi^{\frac{m}{2}}f_{P}^{2}}{\sqrt[4]{\lambda(P)}%
}\left(  A(P)+B(P)+C(P)\right).
\]
\begin{align*}
A(P) &=&\sum_{1\leq i<j<k\leq m}\frac{c_{ijk}^{2}}{16\sqrt{\lambda_{i}%
(P)\lambda_{j}(P)\lambda_{k}(P)}}\frac{1}{<e_{i}+e_{j}+e_{k}, \sqrt{\lambda
(P)}>} \\
B(P)&=&\sum_{0\leq i<j\leq m}\left[  \frac{c_{iij}^{2}}{8\lambda_{i}%
(P)\sqrt{\lambda_{j}(P)}}+\frac{c_{ijj}^{2}}{8\lambda_{j}(P)\sqrt{\lambda
_{i}(P)}}\right], \\
C(P)&=&\sum_{0\leq i<j\leq m}\frac{c_{iij}^{2}+c_{ijj}^{2}}{16\lambda
_{i}(P)\lambda_{j}(P)}+\sum_{i=1}^{m}\frac{c_{iii}^{2}}{8\lambda_{i}^{2}(P)}%
\end{align*}
\begin{theorem}
\label{Th-hk copy(1)} The expansion of the eigenvalue in power of
$\sqrt{\varepsilon}$ is up to order three
\[
\lambda_{\epsilon}=\min_{V}\text{ }c+\Lambda\sqrt{\varepsilon}+\theta
\varepsilon+o(\varepsilon)
\]
where
\[
C_{\min}=\{P\hbox{ such that }c(P)=\min c\},
\]%
\[
\Lambda=\inf\text{ }\left[  \sum_{n=1}^{m}\sqrt{\lambda_{n}(R)}|R\in C_{\min
}\right]  ,
\]%
\[
C_{\min\min}=\{P\hbox{ such that }\Lambda(P)=\sum_{n=1}^{m}\sqrt{\lambda
_{n}(P)}|P\in C_{\min}\}
\]
and
\begin{align}
\theta &  =\min_{P\in C\min\min}\frac{\pi^{\frac{m}{2}}Kf_{P}^{2}}%
{\sqrt[4]{\lambda(P)}}\left[  -\frac{R(P)}{4}-\frac{1}{12}\sum_{i, j}%
R_{ijij}(P)\sqrt{\frac{\lambda_{i}(P)}{\lambda_{j}(P)}+}\right. \label{truc}\\
&  \left.  \frac{1}{4}\sum_{i\leq j}\frac{c_{iijj}(P)}{\sqrt{{\lambda_{i}%
(P)}{\lambda_{j}(P)}}}+\frac{1}{2}\sum_{i}\frac{c_{iiii}(P)}{\lambda_{i}%
(P)}+A(P)+B(P)+C(P)\right] \nonumber
\end{align}

\end{theorem}

{\noindent\textbf{Remarks.}}
\begin{enumerate}
\item The sequence $u_{\epsilon}$ is in $H^{m}$ but not in $H^{s}$,  for
$s>m=dimV$. This is exactly the critical case in the Sobolev embedding
theorem. That why the limits of $u_{\epsilon}$ as $\epsilon$ goes to zero are
measures but not
functions.
\item The coefficients of the limit measure depends on two factors: one is the
second derivative of the potential at some points of $C_{\min\min}$ and second
on the coefficients $\alpha$,  that measure the ratio of the local maximum
versus the global maximum.
\item Actually,  we conjecture that if two points $P_{1}$ and $P_{2}$ are
always
charged and under some conditions of analycity on the metric,  any small
neighborhood $V_{1}$ of $P_{1}$ is conjugated to a neighborhood $V_{2}$ of
$P_{2}$ by an isometry. Such an isometry should leave invariant the blow-up
equation defined on the manifold and satisfies by $w_{\epsilon}$. If in
$V_{1}$,  there exists a subsequence of local maximum,  then such subsequence
is
also present in $V_{2}$,  which garantee the concentration in both places. In
some sense,  when the concentration occurs at those points for any
subsequences,  it implies the existence of hidden symmetries.
\item Suppose that the potential has only two wells centered at $P_{1}$ and
$P_{2}$ that are charged. Obviously when the coefficients of concentration are
not unique,  the set of possible limit coefficients $\alpha_{1}, \alpha_{2}$,
where $\mu=\alpha_{1}\delta_{P_{1}}+\alpha_{2}\delta_{P_{2}}$,  are restricted
to the intervals:
\[
\alpha_{1}\in\lbrack c_{1}/c_{1}+c2, 1]\hbox{ and }\alpha_{2}\in\lbrack
0, c_{2}/(c1+c2)].
\]
Indeed one of the coefficient,  say $\alpha_{1}$ characterize the convergence
of the absolute maximum.
\end{enumerate}
\subsection{Removing the degeneracy\label{dege}}
In the previous analysis we have defined two local characteristics (quantities
that are functions of $g$ and $c$),  $\sum_{n=1}^{m}\lambda_{n}(P)$ and
$-\mathcal{D}(P)+\mathcal{C}(P)$ which have properties that the minimum of the
first one on $C_{min}$ gives the first coefficient of the developpment of
$\lambda_{\epsilon}$ in the power of $\sqrt{\epsilon}$,
\[
\lambda_{\epsilon}=a_{0}+a_{1}\sqrt{\epsilon}+..+a_{n}\sqrt{\epsilon^{n}}+...
\]
$C_{minmin}$ is the subset of $C_{min}$ where the characteristic achieves its
minimum. The second coefficient of the developpment of $\lambda_{\epsilon}$ is
equal to the minimum of the second characteristic on the set $C_{minmin}$. We
could continue this process and define a sequence of characteristics $\chi
_{2}, \chi_{3}, \chi_{4}, ...$($\chi_{0}=\min_{V_{m}}c$,  $\chi_{1}=\Lambda$
and
$\chi_{2}=\theta)$ and subsets of $C_{min}$,  $C_{3}, C_{4}, C_{5}...$ having
the
following property,  for any integer $n$:

\begin{itemize}
\item $\chi_{n}$ is a function defined on $C_{n}$ and depends only on the
covariant derivatives of $c$ and $g$
\item the n$^{th}$ coefficient $c_{n}$ is equal $\min_{C_{n}} \chi_{n}$.
\item $C_{n+1}=\{P\in C_{n}|\chi_{n}(P)=a_{n}\}.$
\end{itemize}

Since the set $C_{min}$ is finite,  the process becomes stationnary after a
finite number of steps i.e. $C_{n} = C_{n+1}$ and $\chi_{n}$ is constant on
$C_{n}$,  for n larger than $\bar{n}$,  some integer. Now there are two
possibilities. Either $C_{\bar{n}}$ is reduced to a single point,  in which
case,  this is the only point that is charged or the so-called degenerate
case,
see (\cite{Simon1} page 304) where $C_{\bar{n}}$ has more than one point.

The knowledge of the expansion of $\lambda_{\epsilon}$ is not enough to
compute the value of the concentration coefficient. It is an open problem to
compute precisely this value in terms of the geometry and the extracted
subsequence.

Finally we conjecture that for any points $P_{1}, P_{2}$ in $C_{\bar{n}}$,
there exists a isometry $\phi:\Omega_{1}\rightarrow\Omega_{2}$,  where
$\Omega_{i}$ is an open neighborhood of $P_{i}$ \ such that denoting by
$c_{1}, c_{2}$ the restrictions of $c$ to $\Omega_{1}, \Omega_{2}$
respectively,
$c_{2}\circ\phi=c_{1}$ and $w_{\epsilon}|_{\Omega_{2}} = w_{\epsilon}%
|_{\Omega_{1}}\circ\phi$.


\section{Blow up analysis with a gradient vector field}

In this section,  we study the behavior of the sequence of first
eigenfunctions,  when the vector field $b$ is the gradient of a Morse
function,
$b=\nabla\phi$,  with respect to the metric g. It is possible to describe
explicitly the set of limit measures and we will see how the supports of these
limit measures coincide with those suggested by the analysis of the
Topological Pressure (\cite{Kifer90}). More precisely,  the potential $c$ and
the field $b=\nabla\phi$ interact to restrict the limits of the
eigenfunctions,
to certain subsets of the critical set of $b$. The limits depend on the
couple of functions $(c, \phi)$.

Somewhat surprisingly not all the attractors points are charged. It depends on
the topological pressure (see \cite{Kifer90}) and can be understood as
follows:$c$ acts as a killing term (see \cite{Schuss, HMS}) and can destroy
all
the particles near an attractor;

We begin analyzing the problem. The eigenfunction satisfies the partial
differential equation and normalisation condition
\begin{align}
\epsilon\Delta_{g}u_{\epsilon}+(\nabla\phi, \nabla u_{\epsilon})_{g}%
+cu_{\epsilon}  &  =\lambda_{\epsilon}u_{\epsilon}\label{edpfdtgr}\\
\int_{V_{n}}u_{\epsilon}^{2}dvol_{g}  &  =1.
\end{align}
We denote by Sing(b) the set of singular points of the field $b$.When b is the
gradient of a function $\phi$,  these points are just the critical points of
$\phi.$

\subsection{What do we learn from the Topological Pressure}
The topological pressure can be defined as follows:
\begin{equation}
\Pr=\inf_{P\in\operatorname{Si}ng(b)}\{c(P)-\sum_{i=1}^{m}\min(0, Re\lambda
_{i}(P)\} \label{Topo}%
\end{equation}
where $\lambda_{i}(P), 1\leq i\leq m, $ are the eigenvalues of the linear part
of the field $grad\phi$ \ at the point $P$. The definition for Morse-Smale
fields can be found in (\cite{Kifer80}),  for more general vector fields in
(\cite{Kifer90}). Because the infimum in (\ref{Topo}) is taken over a finite
number of points,  it is attained somewhere but not necessarily at an
attractor
(that is clear from the formula).

\noindent It will be useful to perform a gauge transformation:
\begin{align*}
v_{\epsilon}  &  =u_{\epsilon}e^{-\phi/2\epsilon}\\
b  &  =\nabla\phi.
\end{align*}
By this gauge transformation,  equation (\ref{edpfdtgr}) is transformed into
\[
\epsilon\Delta_{g}v_{\epsilon}+(c+\frac{\Delta_{g}\phi}{2}+\frac{||\nabla
\phi||_{g}^{2}}{4\epsilon})v_{\epsilon}=\lambda_{\epsilon}v_{\epsilon}%
\]
with the new condition of normalization $\int_{V}v_{\epsilon}^{2}dvol_{g}=1$.
We will analyze the set of limits of the measures $v_{\epsilon}^{2}dvol_{g}$
as $\varepsilon$ tends to 0.
\subsection{blow-up Analysis}
It has been proved in the second section that the sequence $\sup_{\epsilon
}v_{\epsilon}$ converges to infinity as $\epsilon$ goes to zero and the
sequence $v_{\epsilon}$ converges uniformly to zero on every compact set that
does not intersect the singular set of the field. As in the field- free case,
if P$\in\operatorname{Sing}(b)$ we choose a normal coordinate system (x$_{1}%
$, ...x$_{m}$):U---%
$>$%
$\mathbb{R}$,  centered at P,  defined on a domain U such that:

1)x$_{1}\times$...$\times$x$_{m}(U)$ contains the closed ball B$_{P}(\delta)$
centered at P and having radius $\delta>0.$

2)for all i, j,  1$\leq$ i, j$\leq$ m,  $\frac{\partial^{2}\phi}{\partial
x_{i}\partial x_{j}}(P)=\lambda_{i}(P)\delta_{ij}$.

3) U$\cap$Sing(b)=\{P\}

On the magnified set $\frac{1}{\sqrt{\varepsilon}}$x$_{1}\times$...$\times
$x$_{m}(U)$ , we can define the function $w_{\varepsilon}$:$w_{\varepsilon
}(y)=\frac{v_{\varepsilon}(y\sqrt{\varepsilon})}{\overline{v}_{\varepsilon}}$
where $\overline{v}_{\varepsilon}$=$\underset{c}{\max}$ $v_{\varepsilon}.$

The main theorem of this section is the following:
\begin{theorem}
\begin{itemize}
\item (i)Suppose that the vector field $b$ is gradient-like,  $b=\nabla\phi$,
where $\phi$ has the expansion near each critical point: $\phi(P)=\phi
(P)+\sum_{i=1}^{m}\lambda_{i}(P)(x^{i})^{2}+O(||x||_{g}^{3})$. The set
$\mathcal{M}$ of all possible limit measures of the sequence $v_{\epsilon}%
^{2}dvol_{g}$,  is described by:
\[
\mathcal{P}=\{\mu\in M(V)|\mu=\sum_{P\in S}c_{P}\delta_{P}, c_{P}\geq
0, \sum_{P\in S}c_{P}=1\}
\]
$S$ is set the critical points of $b$,  where the Topological Pressure is
achieved.

\item (ii) When all the points of $S$ are maximally charged,  the coefficients
$c_{P}$ can be computed explicitly,  $c_{P}=\frac{\prod_{1}^{m}|\lambda
_{i}(P)|^{-1/2}}{\sum_{i\in\Lambda}\prod_{1}^{m}|\lambda_{i}(P)|^{-1/2}}$

\item When a partial set $\widehat{S}$ of $S$ is maximally charged,  the
formula changes to:
\[
c_{P}=\frac{\prod_{1}^{m}|\lambda_{i}(P)|^{-1/2}}{\sum_{P\in\widehat{S}}%
\prod_{1}^{m}|\lambda_{i}(P)|^{-1/2}}%
\]
\end{itemize}
\end{theorem}
\begin{lem}
\label{maxpr} For each $\varepsilon$,  let us denote by $\mathcal{M}%
_{\varepsilon}, $ the set of all $\max$imum points of v$_{\varepsilon}$ in the
manifold.There exists a constant $A$ depending only on c and $b$,  such that:
\[
\underset{q\in\mathcal{M}_{\varepsilon}}{\sup}d_{g}^{2}(q, Sing(b))\leq
A\epsilon
\]
It follows from this that the set of limit points of the set of maximum points
of v$_{\varepsilon}$ is contained in $Sing(b)$.
\end{lem}
\textbf{\noindent Proof.}

\noindent If we denote by $Q_{\epsilon}$ a sequence of maximum points of
$v_{\epsilon}.$
Using the Maximal Principle,  applied to equation (\ref{edpfdtgr}),  it is
possible to estimate the velocity at which the sequence of maximal points
$Q_{\epsilon}$ converges to a critical point of the field. Indeed,  we obtain
that
\[
(c(Q_{\epsilon})+\frac{\Delta\phi(Q_{\epsilon})}{2}+\frac{||\nabla
\phi(Q_{\epsilon})||^{2}}{4\epsilon}\leq\lambda_{\epsilon}\leq C
\]
for some constant C. This implies that $||\nabla\phi(Q_{\epsilon})||^{2}%
\leq4C\epsilon$. Now it is easy to see that there exists a constant $\Gamma
$\ depending only on c such that for P $\in$ V,  d$_{g}^{2}$(P, Sing(b))$\leq$
$\Gamma$(c(P)--$\underset{V}{\min}$ c). Take A=4$\Gamma\Lambda.$ Then
$d_{g}^{2}(Q_{\epsilon}, \operatorname{Si}ng(b))\leq A\epsilon.$ The lemma
shows that $Q_{\epsilon}$ tends to $Sing(b)$ at least as fast as
$\sqrt{\epsilon}$. In fact,  theorem below shows that it tends faster.

\begin{theorem}{\bf Concentration with a gradient vector field.}

\label{th-final}(i)For any P$\in C_{\min, }$any sequence of v$_{\varepsilon}%
$'s,  with $\varepsilon$ tending to 0,  contains a subsequence
\{v$_{\varepsilon
_{n}}$\},  such that the sequence of blown-up functions $\frac{v_{\varepsilon
_{n}}(y\sqrt[4]{\varepsilon_{n}})}{\overline{v}_{\varepsilon_{n}}}$ at P
converges to a function w:$\mathbb{R}^{m}$--%
$>$%
$\mathbb{R}_{+}$,  both in the L$^{2}$norm and the C$^{\infty}$ topology.

(ii) w satisfies the equation:
\begin{equation}
\Delta_{E}w+\left(  c(P)+\frac{\Delta\phi}{2}(P)+\sum\lambda_{i}(P)(x^{i}%
)^{2}\right)  w=\lambda w
\end{equation}%
\[
0<w\leq\max_{\mathbb{R}^{m}}w=1.
\]
(iii)When the eigenvalue $\lambda$ is equal to the Topological Pressure at
$P$,  the solution $w$ is nonzero and given explicitly by:
\[
w(x)=\prod_{i=1}^{m}e^{-\mid\lambda_{i}(P)\mid x_{i}^{2}/2}%
\]
in particular,
\[
\int_{\mathbb{R}^{m}}w^{2}(x)=\frac{\pi^{m/2}}{\prod_{i=1}^{m}|\lambda
_{i}|^{1/2}}%
\]
Moreover,
\[
\lim_{\epsilon\rightarrow0}\epsilon^{m/2}\sup_{V}v_{\epsilon}^{2}=K(c, \phi)
\]
where $K(c, \phi)$ is a positive constant depending only on the functions
$c, \phi$.

(iv)If all \ the singular points \ at which the topological pressure is
attained (set denoted by "top") are maximally charged,
\[
K(c, \phi)=\frac{\pi^{m/2}}{%
{\displaystyle\sum}
\left\{  \prod_{i=1}^{m}|\lambda_{i}(P)|^{-1/2}|P\in top\right\}  }.
\]
Moreover
\[
\lim_{\epsilon\rightarrow0}\frac{d_{g}(Q_{\epsilon}, P)}{\epsilon^{1/2}}=0.
\]
\end{theorem}

\bigskip{\noindent\textbf{Remark.}}

\noindent At this stage, let us point out the
difference with the pure potential case,  studied in the last section. There
the scaling factor was $\varepsilon^{1/4}$ and the concentration took place on
certain minimum points of the potential only. In the present situation,  the
scaling factor is $\sqrt{\varepsilon}$ and the concentration is determined by
the couple field-potential ( $\nabla\phi, c)$.

Before proving the theorem,  two lemmas are needed. The first one provides
estimates of the first eigenvalue $\lambda_{\epsilon}$ and the
second, estimates of the decay of renormalized sequence $w_{\epsilon}(x)$.

\begin{lem}
\label{estimm} The first\ eigenvalue satisfies the inequality
\[
0<\lambda_{\epsilon}\leq\text{min \text{\{} }c(P)-\sum_{\{i, \lambda
_{i}(P)<0\}}\lambda_{i}(P)|P\in\operatorname{Si}ng(b)\text{\}}%
\]
where Sing(b) is the set of critical points of the field $b=\nabla\phi$ and
$\lambda_{i}(P), 1\leq i\leq m, $ are the eigenvalues of the field at the
critical point $P$.\bigskip
\end{lem}
\textbf{\noindent}\textbf{Proof.}

\noindent This is a consequence of the variational
approach. Similarly to the case without field,  define the variational
quotient
\[
Q_{\epsilon}(v)=\frac{\epsilon\int_{V}\left[  ||\nabla v||_{g}^{2}%
+c_{\varepsilon}v^{2}\right]  dvol_{g}}{\int_{V}v^{2}dvol_{g}}.
\]
where $c_{\epsilon}=c+\frac{\Delta_{g}\phi}{2}+\frac{\mid\nabla\phi||_{g}^{2}%
}{4\epsilon}$. Then
\[
\lambda_{\epsilon}=\inf_{v\in H^{1}(V)-\{0\}}Q_{\epsilon}(v)>0
\]
In the neighborhood of a singular point $P$,  we have an expansion,
\[
c_{\epsilon}(x)=c(P)+\frac{\Delta_{g}\phi(P)}{2}+\sum_{i=1}^{m}a_{i}%
(P)x^{i}+\sum_{i=1}^{n}\frac{\lambda_{i}^{2}(P)x_{i}^{2}}{\varepsilon}%
+\sum_{\leq1i\leq j\leq k\leq m}c_{ijk}(x)x_{i}x_{j}x_{k}%
\]
where the $c_{ijk}$ are C$^{\infty}$ functions on U$_{P}$. In order to get an
upper bound for the eigenvalue $\lambda_{\epsilon}$,  we estimate the quotient
$Q_{\epsilon}$, for the following test function with compact support,
\begin{align*}
\psi_{\epsilon}  &  =e^{-\sum\mu_{i}(x^{i})^{2}/2}-e^{-\rho/(2\epsilon
)}\hbox{ in }\mathcal{N}_{P}(\rho)\\
&  =0\text{,  otherwise}%
\end{align*}
where $\mu_{i}=\frac{\mid\lambda_{i}(P)\mid}{\varepsilon}$ and $\mathcal{N}%
_{P}(\rho)$ is the connected component of the set \{x%
$\vert$%
$\sum_{i=1}^{m}|\lambda_{i}|x_{i}^{2}\leq\rho\}$ containing 0 and $\rho$ is
taken so small that $\mathcal{N}_{P}(\rho)\cap$ \{x%
$\vert$%
$\sum_{i=1}^{m}x_{i}^{2}=\delta^{2}\}$ is empty. So $\mathcal{N}_{P}(\rho)$ is
contained in B$_{P}$($\delta$). After computations similar to the ones in
Lemma \ref{lm1} of the section \ref{blow},  we obtain
\begin{align*}
Q_{\epsilon}(\psi_{\varepsilon})  &  =c(P)+\frac{\Delta_{g}\phi(P)}{2}%
+\frac{\epsilon}{2}\sum_{i=1}^{n}\mu_{i}+\sum_{i=1}^{n}\frac{\lambda
_{i}(P)^{2}}{2\epsilon\mu_{i}}+o(\epsilon)\\
&  =c(P)+\frac{\Delta_{g}\phi(P)}{2}+\sum_{i=1}^{n}|\lambda_{i}(P)|+o(\epsilon
)
\end{align*}
Using the fact that $\frac{\Delta_{g}\phi(P)}{2}=-\sum_{i=1}^{n}\lambda
_{i}(P)$,  we obtain that
\[
Q_{\epsilon}(\psi_{\varepsilon})\leq c(P)+\sum_{\lambda_{i}(P)<0}\lambda
_{i}(P).
\]
Since a similar test function can be built in the neighborhood of every
critical points,  we obtain the estimate of the lemma,
\[
\lambda_{\epsilon}\leq\min_{\{P\in\operatorname{Si}ng(P)\}}[c(P)-\sum
\{\lambda_{i}(P)|1\leq i\leq m, \lambda_{i}(P)<0\}].
\]
\bigskip

\begin{lem}
\label{88} The blown-up sequence $w_{\epsilon}$,  defined by
\[
w_{\epsilon}(x)=\frac{v_{\epsilon}(\sqrt{\epsilon}x)}{\sup_{V_{n}}v_{\epsilon
}}%
\]
satisfies the following properties:

For all $\varepsilon_{0}\in]0, 1[:$
\begin{enumerate}
\item $\underset{]0, \varepsilon_{0}]}{\sup}$
$\int_{B_{P}(\delta/\sqrt[4]{\varepsilon})}w_{\varepsilon}(y)^{2}dy<+\infty.$
More generally,  for any continuous function f:[0,
1]$\times\mathbb{R}^{m}$---%
$>$%
$\mathbb{R}, (\epsilon, y)$---%
$>$%
f $(\epsilon, y), $\ having at most polynomial growth at infinity in y
uniformly
in $\varepsilon$,  $\underset{]0, \varepsilon_{0}]}{\sup}\int_{B_{P}%
(\delta/\sqrt{\varepsilon})}f(\epsilon, y)w_{\varepsilon}(y)^{2}dy<+\infty$.
\item  the set of restrictions w$_{\varepsilon}^{2}$%
$\vert$%
$B_{P}(\delta/\sqrt{\varepsilon})$, $\varepsilon$ $\in]0, 1]$,  of the
w$_{\varepsilon}^{2}$ to the balls $B_{P}(\delta/\sqrt[4]{\varepsilon})$
satisfies the following condition: for any $\eta>0, $ there exists a compact
K$\subset\mathbb{R}^{m}$ and a $\varepsilon(\eta)>0$ such that
\[
\int_{B_{P}(\delta/\sqrt{\varepsilon})-K}w_{\varepsilon}(y)^{2}dy\leq\eta,
\]
for all $\varepsilon$ $\in]0, \varepsilon(\eta)].$
\end{enumerate}
\end{lem}

\bigskip

{\noindent}\textbf{Proof of Lemma\ref{88}}

\noindent The proof is similar as the one given to prove the decay estimate in
the case
of the pure potential case and is based on the Feynman-Kac integral
representation of the solution. The Feynman-Kac formula gives that
\[
e^{-\mu_{\epsilon}t}v_{\epsilon}(x)=e^{-\mu_{\epsilon}t}E_{x}\left(
v_{\epsilon}(X_{\epsilon}(t))\chi_{(t<\tau_{\epsilon}^{x})}e^{-\int_{0}%
^{t}c_{\epsilon}(X_{\epsilon}(s)ds)}\right)  +E_{x}\left(  e^{-\mu_{\epsilon
}\tau_{\varepsilon}^{x}}v_{\epsilon}(X_{\epsilon}(\tau_{\varepsilon}^{x}%
))\chi_{(t>\tau_{\epsilon}^{x})}e^{-\int_{0}^{\tau_{\epsilon}^{x}}c_{\epsilon
}(X_{\epsilon}(t))ds}\right)
\]
where we recall that $c_{\epsilon}=c+\frac{\Delta_{g}\phi}{2}+\frac
{||\nabla\phi||_{g}^{2}}{4\epsilon}$ and X$_{\varepsilon}$ is defined in
relation (\ref{sde}).$\tau_{\varepsilon}^{x}$ has the same meaning as in
Section \ref{blow}). Since the function $c+\frac{\Delta_{g}\phi}{2}$ is
bounded,  it is enough to estimate $E_{x}(e^{-\int_{0}^{t}\frac{\mid\nabla
\phi\mid^{2}}{4\epsilon}(X_{\epsilon}(s))ds})$. Using the scale change
$x\rightarrow\epsilon^{1/2}x$,  this term can be estimated as in Proposition
\ref{estimp} and the results are similar. Indeed,  because $\frac{||\nabla
\phi(\epsilon^{1/2}y)||^{2}}{4\epsilon}=\sum_{n=1}^{m}\lambda_{n}(P)^{2}%
y_{n}^{2}$,  the estimate is valid in the neighborhood of any critical point.
\quad\hbox{\hskip4pt\vrule width 5pt height 6pt depth 1.5pt}

\bigskip
{\noindent}\textbf{Proof of the theorem.}
\noindent Consider the renormalized sequence $w_{\epsilon}(y)=\frac
{v_{\epsilon}(y\sqrt{\epsilon})}{\bar{v_{\epsilon}}}$,  $w_{\epsilon}$
satisfies the partial differential equation:
\begin{align}
&  \Delta_{g_{\epsilon}}w_{\epsilon}(y\sqrt{\epsilon})+\left(  c(y\sqrt
{\epsilon})+\frac{\Delta_{g}\phi(y\sqrt{\epsilon})}{2}+\frac{||\nabla
\phi||_{g}^{2}(y\sqrt{\epsilon})}{4\epsilon}\right)  w_{\epsilon}%
(y\sqrt{\epsilon})=\lambda_{\epsilon}w_{\epsilon}\text{ }(y\sqrt{\epsilon
})\text{in }\nonumber \\ & \frac{1}{\sqrt{\varepsilon}}x_{1}\times...\times
x_{m}%
(U_{P})\label{fdtgrad}\\\nonumber
&  0<w_{\epsilon}\leq1
\end{align}
To study the equation \ref{fdtgrad},  note that the coefficients of the
partial
differential equation \ref{fdtgrad} converges uniformly on every compact set
as $\epsilon$ goes to zero: $g_{\epsilon}(x)=g(y\sqrt{\epsilon})$ converges to
the Euclidean metric (see the previous section for more details about the type
of convergence,  usually in $C^{2, \alpha}$ of any compact).

Since the coefficients of the partial differential equation \ref{fdtgrad}
remain bounded as $\epsilon$ goes to zero. Indeed,  due the velocity of
convergence of the sequence $Q_{\epsilon}$\ref{th-final},
\begin{align*}
\lim_{\epsilon\rightarrow0}\left(  c(y\sqrt{\epsilon})+\frac{\Delta_{g}%
\phi(y\sqrt{\epsilon})}{2}\right)   &  =c(P)+\frac{\Delta_{g}\phi(P)}{2}\\
\lim_{\epsilon\rightarrow0}\frac{||\nabla\phi(y\sqrt{\epsilon})||_{g}^{2}%
}{4\epsilon}  &  =\sum_{n=1}^{m}\frac{\lambda_{n}^{2}(P)y_{n}^{2}}{4}%
\end{align*}
Classical elliptic estimates show that $w_{\epsilon}$ converges uniformly on
every compact set of $\mathbb{R}^{m}$ to a solution $w$ of
\begin{align}
&  \Delta_{E}w(y)+\left(  c(P)+\frac{\Delta_{g}\phi}{2}(P)+\sum_{n=1}%
^{m}\lambda_{n}^{2}(P)y_{n}^{2}\right)  w(y)=\lambda w(y), \text{ on }%
\mathbb{R}^{m}\label{edpp}\\
&  0<w\leq\max_{\mathbb{R}^{m}}w \leq 1.
\end{align}
Let $Q^{*}$ be a limit point of the sequence $\frac{Q_{\epsilon}}%
{\sqrt{\epsilon}}$,  since $w_{\epsilon}$ converges uniformly on every compact
to $w$. $w_{\epsilon}(\frac{Q_{\epsilon}}{\sqrt{\epsilon}})$ converges to
$w(Q^{*})$. Hence $w(Q^{*})=1$. Because $w$ belongs to $L_{2}(\hbox{\bb
R}^{n}%
)$ by Lemma \ref{88} and is not zero, $w$ an eigenfunction of $L_{Q}$ in
$L_{2}(\hbox{\bb R}^{n})$ associated to the eigenvalue $\lambda$.

Moreover, using the results of lemma \ref{estimm},
\begin{eqnarray} \label{TPP}
\lambda-(c(P)+\frac{\Delta\phi}{2}(P))\leq\sum_{n=1}^{m}|\lambda_{n}(P)|
\end{eqnarray}
We will now prove, in order for the function $w$ to be nonzero,  necessarily
$\lambda$ is equal to the Topological Pressure.

In order to prove that $\mu=\lambda-(c(P)+\frac{\Delta\phi}{2}(P))=\sum
_{n=1}^{m}|\lambda_{n}(P)|$,  one must study the spectrum of the self-adjoint
operator defined by $L_{P}=\Delta_{E}+\sum_{n=1}^{m}\frac{\lambda_{n}%
^{2}(P)x_{n}^{2}}{4}$in $L_{2}(\hbox{\bb R}^{n})$. This operator is compact
and has a discrete spectrum. For all these,  see \cite{RS}. Using the
arguments
in the proofs of the main theorem in Section \ref{refharmonique},  we know
that
the eigenvalues and eigenfunctions of the operator defined by $L_{P}$ in
$L_{2}(\hbox{\bb R}^{n})$ are respectively
\[
\sum_{1}^{n}(2\alpha_{i}+1)\frac{|\lambda_{i}(P)|}{2}, \, \alpha\in\hbox{\bb
Z}_{+}^{n}%
\]
and
\[
H_{\alpha}(x)=\prod_{1}^{n}h_{\alpha_{i}}(y_{i})e^{-\frac{\lambda_{i}(k)}%
{2}(y_{i})^{2}}\hbox{ where }h_{k}(x)=e^{x^{2}/4}\frac{d^{k}}{dx^{k}%
}\{e^{-x^{2}/4}\}.
\]
The lowest eigenvalue of $L_{P}$ is
\[
\Lambda_{1}=\sum_{1}^{n}|\lambda_{i}(P)|.
\]
Inequality \ref{TPP} shows that the eigenvalue $\lambda-(c(P)+\frac{\Delta
\phi}{2}(P))$ of $L_{P}$ in $L_{2}(\hbox{\bb R}^{n})$ is at most equals to
$\lambda_{1}$. Hence it is equal to $\lambda_{1}$ and the corresponding
eigenfunction $w$ is a positive multiple of $H_{0}$. Since the only maximum
point of $H_{0}$ is $0$,  $Q^{\ast}=0$ and $w=H_{0}(0)$.

In particular,  the previous analysis shows that
\[
\lim_{\epsilon\rightarrow0}\lambda_{\epsilon}=(c(P)+\frac{\Delta\phi}%
{2}(P))+\Lambda_{1}=\hbox{ Topological Pressure}.
\]
This implies that a point P is not charged if the Topological Pressure is not
attained at P.

Using the exponential decay of the last lemma,  at the point $P$,  where the
Topological Pressure is attained,  we have%
\[
\int_{B_{P}(\delta)}v_{\epsilon}^{2}dvol_{g}=\epsilon^{m/2}\bar{v}_{\epsilon
}^{2}\int_{B_{Q_{k}}(\delta/\epsilon)}w_{\epsilon}^{2}dvol_{g_{\epsilon}}.
\]
Since the sequence $w_{\epsilon}$ converges in $L_{2}$,  this implies that the
sequence $\epsilon^{m/2}\bar{v}_{\epsilon}^{2}$ converges and we get,
\[
\lim_{\epsilon\rightarrow0}\epsilon^{m/2}\bar{v}_{\epsilon}^{2}=C(c, \phi).
\]
The concentration coefficient $c_{k}$ can be computed,  if we denote
\[
c_{P}=\lim_{\epsilon\rightarrow0}\int_{B_{P}(\delta)}v_{\epsilon}^{2}%
dvol_{g}=K(c, \phi)\int_{\mathbb{R}^{m}}w_{Q}^{2}.
\]
Using the $L_{2}$ normalization conditions,  the Blow up analysis and the fact
that the sequence $v_{\epsilon}$ converges to zero on any compact sets that
does not intersect the critical points where the topological pressure is
attained,  we get,
\[
1=K(c, \phi)\sum_{P\in S}\int_{\mathbb{R}^{m}}w_{P}^{2}+o(1)
\]
which proves the last part of the theorem. We shall omit the proof of
statement\textit{ (iv) }which is similar to the one given at the end of
section\ref{blow}.

\bigskip\textbf{\noindent Remarks:}

\begin{enumerate}
\item Can one find explicitly the asymptotic expansion of $\lambda_{\epsilon}%
$,
\begin{align}
\lambda_{\epsilon} = \sum_{k=0}^{\infty} c_{k}\epsilon^{k/2}.
\end{align}

\item We proved earlier that $c_{0}$ is equal to the Topological Pressure and
captures enough information to locate the supports of the possible limit
measures. This is surprising because it was not true in the pure potential
case.See the section(\ref{dege})
\end{enumerate}

\section{The general case for the first eigenfunction problem} \label{general}
In this section,  we shall consider a special class of vector fields $b$ such
that the limits of the first eigenfunctions $u_{\epsilon}$ concentrate on the
limit sets of $b$. In particular if the limit sets of the vector field
contains limit cycles,  we prove that the weak limits of $u_{\epsilon}$ as
$\epsilon$ tends to zero,  concentrate on these limit cycles.
when a

Here we will consider the following class of vector fields : $b=grad_{g}%
\mathcal{L}+\Omega$ where $\Omega$ is a Morse-Smale vector field and
$\mathcal{L}$ is a special type of Lyapunov function of $\Omega$. We will
study the behavior of the first eigenfunction for the operator $L_{\epsilon}$
the drift of which is belongs to this class. We will need the following lemma
proved in the appendix.

\begin{lem}
\label{1}Given a Morse Smale field $\Omega$ there exists $C^{\infty}$ Lyapunov
functions for $\Omega$,  taking the value zero on all the repulsors of
$\Omega$
and such that the function $\Psi(\mathcal{L})=\frac{1}{4}({||\nabla
\mathcal{L}||}_{g}^{2}+\frac{1}{2}<grad \mathcal{L}, \Omega>_{g})$ is positive
except on the recurrent set of $\Omega$ where it is zero.
\end{lem}



\begin{Defi}
\label{d1} A Lyapunov function with the property stated in lemma \ref{1} will
be called a special Lyapunov function.
\end{Defi}

The main theorem can now be stated:

\begin{theorem}
\label{thfdtpr} On a compact orientable Riemannian manifold $V_{m}$,  consider
a Morse-Smale vector field $\Omega$ whose recurrent set consists of the
stationary points $P_{1}, ..., P_{M}$ and of the periodic orbits $\Gamma_{1}%
$, ..,  $\Gamma_{N}$. $\mathcal{L}$ denotes a special Lyapunov function.

For $\epsilon>0$ let $\lambda_{\epsilon}$,  $u_{\epsilon}$ denotes
respectively
the first eigenvalue and the associated eigenfunction of the operator
\begin{align}
L_{\epsilon}= \epsilon\Delta_{g}+b\nabla+c\, , \text{ on }V_{m}%
\end{align}

Then the weak limits of the family of probability measures%

\[
\ \frac{e^{-\mathcal{L}/\epsilon}u_{\epsilon}^{2}dV_{g}}{\int_{V}%
e^{-\mathcal{L}/\epsilon}u_{\epsilon}^{2}dV_{g}}%
\]
are of the form
\[
\nu=\sum_{j=1}^{N}\mu_{j}+\sum_{i=1}^{M}\gamma_{P_{i}}\delta_{P_{i}}%
\]
where the measures $\mu_{j}$ are supported by the limit cycle $\Gamma_{j}$ and
the $P_{i}$'s $1\leq i\leq N$ are the critical points of $b$. the
$\gamma_{P_{i}}$ are scalars.
\[
1=\sum_{j=1}^{N}\int_{\Gamma_{j}}\mu_{j}+\sum_{i=1}^{M}\gamma_{P_{i}}.
\]

\end{theorem}


\bigskip

{\noindent\textbf{Remark.}}

In a future paper\cite{HK3},  we shall prove that for an appropriate choice of
Lyapunov
function, no concentration can occur on a set of isolated points on the cycle.

point where the topological pressure is
found

\subsubsection{Proof of theorem 2}

The proof of theorem 2 will occupy the next two sections. Consider the
following transformation $u\longrightarrow v.:$
\begin{align}
v  &  =e^{-\frac{\phi}{2\epsilon}}u\\
u  &  =e^{\frac{\phi}{2\epsilon}}v
\end{align}
and the second order differential operator $L_{\varepsilon}^{^{\prime}}$:%
\[
L_{\varepsilon}^{^{\prime}}v=\varepsilon e^{-\frac{\phi}{2\varepsilon}%
}L_{\varepsilon}(e^{\frac{\phi}{2\epsilon}}v)
\]
\bigskip It is easy to see that:
\[
\Delta_{g}u=e^{\frac{\phi}{2\epsilon}}\left[  (\Delta_{g}v-\frac{1}{\epsilon
}<\nabla\phi, \nabla v>_{g}+v_{\epsilon}\left(  \frac{\Delta_{g}\phi}%
{2\epsilon}-\frac{||{\nabla\phi||}_{g}^{2}}{4\epsilon^{2}}\right)  \right]  ,
\]
and
\[
<\nabla u, b>_{g}=e^{\frac{\phi}{2\epsilon}}\left[  <\nabla v, b>_{g}%
+v\frac{<\nabla\phi, b>_{g}}{2\epsilon}\right]
\]
Finally,
\begin{equation}
L_{\epsilon}^{\prime}(v)=e^{\frac{\phi}{2\epsilon}}\left[  \epsilon\Delta
_{g}v+<b-\nabla\phi, \nabla v>_{g}+v\left(  c+\frac{<\nabla\phi, b>_{g}%
}{2\epsilon}+\frac{\Delta_{g}\phi}{2}-\frac{||{\nabla\phi||}_{g}^{2}%
}{4\varepsilon}\right)  \right]
\end{equation}
Setting $b_{\phi}=b-\nabla\phi$,  we obtain :
\[
\epsilon L_{\varepsilon}^{^{\prime}}v=\epsilon^{2}\Delta_{g}v+\varepsilon
<b_{\phi}, \nabla v>_{g}+v\left(  \varepsilon(c+\frac{\Delta_{g}\phi}{2}%
)+\frac{<\nabla\phi, b>_{g}}{2}-\frac{||{\nabla\phi||}_{g}^{2}}{4}\right)
\]

If we set $v_{\varepsilon}=e^{-\frac{\phi}{2\epsilon}}u_{\varepsilon}$, the
equation
\[
\epsilon\Delta_{g}u_{\varepsilon}+<b, \nabla u_{\varepsilon}>+cu_{\varepsilon
}=\lambda_{\varepsilon}u_{\varepsilon}%
\]
becomes :
\begin{equation} \label{tf2}
\epsilon^{2}\Delta_{g}v_{\epsilon}+\epsilon<b_{\phi}, \nabla v_{\epsilon}%
>_{g}+c_{\epsilon}v_{\epsilon}=\epsilon\lambda_{\epsilon}v_{\epsilon}\text{,
on }V
\end{equation}
where $c_{\epsilon}=\epsilon(c+\frac{\Delta\phi}{2})+\frac{<\nabla\phi
, b_{\phi}>_{g}}{2}+\frac{||{\nabla\phi||}_{g}^{2}}{4}$.

We set $b_{\phi}=\Omega$.
$c-\frac{\Delta\phi}{2}-\frac{divb_{\phi}}{2}\geq0$.
Taking $\varphi=\mathcal{L}$,  a special Lyapunov function for $\Omega$ in the
previous paragraph,  we will prove that $\epsilon\lambda_{\epsilon}$ tends to
the minimum of the function $\Psi_{\mathcal{L} }=\frac{(\nabla\mathcal{L}
)^{2}}{4}+\frac{( \Omega, \nabla\mathcal{L} )}{2} $ when $\epsilon$ goes to
zero.


\begin{lem}
\label{lee} Under the assumptions of Theorem \ref{thfdtpr} on the vector
field
\begin{align}
\lim_{\epsilon\rightarrow0}\epsilon\lambda_{\epsilon}=0=\underset{V}{\min}%
\Psi_{\mathcal{L}}%
\end{align}

\end{lem}


\medskip\noindent\textbf{Proof:} Multiplying equation (\ref{tf2}) by
$v_{\epsilon}$ and integrating on V,  we obtain:
\begin{equation}
\int_{V}[\epsilon^{2}{||\nabla v_{\epsilon}||}_{g}^{2}+(-\frac{\epsilon
div(b_{\phi})}{2}+c_{\epsilon})v_{\epsilon}^{2}]dvol_{g}=\epsilon
\lambda_{\epsilon}\int_{V}v_{\epsilon}^{2}dvol_{g} \label{tf3}%
\end{equation}
Normalize v$_{\varepsilon}$ so that $\int_{V}v_{\epsilon}^{2}dvol_{g}$=1.From
equation (\ref{tf3}),  we obtain that $\lim_{\epsilon\rightarrow0}%
\epsilon\lambda_{\epsilon}\geq\underset{V}{\min}\Psi_{\phi}$. An upper bound
of the function $\epsilon\lambda_{\epsilon}$,  in the case when $\phi
=\mathcal{L}, $ will be computed using the following two results:

1) Let $S$ be a regular sub-domain of $V$ and $\lambda_{\epsilon}^{\prime}(S)$
be the first eigenvalue of $L_{\epsilon}^{\prime}$ on $S$ \ for the
homogeneous Dirichlet problem. If there exists a positive number $A$ and a
positive function $\psi$ on $S$,  zero on the boundary of $S$,  such that
$L_{\epsilon}^{\prime}\psi\leq\epsilon A\psi$,  then $\lambda_{\epsilon
}^{\prime}(S)\leq\epsilon A$ (see \cite{DF})).

2) If $\lambda_{\epsilon}$ is the first eigenvalue of $L_{\epsilon}$ on $V $,
then $\varepsilon\lambda_{\epsilon}\leq\lambda_{\epsilon}^{\prime}(S)$.Note
that $\varepsilon\lambda_{\epsilon}$ is the first eigenvalue of
$L_{\varepsilon}^{\prime}$ on V (\cite{DF}).

We proceed here with the proof in the case of a limit cycle and leave the case
of a singular point,  the easier one,  to the reader. We construct a test
function $\psi$ as follow. Let $\omega$ be a stable limit cycle of b. Let
$\Gamma_{\delta}$ be the connected component containing $\omega$,  of the set
$\{P\in V$ $|$ $\mathcal{L}(P)\leq\mathcal{L}(\omega)+\delta\}$. For $\delta$
sufficiently small,  $\Gamma_{\delta}$ is a regular sub-domain of $V$ and a
neighborhood of $\omega$ not intersecting any other recurrent set of
$b_{\mathcal{L}}$. For simplicity we set $b_{\mathcal{L}}=B.$

Define the function $\psi:$ $\Gamma_{\delta}->\mathbb{R}, $ as follow :%
\[
\psi=e^{-\mathcal{L}/(2\epsilon)}-e^{-\delta/(2\epsilon)}%
\]
\bigskip Denoting \ by $\nabla$ the gradient with respect to g, we have:%
\[
\, \epsilon^{2}\Delta_{g}\psi=e^{-\mathcal{L}/(2\epsilon)}\left[
\frac{\varepsilon\Delta_{g}\mathcal{L}}{2}-\frac{||{\nabla\mathcal{L}||}%
_{g}^{2}}{4}\right]
\]%
\[
L_{\epsilon}^{\prime}(\psi)=\epsilon c\psi-e^{-\delta/(2\epsilon)}\left[
\frac{\epsilon\Delta_{g}\mathcal{L}}{2}+\frac{<\nabla\mathcal{L}, \Omega>_{g}%
}{2}+\frac{{||\nabla\mathcal{L}||}_{g}^{2}}{4}\right]
\]
Let $R$ be a positive number,
\[
L_{\epsilon}^{\prime}\psi-\epsilon R\psi=e^{-\delta/(2\epsilon)}\left[
\epsilon(c-R)(e^{\frac{\delta-\mathcal{L}}{2\epsilon}}-1)-\frac{\epsilon
\Delta_{g}\mathcal{L}}{2}-\frac{<\nabla\mathcal{L}, B>_{g}}{2}-\frac
{{||\nabla\mathcal{L}||}_{g}^{2}}{4}\right]
\]
Below we shall determine $R$ and $\delta$ so that $L_{\epsilon}^{\prime}%
\psi-\epsilon R\psi\leq0$. Assume this done,  using results 1 and 2 above,
with
$S=\Gamma_{\delta}$ we get that $\epsilon\lambda_{\epsilon}\leq\lambda
_{\epsilon}^{\prime}(S)\leq\epsilon R$. Hence $\lim_{\epsilon\rightarrow
0}\epsilon\lambda_{\epsilon}=0$. We use Lemma \ref{lya} to determine $R$ and
$\delta$. We get that up to terms of order four at least,  using appendix I,
\[
L_{\epsilon}^{\prime}\phi-\epsilon R\phi=e^{-\delta/(2\epsilon)}%
(\epsilon(c-R)(e^{\frac{\delta-\mathcal{L}}{2\epsilon}}-1)-{\epsilon}%
tr_{g}A-(1-\mu/2)\left\Vert {Ax}\right\Vert ^{2})
\]

We set $\delta=2\alpha\epsilon$ where $\alpha$ will be determined later. There
are two cases. If $\mathcal{L}\leq\alpha\epsilon, $ then $e^{\frac
{\delta-\mathcal{L}}{2\epsilon}}-1\geq e^{\frac{\alpha}{2}}-1$ and we can
choose $\alpha$ so that $(e^{\frac{\alpha}{2}}-1)$ is arbitrarily large.
$\left\Vert {Ax}\right\Vert ^{2}\leq C\epsilon$ where $C>0$ depends only on
the matrix A and $\alpha$.  When $\epsilon$ converges to zero,  for $R$ big
enough and $\mathcal{L}\leq\alpha\epsilon$:%
\[
L_{\epsilon}^{\prime}\phi-\epsilon R\phi\leq\epsilon\left[  (c-R)(e^{\frac
{\alpha}{2}}-1)+|tr_{g}A|\right]  \leq0
\]
When $\mathcal{L}\geq\alpha\epsilon$,  then
\[
(c-R)(e^{\frac{\delta-L}{2\epsilon}}-1)\leq0\text{ for R large enough}%
\]
and:
\[
L_{\epsilon}^{\prime}\phi-\epsilon R\phi\leq\epsilon(tr_{g}A-(1-\frac{\mu}%
{2})a_{1}\alpha).
\]
Because $\mathcal{L(}x\mathcal{)}\geq\alpha\epsilon$, it implies that
$\left\Vert {Ax}\right\Vert ^{2}\geq a_{1}\alpha\epsilon$, where $a_{1}$
depends only on the matrix $A$. Finally $L_{\epsilon}^{\prime}\phi-\epsilon
R\phi\leq0$ if $\frac{trA}{(1-\frac{\lambda}{2})a_{1}}\leq\alpha$. Hence for R
large enough,  $L_{\epsilon}^{\prime}\phi-\epsilon R\phi\leq0$ in
$\Gamma_{\delta}$. Using the results of (\cite{Fr}),  we obtain that
$\epsilon\lambda_{\epsilon}\leq\lambda_{\epsilon}(\Gamma_{\epsilon}%
)\leq\epsilon R$. This ends the proof of Lemma \ref{lee}

Now we will prove that any weak limit of  $v_{\epsilon}^{2}dvol_{g}/\int
_{V}v_{\epsilon}^{2}dvol_{g}$ as $\epsilon$ goes to zero,  concentrates on the
limit sets of $b$.
\begin{lem}
\label{3} All weak limits of measures $v_{\epsilon}^{2}dvol_{g}/\int
_{V}v_{\epsilon}^{2}dvol_{g}$ as $\epsilon$ goes to zero,  are concentrated on
the minimum set of the function $\Psi_{\mathcal{L}}$.
\end{lem}
\bigskip

{\bf \noindent Remark.}

\noindent By construction $\Psi_{\mathcal{L}}$ attains its
minimum on the limit sets of $\Omega$ only and this minimum is zero (note that
on a repulsive set,  L is maximal,  and $\Psi_{\mathcal{L}}$ is also zero).
\bigskip

{\noindent\textbf{Proof of the Lemma \ref{3}.}}

Proceeding as in the self-adjoint case,  we multiply equation (\ref{tf2}) by
$v_{\epsilon}\phi$,  where $\phi$ is an arbitrary test function. We obtain
after integration by parts:
\[
\epsilon^{2}\int_{V}\text{ }\left[  \phi{\ ||\nabla v_{\epsilon}||}_{g}%
^{2}+(c_{\epsilon}\phi-\epsilon^{2}\Delta\phi/2-\epsilon div(\Omega
\phi))v_{\epsilon}^{2}\right]  =\epsilon\lambda_{\epsilon}\int_{V}\phi
v_{\epsilon}^{2}%
\]
$c_{\epsilon}$ converges to $\Psi_{\mathcal{L}}$ and $\epsilon\lambda
_{\epsilon}$ to $0=\underset{V}{\min}\Psi_{\mathcal{L}}$ as $\epsilon$ goes to
zero.

Let $\mu$ be a weak limit of the measures $v_{\epsilon}^{2}dvol_{g}$ then
\begin{equation}
\int_{V}\phi(\underset{V}{\min}\Psi_{\mathcal{L}}-\Psi_{\mathcal{L}})d\mu=0
\label{zzz}%
\end{equation}
Since equation (\ref{zzz}) is true for any test function $\phi$,  $d\mu=0$ on
the open set where $\Psi_{\mathcal{L}}\neq\underset{V}{\min}\Psi_{\mathcal{L}%
}$. This shows that the support of $\mu$ is contained in the limit set of
$\Omega$. Let $S_{1}, ...S_{p}$ be the stationary points of $\Omega$ and let
$\Gamma_{1}, ...\Gamma_{q}$ be the limit cycles. We have the decomposition:
\[
\mu=\sum_{j=1}^{p}\gamma_{k}^{2}\delta_{S_{k}}+\sum_{k=1}^{q}\mu_{k}%
\]
where $\mu_{k}$ is a measure supported by $\Gamma_{k}$.

\subsubsection{Remarks on the limit measures}

We end this section with two remarks. An interesting question is whether the
repulsive sets can be charged by limit measures. Were it to hapen this is
similar to the fact,  well-known in the large deviations theory, that the
trajectories of a stochastic system can exit with a positive probablity the
basin
of attraction of an attractor of its drift. The boundary of the basin plays
the role of the repulsive sets here. In other words a particle moving
according to such a system can escape from the basin of attraction.

The second remark deals with the relations between the limit measures whose
existence was shown above,  and the measures, called equilibrium measures,
which
maximize the topological pressure $P$,  defined in the introduction. More
precisely $P$ is given by the formula:%
\begin{equation}
P=\sup_{\nu\in M}\left[  h_{\nu}(F^{1})+\int_{V}(c+\phi^{u})d\nu\right]
\end{equation}
where M is the set of probablity measures invariant by the flow of the drift.
A
measure $\mu\in M$ is an equilibrium measure if it maximizes $h_{\nu}%
(F^{1})+\int_{V}(c+\phi^{u})d\nu.$ For more explanations see(\cite{Kifer90}).
\subsection{Final Remark}
The characterization of the set of possible limit measure in the
nonvariational case is studied in \cite{HK3},  where we prove that the
concentration can occur along some submanifolds of the recurrent sets of a
hyperbolic field b. In some cases,  we are able to show that the measure is
absolutely continuous with respect to the Hausdorff measure induced on the
set. The order at which the Lyapunov function vanishes,  in the neighborhood
of
the recurrents,  play the role of a filter that allows the eigensequence to
concentrate only  along a subset where the topological pressure is achieved.

\section{Appendix 1}

In this appendix,  we give the construction of Lyapunov functions for
Morse-Smale vector field satisfying the condition $\frac{1}{4}({\ ||\nabla
\mathcal{L}||}_{g}^{2}+2<\nabla\mathcal{L}, b>_{g})>0$. We start by a local
construction near the recurrent sets and then give a global construction on a
compact Riemannian manifold.
\subsection{Local construction of Lyapunov functions}
\begin{lem}
\label{lya} Given a Riemannian manifold (V, g), of dimension n and a vector
field b on V,  for any hyperbolic stationary point of b or hyperbolic periodic
orbit there exists a local Lyapunov function $\mathcal{L}$ at that point or
periodic orbit
such that on the domain of definition of $\mathcal{L}$ outside the set of
hyperbolic
points and periodic orbits,  $\Psi(L)=\frac{1}{4}({\ ||\nabla\mathcal{L}||}%
_{g}^{2}+2<\nabla\mathcal{L}, b>_{g})>0$.
\end{lem}
\bigskip

\noindent Since $\Omega$ is a Morse-Smale vector field,  the limit sets are
contained in
the union of critical points and limit cycles. We refer to Kamin (see
\cite{K, K1} for the construction of $\mathcal{L}$ in the case of a critical
attractive point of the field.

First we construct $\mathcal{L}$ near an attractive orientable periodic orbit
$\Gamma$. Then the normal bundle $\mathcal{N}$ to $\Gamma$ is trivial. Denote
by T the minimal period of $\Gamma.$ Let p:$\mathbb{R}\rightarrow V$,  be a
periodic trajectory of period T,  the image of which is $\Gamma.$ We consider
a
Fermi coordinate system $(\theta, x_{1}, .., x_{n-1})$ in a neighborhood
$\mathcal{U}$ of the cycle. $\theta$ is the cyclic coordinate and
$(e_{1}(\theta), .., e_{n-1}(\theta))$ is the corresponding orthonormal frame
field trivializing the normal bundle of $\Gamma$ i.e.dx$^{i}(e_{j}%
)=\delta_{ij}$, the Kronecker symbol, 1$\leq i, j\leq n-1$. If $\xi$ is a
point
in the domain $\mathcal{U}$ of this coordinate system,
\[
\xi=\exp_{p(\theta(\xi))}[\sum_{i=1}^{n-1}x_{i}(\xi)e_{i}(\theta(\xi))]
\]
In these coordinates:%
\[
g=\sum_{i, j=0}^{n-1}g_{ij}dx^{i}dx^{j},
\]
where we set dx$_{0}=d\theta.$Along the cycle,  $g_{ij}(\theta,
0)=\delta_{ij}%
$, 1 $\leq i, j\leq n-1, $g$_{0j}(\theta, 0)=g_{j0}(\theta, 0)=0, 1\leq j\leq
n-1$. The Christoffel symbols associated to the coordinates $(x_{1}%
, .., x_{n-1})$are zero along $\Gamma$ : $\Gamma_{ij}^{k}(\theta, 0)=0$,  for
$i, j, k\in\{1..n-1\}$. In a neighborhood of $\Gamma$,  $g_{ij}(\theta
, x)=\delta_{ij}+O(d(x))$. In this coordinates the equation $\frac{T\xi}%
{dt}=\Omega(\xi)$ can be written as follows:
\begin{align}
\dot{\theta}  &  =1+O(d(x))\\
\dot{x}  &  =B(\theta)x+O(d^{2}(x))
\end{align}
The solutions are given by
\[
\theta(t)=\theta(\xi(t)), x_{i}(t)=x_{i}(\xi(t)), 1\leq i\leq n-1,
\]
where $\xi$ is trajectory of $b$ and d(x) is the distance from x to $\Gamma$,
$B(\theta)$ is a (n-1)x(n-1) matrix-valued function.

Consider the solution X:$\mathbb{R} \rightarrow GL[n-1;\mathbb{R]}$ of the
matrix equation $\dot{X}=B(\theta)X$ such that $X(0)=Id_{n-1}$. Then there
exist a (in general complex) matrix D and a matrix function P:$\theta
\in\mathbb{R}\rightarrow P$($\theta$)$\in$M[(n-1)x(n-1);$\mathbb{C]},$  such
that $X(\theta)=P(\theta)e^{\theta D}$ for all $\theta\in\mathbb{R}.$ Because
the orbit $\Gamma$ is hyperbolic and attractive,  the real parts of the
eigenvalues of D are negative: the eigenvalues of D are exactly the
characteristic multipliers of the orbit.

Denote by D* the complex conjugate transpose of D. Let $\mu$ be a strictly
positive parameter. Consider the matrix Lyapunov equation in the unknown
matrix A:
\[
AD+D^{\ast}A=-\mu A^{2}.
\]
It is well known that there exists a symmetric positive definite (n-1)x(n-1)
matrix A solution of this equation,  given by the formula:
\[
A^{-1}=\mu\int_{0}^{+\infty}e^{tD^{\ast}}e^{tD}dt.
\]
We define a local Lyapunov function $\mathcal{L}$:$\mathcal{U}\rightarrow
\mathbb{R}$ as follows: if $\ \xi\in\mathcal{U}$;
\begin{align}
\xi=\exp_{p(\theta(\xi
)}[\sum_{k=1}^{n-1}x_{k}(\xi)e_{k}(\theta(\xi))],
\end{align}
then: $\mathcal{L} =(AX, \overline{X})_{\mathbb{R}^{(n-1)}}$where X
=P($\theta(\xi)^{-1}($%
x$_{1}(\xi), ...., $x$_{(n-1)}(\xi))$ and the bar denotes the complex
conjugate.
\noindent Let $\xi$: t$\in\mathbb{R}_{+}\rightarrow\xi$(t)$\in$V,  be a
positive semi-
trajectory of b contained in\ $\mathcal{U}$ \ and let X(t)=P($\theta
(\xi))^{-1}$(x$_{1}$($\xi(t)$), ..., x$_{(n-1)}$($\xi(t)$)).
\[
\frac{dX(t)}{dt}=BX(t)+O(||X(t)||^{2})
\]
\begin{align*}
{\frac{d\mathcal{L}(\xi(t))}{dt}}  &  ={}(A\frac{dX(t)}{dt}, \overline
{X(t)})+(AX(t), \overline{\frac{dX(t)}{dt}})\\
\frac{d\mathcal{L}(\xi(t))}{dt}  &  ={}(AD+D^{\ast}A)X(t), \overline
{X(t)})+O(||X(t)||^{3})\\
\frac{d\mathcal{L}(\xi(t))}{dt}  &  =-\lambda(AX(t), A\overline{X(t)}%
)+O(||X(t)||^{3})
\end{align*}
where for U, V$\in\mathbb{C}^{(n-1)}$ , \ \ (U,
V)=$\sum_{k=1}^{(n-1)}U_{k}V_{k}%
$. Hence L is decreasing along the trajectories of the field b in a
sufficiently small neighborhood of $\Gamma$.
\[
(\nabla\mathcal{L}(\xi(t)), b(\xi(t)))=(AX(t),
A\overline{X(t)})+O(||X(t)||^{3}%
)
\]%
\[
(\nabla\mathcal{L}(\xi(t)), \nabla\mathcal{L}(\xi(t))=4(AX(t), A\overline
{X(t)})+O(||X(t)||^{3})
\]
and
\[
\Delta\mathcal{L}=-2tr(A).
\]
Recall that $\Psi(L)=\frac{|\nabla L|^{2}}{4}+\frac{(\nabla L, \Omega)}{2}$
and
thus
\[
\Psi()=(1-\lambda/2)(AX(t), A\overline{X(t)})+O(||X(t)||^{3}.
\]
For $1-\lambda/2>0$,  $\Psi(L)$ is strictly positive in the neighborhood of
the cycle.
\bigskip

\noindent Assume now that we have a periodic orbit $\gamma$ of b with minimal
period T
which reverses the orientation. Let $\Pi:\overset{\symbol{126}}{V}\rightarrow
V$ be the covering of V associated to the cyclic subgroup of $\pi_{1}$(V)
generated by 2[$\gamma$] ([$\gamma$] = homotopy class of $\gamma)$. This is a
Galois covering space with group$\mathbb{\ Z}_{2}.$ Denote by $i$ the
nontrivial
deck transformation of this covering. g and b have unique lifting to
$\overset{\symbol{126}}{V}$ still denoted by g and b. $\Gamma$=$\Pi
^{-1}(\gamma)$ is a periodic orbit of the lifting of b (i.e) b of minimal
period 2T. We can apply the preceding construction to $\Gamma$ but with extra
care here because of the deck involution \textit{i.} g and b are invariant by
\textit{i.} We choose the neighborhood $\mathcal{U}$ invariant by \textit{i}
and the Fermi coordinate system so that:
\[
Tie_{k}(\theta)=e_{k}(\theta+T)\text{ \thinspace\ for 0 }\leq\text{k}%
\leq(n-1)
\]

Then the previous construction produces a Lyapunov function $\overset
{\symbol{126}}{L}$ for the lifting of b in a neighborhood of $\Gamma$.This
function is invariant by the Galois group of the covering and hence can be
pushed down by $\Pi$ to a Lyapunov function L of b in a neighborhood of
$\gamma$:L=$\overset{\symbol{126}}{L}\circ\Pi$.

For repulsive orbits the same construction applies changing b into --b.
Finally the general case of a general hyperbolic periodic orbit $\Gamma$ can
be easily handled by noticing that $\Gamma$ has a basis of open neighborhoods
$\mathcal{U}$ such that $\mathcal{U}$ is diffeomorphic to the fiber product
over $\Gamma$ of$\mathcal{\ U\cap}$W$^{s}(\Gamma)$ and$\mathcal{\ U\cap}%
$W$^{u}(\Gamma)$ and then ''patching up'' the Lyapunov functions constructed
above for the restrictions of b to the stable and unstable manifolds
W$^{s}(\Gamma)$ , \ W$^{u}$($\Gamma)$ of $\Gamma$.

More precisely,  due to the transversal intersection of the stable manifold
with the unstable manifold at the hyperbolic set,  a Lyapunov function is
built
in each set as follows: $N^{s}$ (resp. $N^{u}$ ) denotes the normal bundle of
the stable (resp. unstable) manifold at $\Gamma$. Inside each normal fiber,  a
Lyapunov function is found. The coordinate system at a point $\ \xi
\in\mathcal{U}$ is such that $\xi=\exp_{p(\theta(\xi)}[\sum_{k=1}^{n-1}%
x_{k}(\xi)e_{k}(\theta(\xi))]=\exp_{p(\theta(\xi)}[X_{s}+X_{u}]$ then as in
the first part of the lemma,  we can defined find two matrices $A_{u}$ and
$A_{s}$ solving the matrix Lyapunov equation in each subspace $N^{u}$ and
$N^{s}$ respectively. If we denote $n_{u}=dimN^{u}$ and,  $n_{s}=dimN^{s}$,
then $n_{u}+n_{s}-1=n$ and we can now define the Lyapunov function
$\mathcal{L}$ by : $\mathcal{L}(\xi)=(A_{s}Y_{s}, \bar{Y}_{s})_{\mathbb{R}%
^{(n_{s})}}-(A_{u}Y_{u}, \bar{Y}_{u})_{\mathbb{R}^{(n_{u})}}$ where
Y$_{u}=P_{u}$($\theta(\xi)^{-1}($x$_{1}(\xi), ...., $x$_{(n_{u})}(\xi))$ and
Y$_{s}=P_{s}$($\theta(\xi)^{-1}($x$_{n_{u}+1}(\xi), ...., $x$_{(n-1)}(\xi))$.
Here $P_{u}$ and $P_{s}$ are two matrices defined as in the first part of the
Lemma for the solution of the matrix equation in each normal bundle. From the
construction,  it follows that $\mathcal{L}$ is positive and decays along the
trajectory of the vector field.

We can repeat this type of construction locally in a any neighborhood of a
limit set of the field. Indeed since the limit set is hyperbolic,  the
construction of the function $\mathcal{L}$ is possible on each attractive or
repulsive fiber and due to the transversal intersection we can match the
construction and the function $\mathcal{L}$ vanishes only on the limit set in
these neighborhoods. Outside the limit set,  there is a large choice of
extensions of the function $\mathcal{L}$.
\subsection{Global construction of Lyapunov functions}
In this section we prove the existence of a global Lyapunov function,
assuming
the existence of a local one (proved in the last section).

$M$ is a $C^{\infty}$ compact Riemannian manifold and $\phi_{t}$,  a
Morse-Smale flow. $F$ is the vector field on V,  generator of the flow. Let us
denote by $\Omega$ the limit set of the flow. $\Omega$ is endowed with a
partial order as follows: if $\omega_{1}, \omega_{2}$belong to $\Omega
, \omega_{1}\succ\omega_{2}$ if there exists a trajectory the $\alpha-$\ limit
set of which is $\omega_{1}$ \ and the $\omega-$limit set is $\omega_{2}.$
This
partial order determines a filtration of $\Omega$: $\Omega=\Omega_{0}%
\supset\Omega_{1}\supset...\supset\Omega_{m}$ as follows: Let $\Omega_{0,
max}$
be the set of all maximal elements of $\Omega$. Set $\Omega_{1}=\Omega
-\Omega_{0, max}$. Let $\Omega_{1, max}$ be the set of the maximal elements of
$\Omega_{1}$. Set $\Omega_{2}=\Omega_{1}-\Omega_{1, max}$ and so on. When
$\Omega_{n}$ has been defined,  let $\Omega_{n, max}$ be the set of all
maximal
elements of $\Omega_{n}$. Set $\Omega_{n+1}=\Omega_{n}-\Omega_{n, max}$. This
filtration ends at a certain m. $\Omega_{m}$ is the set of all minimal
elements of $\Omega$.
\subsubsection{Local Lyapunov functions}
A strict Lyapunov function for $F$ on an open subset $O$ of $M$ is a
$C^{\infty}$ function $\mathcal{L}:O\rightarrow M$ such that $d\mathcal{L}%
(F)<0$ in $O-\Omega$ . For $\omega\in\Omega$ there exists two relatively
compact neighborhoods $O_{\omega}$ and $\tilde{O}_{\omega}$ of $\omega$ in
$M_{n}$ such that the closure $\bar{O}_{\omega}$of $O_{\omega}$ in M,  is
contained in $\tilde{O}_{\omega}$ and a strict Lyapunov function
$\mathcal{L}_{\omega}:\tilde{O}_{\omega}\rightarrow\mathbb{R}$ such that:

(i) $\tilde{O}_{\omega}\cap\tilde{O}_{\omega^{\prime}}=\varnothing, $ for
$\omega, \omega^{\prime}\in\Omega$,  $\omega\neq\omega^{\prime}$.

(ii) The boundary $\partial O_{\omega}$ is the union of $\partial_{s}%
O_{\omega}\cup\partial_{u}O_{\omega}\cup\partial_{l}O_{\omega}$ where the
three components are compact codimension 1 submanifolds with boundaries
$\partial\partial_{s}O_{\phi}$,  $\partial\partial_{u}O_{\omega}$,
$\partial\partial_{l}O_{\omega}$ such that $\partial\partial_{s}O_{\omega}%
\cup\partial\partial_{u}O_{\omega}=\partial\partial_{l}O_{\omega}$ and
$\partial_{s}O_{\omega}\cap\partial_{u}O_{\omega}=\emptyset.$

(iii) $\partial_{s}O_{\omega}$ and $\partial_{u}O_{\omega}$ are transversal to
the flow and $\partial_{l}O_{\omega}$is foliated by arcs of trajectories of
the flow $\phi_{t}$ linking $\partial_{s}O_{\omega}$ to $\partial_{u}%
O_{\omega}.$

(iv) $L_{\omega}$ is constant on $\partial_{s}O_{\omega}$ and $\partial
_{u}O_{\phi}.$

(v) $\partial_{s}O_{\omega}\cap W^{s}(\omega)$ (resp. $\partial_{u}O_{\omega
}\cap W^{u}(\omega)$ ) is a compact $C^{\infty}$ sub-manifold contained in
$W^{s}(\omega)$ (resp. $W^{s}(\omega)$) which is a section of the restriction
of the flow to $W^{s}(\omega)$ (resp. $W^{u}(\omega)$). If $\omega$ is maximal
in $\Omega$,  $\partial_{l}O_{\omega}=\emptyset=\partial_{s}O_{\omega}$. If
$\omega$ is minimal in $\Omega, $ $\partial_{l}O_{\omega}=\emptyset
=\partial_{u}O_{\omega}$.
\subsubsection{Construction of the global Lyapunov function}
The construction is inductive. To fix the ideas,  we can always assume that,
if
$\omega$ is neither maximal nor minimal, $\mathcal{L}_{\omega}(\partial
_{s}O_{\omega})=1$ ,  $\mathcal{L}_{\omega}(\partial_{u}O_{\omega})=-1$
$\mathcal{L}_{\omega}(\omega)=0$ and $\mathcal{L}_{\omega}(\bar{O}_{\omega
})=[-1, 1]$.

In the case $\omega$ is maximal ,  we can assume that $O_{\omega}$ is an open
ball whose boundary $\partial_{u}O_{\omega}$\{$\mathcal{L}_{\omega}(\omega)=0$
\} is a sphere that $\mathcal{L}_{\omega}(\omega)=-1$ and $\mathcal{L}%
_{\omega}(\bar{O}_{\omega})=[-1, 0]$.

In the case $\omega$ is minimal ,  we can assume that $O_{\omega}$ is an open
ball whose boundary $\partial_{s}O_{\omega}$ is a sphere \{$\mathcal{L}%
_{\omega}(\omega)=0$ \},  that $\mathcal{L}_{\omega}(\omega)=1$ and
$\mathcal{L}_{\omega}(\bar{O}_{\omega})=[0, 1]$.

The construction starts as follows: Let $M_{0}=\{\cup\bar{O}_{\omega}%
|\omega\in\Omega_{0}^{max}\}$. $M_{0}$ is a compact manifold with boundary\{
$\cup\partial_{u}O_{\omega}|\omega\in\Omega_{0}^{max}\}$. Define
$\mathcal{L}_{0}:M_{0}\rightarrow\mathbb{R}, $ as follows $\mathcal{L}{_{0}%
}_{|\bar{O}_{\omega}}=\mathcal{L}_{\omega}$. It is clear that

(i) $M_{0} \cap\cup\{ \tilde{O}_{\omega},  \omega\in\Omega_{1} \} =\emptyset$

(ii) $M_{0}$ is a sub-manifold of codimension $0$ with boundary $\partial
M_{0}$.

(iii) $\mathcal{L}_{0}:$ is a strict Lyapunov function on $M_{0}$ and
$\mathcal{L}_{0}(\partial M_{0})=-1$

Assume now that we have constructed a sub-manifold $M_{n}$of codimension $0$
in $M$ with boundary $\partial M_{n}$,  and a strict Lyapunov function
$\mathcal{L}_{n}:M_{n}\rightarrow\mathbb{R}$,  such that

(i) $M_{n}$ contains $\cup_{k=0}^{n} \Omega^{max}_{k}$ in its interior and
$M-M_{n} \supset\cup\{ \bar{O}_{\omega} | \omega\in\Omega_{n+1} \}$

(ii) $L_{n}$ is a constant on $\partial M_{n}$ and equal to c$_{n}, $ say.

For each $\omega\in\Omega_{n+1}^{max}$ let $V_{\omega}=\partial M_{n}\cap
\cup\{\phi_{-t}(\partial_{s}O_{\omega})|t\geq0\}$. Then $V_{\omega}$ is a
compact sub-manifold of $\partial M_{n}$ of codimension $0$ with boundary
$\partial V_{\omega}=\partial M_{n}\cap\{\phi_{-t}(\partial\partial
_{s}O_{\omega})|t\geq0\}$ in $\partial M_{n}$.

Also

(i) $V_{\omega} \cap V_{\omega^{\prime}} = \emptyset$ for $\omega
, \omega^{\prime}\in\Omega^{max}_{n+1}$ and $\omega\neq\omega^{\prime}$

(ii) $V_{\omega}$ contains the manifold $\partial M_{n} \cup W^{s}_{\omega}$
in its interior.

There is for each $\omega\in\Omega^{max}_{n+1}$ a $C^{\infty}$ function
$T_{s, \omega} : V_{\omega} \rightarrow]0,  + \infty[$ such that for any $x
\in
V_{\omega} $,  $\phi(t, x) \notin\bar{O}_{\omega},  0\leq t\leq T_{s,
\omega}(x)$
and $\phi( T_{s, \omega}(x), x) \in\partial_{s} O_{\Omega}$.

Let $\epsilon$ be a small positive number $\epsilon<inf_{x\in V_{\omega}%
}T_{s, \omega}(x)$ and let $D_{\omega}^{1}\subset\mathbb{R}\times V_{\omega}$
be the subset $\{(t, x)|x\in V_{\omega}, -\epsilon\leq t\leq T_{s,
\omega}(x)\}$.
$\phi$ maps $D_{\omega}^{1}$ diffeomorphically into $M_{n}$. $D_{\omega}^{1}$
and $\phi(D_{\omega}^{1})$ are manifolds with corners.

We can choose a $C^{\infty}$ function $\hat{c}_{\omega}: D^{1}_{\omega}
\rightarrow]0,  + \infty[ $ such that

(i) if $x\in V_{\omega}$ and $-\epsilon\leq t\leq0$,  $\hat{c}_{\omega
}(t, x)=-\frac{d}{dt}\mathcal{L}_{n}(\phi_{t})$

(ii) if $x\in V_{\omega}$ and $T_{s, \omega}(x)-\epsilon\leq t\leq T_{s,
\omega
}(x)$,  $\hat{c}_{\omega}(t, x)=-\frac{d}{dt}\mathcal{L}_{\omega}(\phi_{t})$.
$\mathcal{L}_{\omega}$ is defined on a neighborhood of $\partial_{s}O_{\Omega
}$.

Let
\begin{align}
\Gamma= \sup_{\omega\in\Omega^{max}_{n+1} \,  ,  x \in V_{\omega}} \int
_{0}^{T_{s, \omega}(x) } \hat{c}_{\omega}(t, x)dt.
\end{align}
Define now a function $\lambda_{\omega}:V_{\omega} \rightarrow]0,  + \infty[ $
by the formula $\lambda_{\omega}(x) = \frac{T_{s, \omega}(x) -\epsilon}%
{\Gamma-\int_{0}^{T_{s, \omega}(x) }\hat{c}_{\omega}(t, x)}dt $ It is a
$C^{\infty} $ function.

Let u be a $C^{\infty}$ function $\mathbb{R}\rightarrow]0, +\infty\lbrack$
such
that supp$u\in]0, 1[$ and $\int_{0}^{1}u=1$. Let us define the function
$a_{\omega}:D_{\omega}^{1}\rightarrow]0, +\infty\lbrack, $ as follows
$a_{\omega}(t, x)=\lambda_{\omega}(x)u(\frac{t}{T_{s, \omega}(x)-\epsilon}) $.
It is a $C^{\infty}$ function. Define $c_{\omega}:D_{\omega}^{1}%
\rightarrow]0, +\infty\lbrack$ as the sum $\hat{c}_{\omega}+a_{\omega}$ then

(i) $\int_{0}^{T_{s, \omega}(x) } c_{\omega}(t, x)dt= \Gamma$

(ii) $c_{\omega}(t, x)=-\frac{d}{dt}\mathcal{L}_{n}(\phi_{t}(x))$ for all
$x\in
V_{\omega}$ and $-\epsilon\leq t\leq0$.

(iii) $c_{\omega}(t, x)=-\frac{d}{dt}\mathcal{L}_{\omega}(\phi_{t}(x))$ for
all
$x\in V_{\omega}$ and $T_{s, \omega}(x)-\epsilon\leq t\leq T_{s, \omega}(x)$.

(iii) implies that for all $x\in V_{\omega}$ all t such that $T_{s, \omega
}(x)-\epsilon\leq t\leq T_{s, \omega}(x)$,
\begin{align}
&  \int_{0}^{t}c_{\omega}(s, x)ds=\int_{0}^{T_{s, \omega}(x)}c_{\omega
}(t, x)dt+\int_{T_{s, \omega}(x)}^{t}c_{\omega}(t, x)dt=\Gamma_{\omega}%
-\int_{T_{s, \omega}(x)}^{t}\frac{d}{dt}\mathcal{L}_{\omega}(\phi_{t}(x))\\
&  =\Gamma_{\omega}+\mathcal{L}_{\omega}(\phi_{T_{s, \omega}}(x),
x)-\mathcal{L}%
_{\omega}(\phi_{t}(x), x)=\Gamma_{\omega}+1-\mathcal{L}_{\omega}(\phi_{t}(x),
x)
\end{align}

For each $\omega\in\Omega_{n+1}^{max}$ let $U_{\omega}=\phi(D_{\omega}%
^{1})\cup\bar{O}_{\omega}$. Define a function $\widehat{\mathcal{L}}_{\omega
}:U_{\omega}\rightarrow\mathbb{R}$ as follows: if $y\in\phi(D_{\omega}%
^{1}), y=\phi_{t}(x), -\epsilon\leq t\leq T_{s, \omega}(x)$,  $\widehat
{\mathcal{L}}_{\omega}(y)=c_{n}-\int_{0}^{t}c_{\omega}(s, x)ds$. If $y\in
\bar{O}_{\omega}$,  $\widehat{\mathcal{L}}_{\omega}(y)=c_{n}-\Gamma_{\omega
}-1+\mathcal{L}_{\omega}(y)$ It is easy to see that $\widehat{\mathcal{L}%
}_{\omega}$ is a strict Lyapunov function on $U_{\omega}$ and $\widehat
{\mathcal{L}}_{\omega}(V_{\omega})=c_{n}.$
\medskip

We will now proceed with the construction of $M$. For each
$\omega\in\Omega_{n+1}^{max}$,  let $CV_{\omega}$ be a collar for
$\partial V_{\omega}$ in $V_{\omega}$ such that $W^{s}(V_{\omega})\cap\partial
M_{n}\subset V_{\omega}-\overline{\bar{CV}_{\omega}}$. There exists a
$C^{\infty}$ function $T_{u, \omega}:CV_{\omega}\rightarrow]0, +\infty\lbrack$
such that $\phi(t, x)\notin\partial_{u}O_{\omega}$ if $-\epsilon\leq
t<T_{u, \omega}$ and $\phi(T_{u, \omega}(x), x)\in\partial_{u}O_{\omega}$. Let
\[
N_{n}=(\partial M_{n}-\cup\{V_{\omega}, \omega\in\Omega_{n+1}^{max}%
\})\cup\{CV_{\omega}|w\in\Omega_{n+1}^{max}\}
\]
$N_{n}$ is a submanifold of codimension 0 with boundary in $\partial M_{n}$.
It is easy to construct a $C^{\infty}$ function $T_{u}:N_{n}\rightarrow
]0, +\infty\lbrack$ such that ${T_{u}}=T_{u, \omega}$ for any $\omega\in
\Omega_{n+1}^{max}$. Let $D^{u}=\{(t, x), x\in N_{n}0\leq t\leq T_{u}(x)\}$.
$\phi$ maps $D^{u}$ diffeomorphically on a manifold with corners of dimension
0 in $M$. Then we can construct a $C^{\infty}$ function $\tilde{c}%
:N_{n}\rightarrow]0, +\infty\lbrack$ such that:
\[
\tilde{c}(t, x)=-\frac{d\widehat{\mathcal{L}}_{\omega}}{dt}%
\]

Let $\tilde{\gamma}:N_{n}\rightarrow]0, +\infty\lbrack$ be the function
$\tilde{\gamma}(x)=\int_{0}^{T_{u}(x)}\tilde{c}(t, x)dt$. $\tilde{\gamma}$ is
a
$C^{\infty}$ function. We define $c:N_{n}\rightarrow]0, +\infty\lbrack$ as
follows:
\[
{c}(t, x)=\frac{\tilde{c}(t, x)}{\tilde{\gamma}(x)}(\Gamma+2)
\]
then $c$ is a $C^{\infty}$ function and $\int_{0}^{T_{u}(x)}{c}(t, x)=\Gamma
+2$. If $\omega\in\Omega_{n+1}^{max}$ and $x\in CV_{\omega}$,
\begin{equation}
\tilde{\gamma}(x)=\int_{0}^{T_{u}(x)}{c}(t, x)=\int_{0}^{T_{u}(x)}%
-\frac{d\widehat{\mathcal{L}}_{\omega}(\phi_{t}(x))}{dt}=\widehat{\mathcal{L}%
}_{\omega}(x)-\widehat{\mathcal{L}}_{\omega}(\phi_{T_{u}}(x))
\end{equation}

\begin{equation}
\tilde{\gamma}(x)=c_{n}-(c_{n}-1+\Gamma+\widehat{\mathcal{L}}_{\omega}%
(\phi_{tT_{u}}(x))=\Gamma+2
\end{equation}
Hence for $\omega\in\Omega_{n+1}^{max}, c\in CV_{\omega}, -\epsilon\leq t\leq
T_{u}(x)$,  $c(t, x)=\tilde{c}(t, x)=-\frac{d\widehat{\mathcal{L}}_{\omega}%
(\phi_{t}(x))}{dt}.$

Let
\[
M_{n+1}=M_{n}\cup\phi(D^{u})\cup\{\cup U_{\omega}|\omega\in\Omega_{n+1}%
^{max}\}.
\]
Define $\mathcal{L}_{n+1}:M_{n+1}\rightarrow\mathbb{R}$ as follows: if $y\in
M_{n}$,  $\mathcal{L}_{n+1}(y)=\mathcal{L}_{n}(y)$ if $y\in D^{u}$,  $y=\phi
_{t}(x), x\in N_{\omega}$,  $\mathcal{L}_{n+1}(y)=c_{n}-\int_{0}^{t}%
c(\tau, x)d\tau$ and if $y\in U_{\omega}$,  $\omega\in\Omega_{n+1}^{max}$,
$\mathcal{L}_{n+1}(y)=\widehat{\mathcal{L}}_{\omega}(y)$.

This function is well defined and $C^{\infty}$. If $y \in\phi(D^{u})
\cap_{\omega\in\Omega^{max}_{n+1}} \cup U_{\omega} $,  then $y= \phi_{t}(x),
x
\in CV_{\omega}$ for $0\leq t\leq T_{u}(x) = T_{u,  \omega}(x)$,  but then%

\[
c_{n}-\int_{0}^{t}c(\tau, x)d\tau=c_{n}+\int_{0}^{t}\frac{d\mathcal{L}_{\omega
}(\phi_{v}(x))}{dv}=c_{n}+\widehat{\mathcal{L}}_{\omega}(\phi_{t}%
(x))-\widehat{\mathcal{L}}_{\omega}(x)=\widehat{\mathcal{L}}_{\omega}(y)
\]

By construction,  $\mathcal{L}_{n+1}$ is a strict Lyapunov function on
$M_{n+1}$. $M_{n+1}$ is a manifold of codimension 0 in $M$ with boundary
$\partial M_{n}=\cup\{\partial_{u}O_{\omega}|\omega\in\Omega_{n+1}^{max}%
\}\cup\tilde{N}$ where $\tilde{N}=\{\phi(T_{u}(x), x), x\in N_{n}\}$. If
$y\in\partial_{u}O_{\omega}\cap\tilde{N}$ then $y=\phi(T_{u}(x), x)$,  for
some
$x\in CV_{\omega}.T_{u}(x)=T_{u, \omega}(x)$. We have $\partial M_{n+1}%
=\mathcal{L}_{n+1}^{-1}(c_{n+1})$,  where
$c_{n+1}=c_{n+1}-\Gamma-2$.\quad\hbox{\hskip4pt\vrule width 5pt height 6pt
depth 1.5pt}

\bigskip

Finally,  we remark that $b$ MS is equivalent to $\Omega$ MS,  when
\[
b=\Omega+\nabla\mathcal{L}.
\]

\subsection{Appendix 2}
In this appendix we are going to provide the estimates needed in (*). On the
Riemannian manifold (V, g) consider an operator of the form:%

\[
L=\varepsilon\Delta_{g}+\varepsilon\theta(b)+(\psi+\sqrt{\varepsilon
}c_{\varepsilon}),
\]
where $\psi$ is a non negative function on V and c$_{\varepsilon}%
, \varepsilon\in\lbrack0, 1], $ is an arbitrary function on V,  parametrized
by
$\varepsilon$, smooth in all the variables .

We assume that we have a family\{u$_{\varepsilon}$%
$\vert$%
$\varepsilon\in]0, 1]$\} of smooth functions on V such that: u$_{\varepsilon
}\geq0$ for all $\varepsilon$ and%
\[
\varepsilon\Delta_{g}u_{\varepsilon}+\varepsilon\theta(b)u_{\varepsilon}%
+(\psi+\sqrt{\varepsilon}c_{\varepsilon})u_{\varepsilon}=0
\]
Then we can state a lemma:
\begin{lem}
There exists constants C,  $\varepsilon_{0}$%
$>$%
0,  depending only on g, b, $\psi$,  and $\underset{[0, 1]\times V}{\max
}|c_{\varepsilon}|$ $($but independent of $\varepsilon!)$ and a universal
mapping $\gamma:\hbox{\bb N}_{+}$---%
$>$%
$\hbox{\bb N}_{+}$ , such that for each integer $n\in\hbox{\bb N}^{+}$ ,  for
all x $\in V_{\psi}(=\{z||$ $z\in V, \psi(z)\neq0\})$, all $\varepsilon
\in]0, \varepsilon_{0}]$:%
\[
u_{\varepsilon}(x)\leq(\underset{V}{\max}u_{\varepsilon})\frac{\gamma
(n)(C\varepsilon)^{\frac{n}{3}}}{\psi^{n}}%
\]
It follows that at any x$\in V_{\psi}$,  u$_{\varepsilon}$(x) tends to zero
faster than any power of $\varepsilon$ and the convergence is uniform on any
compact in V$_{\psi}$.
\end{lem}
\bigskip

{\noindent \bf Proof.}

\noindent The proof will use be an induction on n. Given any smooth functions
f, h on V we
have:
\begin{eqnarray}
f[\varepsilon\Delta_{g}h+\varepsilon\theta(b)h]  &  =\varepsilon\Delta
_{g}(fh)+\varepsilon\theta(\widehat{b})(fh)-fh\left[  \varepsilon\frac
{\Delta_{g}f}{f}+2\varepsilon\frac{||\nabla f||_{g}^{2}}{f^{2}}+\varepsilon
\frac{\theta(b)f}{f}\right] \label{clef}\\
f[\varepsilon\Delta_{g}h+\varepsilon\theta(b)h]  &  =\varepsilon\Delta
_{g}(fh)+\varepsilon\theta(\widehat{b})(fh)-fh\varepsilon\left[  \frac
{\Delta_{g}f}{f}+\frac{\theta(\widehat{b})f}{f}\right] \\
&  on\text{ V}_{f}, \nonumber
\end{eqnarray}
where $\widehat{b}=b+2\frac{\nabla f}{f}, V_{f}=\{x||$ x$\in V, f(x)\neq0\}.$
For each n$\in\mathbb{Z}_{+}$, set v$_{n}=\psi^{n}u_{\varepsilon}.$ Write:%
\begin{align}
\psi^{n}\left[  \varepsilon\Delta_{g}u_{\varepsilon}+\varepsilon
\theta(b)u_{\varepsilon}+(\psi+\sqrt{\varepsilon}c)u_{\varepsilon}\right]   &
=\frac{\psi^{n+1}}{\psi}\varepsilon\left[  \Delta_{g}u_{\varepsilon}%
+\theta(b)u_{\varepsilon}\right]  +(\psi+\sqrt{\varepsilon}c)u_{\varepsilon
}\psi^{n}, \text{ }\label{stupid}\\
&  \text{on V}_{\psi}\nonumber.
\end{align}
Applying the formula(\ref{clef}) to (\ref{stupid}) taking f=$\psi^{n+1}$ and
h=u$_{\varepsilon}$, we get, for n$\geq2$:%
\[
\frac{1}{\psi}\varepsilon\left[  \Delta_{g}v_{n+1}+\theta(\widehat{b}%
)v_{n+1}\right]  +v_{n+1}+\sqrt{\varepsilon}c_{\varepsilon}v_{n}%
-(n+1)\varepsilon\lbrack\theta(\widehat{b})\psi)+\Delta_{g}\psi]v_{n-1}%
\]
\begin{equation}
-(n+1)(n+2)\varepsilon||\nabla\psi||_{g}^{2}v_{n-2}=0 \label{trivial}.
\end{equation}
Let M$_{n}=\underset{V}{\max}$ v$_{n}.$ Note that M$_{0}$=$\underset{V}{\max
}u_{\varepsilon}.$ Choose a point P$\in$V such that M$_{n+1}$=v$_{n+1}$(P).
Note
that $\psi$(P)$\neq$0. Evaluate the equation(\ref{trivial}) at P.%
\[
\varepsilon\left[  \Delta_{g}v_{n+1}(P)+\theta(\widehat{b})v_{n+1}\right]
(P)=\varepsilon\Delta_{g}v_{n+1}(P)\geq0.
\]
For n$\geq2$:%
\[
v_{n+1}(P)+\sqrt{\varepsilon}c_{\varepsilon}(P)v_{n}(P)-(n+1)\varepsilon
\lbrack\theta(\widehat{b})\psi)+\Delta_{g}\psi](P)v_{n-1}(P)
\]
\begin{equation}
-(n+1)(n+2)\varepsilon||\nabla\psi||_{g}^{2}(P)v_{n-2}(P)\leq0 \label{cad}%
\end{equation}
The relation(\ref{cad}) implies the inequality:%
\begin{equation}
M_{n+1}\leq\sqrt{C\varepsilon}M_{n}+(C\varepsilon)(n+1)M_{n-1}+(C\varepsilon
)(n+1)(n+2)M_{n-2}, \text{for n}\geq2, \label{it}%
\end{equation}
where :
\[
C=\max\left(  \underset{V\times\lbrack0, 1]}{\sup}|c_{\varepsilon}%
|, \underset{V}{\sup[}\sqrt{|\theta(b)\psi|+|\Delta_{g}\psi|}, ||\nabla
\psi||_{g}]\right).
\]
The relation (\ref{it}) implies that:%
\begin{equation}
\frac{M_{n+1}}{(n+2)!}\leq\sqrt{C\varepsilon}\frac{M_{n}}{(n+1)!}%
+C\varepsilon\frac{M_{n-1}}{n!}+C\varepsilon\frac{M_{n-2}}{(n-1)!}, \text{for
n}\geq2 \label{it1}%
\end{equation}
Clearly M$_{n}\leq$max(1, $\underset{V}{\max}\psi, \underset{V}{\max}\psi^{2}%
$)M$_{0}$, if n=0, 1, 2.The lemma follows easily from this remark and the
reccurence relation(\ref{it1}). \QED

\textsc{\ } \begin{figure}[ptb]
\centering\psfig{figure =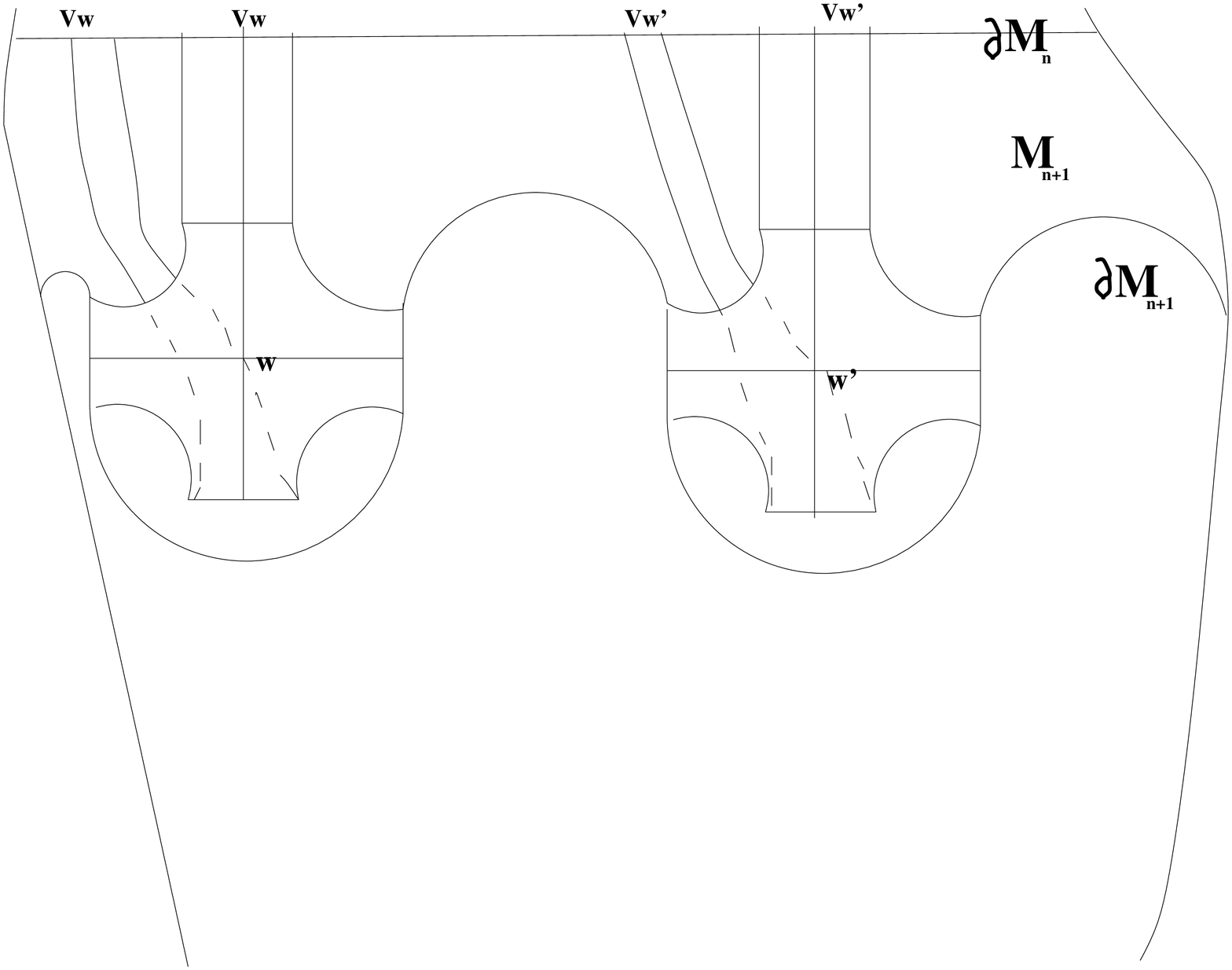, height =95mm, width=95mm}
\caption{ Construction by induction of the global Lyapunov function}%
\label{geodesic}%
\end{figure}


\begin{thebibliography}{99}
                                           %


\bibitem {Aubin}\textsc{T. Aubin}-- Some Nonlinear Problems in Riemannian
Geometry; Springer,  1998

\bibitem {DEF}\textsc{A. Devinatz}-- \textsc{R.Ellis}--\textsc{A. Friedman}--
The Asymptotic behavior of the First Real Eigenvalue of Second Order Elliptic
Operators with a Small Parameter in the Highest Derivative,  II \emph{Indiana
Univ. Math. J. 23,  No. 11,  1974} p991-1011.

\bibitem {DF}\textsc{A. Devinatz}-- \textsc{A. Friedman}-- Asymptotic Behavior
of the Principal Eigenfunction for a Singularly Perturbed Dirichlet Problem
\emph{Indiana Univ. Math. J. 27,  No. 1,  1978} p143-157.

\bibitem {DF1}\textsc{A. Devinatz}--\textsc{A. Friedman}--The Asymptotic
behavior of the solution of a singular perturbed Dirichlet problem
\emph{Indiana Univ. Math. J. 27,  No. 3,  1978} p527-537.

\bibitem {Dobro}\textsc{S. Dobrokotov}--\textsc{V. Kolokol'tsov}%
--\textsc{V. Maslov}--,  splitting of the low energy levels of the
Schr\"odinger equation and the asymptotic behavior of the fundamental solution
of the equation $hu_{t}=\frac{1}{2} h^{2} \Delta u -v(x)u$. Theor. and Math.
Phys.,  87, 561-99 (1991).

\bibitem {DoV}\textsc{M. Donsker} \textsc{S. Varadhan} On a variational
formula for the principal eigenvalue for the operators with maximum principle,
Proc. Nat. Acad. Sci. USA,  72,  1975, p780-783.

\bibitem {Fl}\textsc{W. Fleming}-- Controlled Markov Processes,  1986,  Pisa.

\bibitem {FW}\textsc{M. Freidlin}--\textsc{A. Wentzell}--Random Perturbations
of Dynamical Systems,  Springer-Verlag,  1984.

\bibitem {Fr}\textsc{A. Friedman}--The Asymptotic behavior of the First Real
Eigenvalue of Second Order Elliptic Operators with a Small Parameter in the
Highest Derivative,  \emph{Indiana Univ. Math. J. 22,  No. 10,  1973}
p1005-1015.

\bibitem {GT}\textsc{D. Gilbarg }-- \textsc{N. Trudinger }--Elliptic Partial
differentiable equations of second order; Springer-Verlag 1983,  second
edition.

\bibitem {Helffer}\textsc{B. helffer}--\textsc{J. sj\"ostrand}--Multiple Wells
in the semi-classical limit I. Comm. in P.D.E,  9(4), 1984 p337-408.

\bibitem {Helffer2}\textsc{B. helffer}-- Semiclassical analysis for the
Schrodinger operator and applications, Springer Lectures Notes in Math. 1336,
1988.

\bibitem {HK1}\textsc{D. Holcman,  I. Kupka}-- Singular Perturbation and first
order PDE on manifolds. C. R. Acad. Sci. Paris S\'{e}r. I Math. 333 (2001),
no. 5,  465--470.

\bibitem {HK3}\textsc{D. Holcman,  I. Kupka}-- Semi-classical limit of the
first eigenfunction and concentration on the recurent sets of a dynamical
system,  pre-print.

\bibitem {HMS}\textsc{D.Holcman,  A. Marchevska,  Z. Schuss}-- The survival
probability of diffusion with killing,  pre-print.

\bibitem {H}\textsc{C. Holland} --Stochastically perturbated limit cycles,
\emph{J. Appl. Prob. 15,  311-320 } 1978.

\bibitem {K}\textsc{S. Kamin}--Exponential descent of solutions of elliptic
singular perturbation problems,  \emph{Comm P.D.E,  9(2) 1984}p197-213.

\bibitem {K1}\textsc{S. Kamin}-- Singular perturbation problems and
Hamilton-Jacobi \emph{Integral Equations and Operator Theory,  Vol. 9,  1986}
p95-105.

\bibitem {Kifer}\textsc{Y. Kifer-}Stochastic stability of the
topological pressure. J. Analyse Math. 38 (1980),  255--286.

\bibitem {Kifer80}\textsc{Y. Kifer}--On the principal eigenvalue in a singular
perturbation problem with hyperbolic limit points and circles. J. Differential
Equations 37 (1980),  no. 1,  108--139.

\bibitem {Kifer88}\textsc{Y. Kifer}-- Random Perturbations of Dynamical
Systems,  Birkauser,  1988.

\bibitem {Kifer90}\textsc{Y. Kifer}--Principal eigenvalues,  topological
pressure,  and stochastic stability of equilibrium states. Israel J. Math. 70
(1990),  no. 1,  1--47.

\bibitem {Kra}\textsc{M. Krasnoselskii}-- Positive solutions of Operator
Equations,  Noordhoff,  Gromingen,  1967.

\bibitem{Krylov} \textsc{N. Krylov} Lectures on Elliptic and Parabolic
Equations in Holder Spaces (Graduate Studies in Mathematics,  V. 12.

\bibitem {pinsky} \textsc{R. Pinsky}, Positive harmonic functions and
diffusion.
Cambridge Studies in Advanced Mathematics,  45. Cambridge University Press,
Cambridge,  1995.

\bibitem {R}\textsc{C. Robinson}-- Dynamical Systems,  CRC press \emph{1995}.

\bibitem {RS}\textsc{M. Reed,  B. Simon}--Methods of modern mathematical
Physics, Academic Press, 1987.

\bibitem {RW}\textsc{L. Rogers,  D. Williams}--Diffusions, Markov processes
and Martingales, Wiley, 1987.

\bibitem {Schuss}\textsc{Z. Schuss},  \emph{Theory and Applications of
Stochastic Differential Equations},  Wiley Series in Probability and
Statistics. John Wiley Sons,  Inc.,  New York,  1980.

\bibitem {Si}\textsc{B. Simon}-- Semiclassical analysis of low lying
eigenvalues I. Non-degenerate Minima: Asymptotic expansion,  Ann. Inst. Henri
Poincar\'{e},  Vol XXXVIII, 3,  (1983), 295-307.

\bibitem {Simon1}\textsc{B. Simon}-- Semiclassical analysis of low lying
eigenvalues. II tunneling. Ann. of math 120,  (1984),  89-118.

\bibitem {Si2}\textsc{B. Simon}-- Semi-classical analysis of low lying
eigenvalues,  IV. The flea on the elephant,  J. Funct. Anal. 63 (1985),
123-136.

\bibitem {Sm}\textsc{S. Smale}--Differential Dynamical Systems,  {Bull. of
A.M.S},  1967 p747-830.

\bibitem {St}\textsc{D. Stroock}--An Introduction to the Theory of Large
Deviations,  Springer, 1984.
\end{thebibliography}
\end{document}